\documentclass[12pt,preprint]{aastex}     %AASTEX 5
\usepackage{epsfig}
\usepackage{color,colordvi}
\usepackage{graphicx}  
\def \deeg {$^\circ$}
\def \ebv {E(B-V)}
\def \kms {${\rm km~s}^{-1}$}
\def\HII{H$^\mathrm{+}$}
\def\HI{H$^\mathrm{o}$}
\def\NHtwo{N(H$_\mathrm{2}$)}
\def\NHII{N(H$^\mathrm{+}$)}
\def\NHI{N(H$^\mathrm{o}$)}
\def\OI{O$^\mathrm{o}$}

\def\cc{cm$^{-3}$}
\def\cmtwo{cm$^{-2}$}
\def\elat{$\beta$}
\def\elon{$\lambda$}
\def\glat{$b$}
\def\glon{$\ell$}

\def\nHI{n(H$^\mathrm{o}$)}
\def\nHeI{n(He$^\mathrm{o}$)}

\def\nel{n$_\mathrm{e}$}

\newcommand{\BV}[1]{B$_\mathrm{chm}$--V$_\mathrm{chm}~$}
\newcommand{\Bchm}{B$_\mathrm{ism}$}
\newcommand{\Vchm}{V$_\mathrm{ism}$}

\newcommand{\Brot}{$B_\mathrm{rot}$}

\newcommand{\Gmax}{$\Gamma_\mathrm{max}$}
\newcommand{\Gmin}{$\Gamma_\mathrm{min}$}
\newcommand{\Grot}{$\Gamma$(L$_\mathrm{rot}$,B$_\mathrm{rot}$)} 

\newcommand{\HeI}{He$^\circ$}
\newcommand{\Lrot}{$L_\mathrm{rot}$}

\newcommand{\PAgal}{$\theta_\mathrm{gal}$}
\newcommand{\PAi}{$PA_\mathrm{i}$ }
\newcommand{\PAra}{$\theta_\mathrm{ra}$}
\newcommand{\PArot}{$\theta_\mathrm{rot}$}
\newcommand{\PA}{$\theta_\mathrm{PA}$}

\newcommand{\Polpc}{$P_\%$}
\newcommand{\Pol}{$P$}

\newcommand{\absPArot}{$| \theta_\mathrm{rot}| $}
\newcommand{\angmax}{$\alpha _\mathrm{max}$}
\newcommand{\chisqeq}[1]{$\chi^{2}=$ } 
\newcommand{\chisq}[1]{$\chi^{2} ~$} 
\newcommand{\dPA}{$\delta \theta_\mathrm{PA}$}
\newcommand{\dPol}{$dP$}
\newcommand{\glatP}{$b_\mathrm{pole} $} 
\newcommand{\glonP}{$\ell_\mathrm{pole} $} 
\newcommand{\microG}{$\mu$G}
\newcommand{\three}[1]{$\approx  3$} 
\newcommand{\vari}{$var_\mathrm{i}$} 
\newcommand{\var}{$S_{\ell,b}$} 

\slugcomment{}
\shorttitle{Nearby Interstellar Magnetic field }
\shortauthors{Frisch, Berdyugin et al.}

\begin{document}
\renewcommand{\baselinestretch}{1.1}

\title{Mapping the Interstellar Magnetic Field Around the Heliosphere with Polarized Starlight}

\author{P. C. Frisch}
\affil{Department of Astronomy and Astrophysics, University of Chicago, Chicago, IL  60637 USA}
\email{frisch@oddjob.uchicago.edu}
\author{A. B. Berdyugin and V. Piirola}
\affil{Tuorla Observatory and Finnish Centre for Astronomy with ESO, University of Turku, Finland}
\author{A. A. Cole and K. Hill} 
\affil{University of Tasmania, Hobart, Australia}
\author{C. Harlingten}
\affil{Caisey Harlingten Observatory, Norfolk UK}
\author{A. M. Magalh\~{a}es and D. B. Seriacopi and T. Ferrari and N. L. Ribeiro}
\affil{ Inst. de Astronomia, Geofisica e Ci\^{e}ncias Atmosfericas, Universidade de S\~{a}o Paulo, Brazil}
\author{S. J. Wiktorowicz}
\affil{The Aerospace Corporation, El Segundo, CA 90245 USA}
\author{D. V. Cotton\altaffilmark{1}, J. Bailey\altaffilmark{1}, and L. Kedziora-Chudczer\altaffilmark{1}} \affil{School of Physics, UNSW Sydney , NSW 2052, Australia}
\altaffiltext{1}{Australian Centre for Astrobiology, UNSW Sydney , NSW 2052, Australia}

\author{J. P. Marshall}
\affil{Academia Sinica, Institute of Astronomy and Astrophysics, Taipei 10617, Taiwan}

\author{K. Bott\altaffilmark{2}} \affil{NASA Astrobiology Institute Virtual Planetary Laboratory, Box 351580, UW Seattle, WA 98195, USA}
\altaffiltext{2}{University of Washington Astronomy Department, UW Seattle, WA 98195, USA.}

\author{F. P. Santos} \affil{Center for Interdisciplinary Exploration and Research in Astrophysics (CIERA) and Department of Physics
and Astronomy, Northwestern University, Evanston, IL 60208  USA}

\author{C. Heiles} \affil{Department of Astronomy, Campbell Hall, University of California, Berkeley, CA 94720  USA}

\author{D. J. McComas}
\affil{Department of Astrophysical Sciences, Princeton University, Princeton, NJ 08544, USA}

\author{H. O. Funsten}
\affil{Los Alamos National Laboratory, Intelligence and Space Research Division, P.O. Box 1663, Los Alamos, NM 87545}

\author{N. A. Schwadron}
\affil{University of New Hampshire, Space Science Center, Morse Hall Rm 407, Durham, NH 03824, USA}

\author{G. Livadiotis}
\affil{Southwest Research Institute, P.O. Drawer 28510, San Antonio, TX 78228, USA}
\and

\author{S. Redfield} \affil{Astronomy Department and Van Vleck Observatory, Wesleyan University, Middletown, CT 06459-0123, USA}
\pagebreak
\newpage
\begin{abstract}    
Starlight that becomes linearly polarized by magnetically aligned dust
grains provides a viable diagnostic of the interstellar magnetic field
(ISMF).  A survey is underway to map the local ISMF using data
collected at eight observatories in both hemispheres.  Two approaches
are used to obtain the magnetic structure: statistically evaluating
magnetic field directions traced by multiple polarization position
angles, and least-squares fits that provide the dipole component of
the magnetic field.  We find that the magnetic field in the
circumheliospheric interstellar medium (CHM), which drives winds of
interstellar gas and dust through the heliosphere, drapes over the
heliopause, and influences polarization measurements.  We discover a
polarization band that can be described with a great circle that
traverses the heliosphere nose and ecliptic poles.  A gap in the band
appears in a region coinciding both with the highest heliosheath
pressure, found by IBEX, and the center of the Loop I superbubble.
The least-squares analysis finds a magnetic dipole component of the
polarization band with the axis oriented toward the ecliptic poles.  
The filament of dust around the heliosphere and the warm helium breeze flowing through the
heliosphere trace the same magnetic field directions.  Regions along
the polarization band near the heliosphere nose have magnetic field
orientations within 15\deeg\ of sightlines. Regions in the IBEX ribbon
have field directions within 40\deeg\ of the plane of the sky.
Several spatially coherent magnetic filaments are within 15 pc. Most
of the low frequency radio emissions detected by the two Voyager
spacecraft follow the polarization band. The geometry of the
polarization band is compared to the Local Interstellar Cloud, the
Cetus Ripple, the BICEP2 low opacity region, Ice Cube IC59 galactic
cosmic ray data, and Cassini results.
\end{abstract}


\section{Introduction \label{sec:intro} }

Spatial inhomogeneities in the nearby interstellar magnetic field and
gas modulate the heliosphere during the solar motion through space.
The distribution, kinematics and pressure of gas in the cluster of
local interstellar clouds (CLIC) through which the Sun is traveling is
moderately well known, having been studied with interstellar
absorption lines for the past century \citep{Heger:1919}.  Mapping the
magnetic field in the low density very local interstellar medium has
only recently become possible with the development of high-precision
polarimeters.  With the goal of relating the local interstellar
magnetic field to the magnetic field shaping the heliosphere, we are
mapping the configuration of the local interstellar magnetic field
utilizing starlight that is polarized by magnetically-aligned
interstellar dust grains.  The interstellar polarizations discussed in
this paper arise from dust grains located in the
CLIC, some of which are swept up into and around the heliosphere.
This study focuses on the magnetic field configuration within 15 pc.
Interstellar gas in this region is warm, low density, and
kinematically structured with a tendency to flow away from the
direction of the center of the Loop I superbubble
\citep[][FRS11]{Frisch:2011araa}.  The polarized starlight data
utilized here include newly-collected measurements that have been
acquired at eight observatories in both hemispheres.

The relation between the interstellar magnetic field that shapes the
heliosphere and the ambient interstellar magnetic field is key to
understanding the physical properties of the immediate galactic
environment of the Sun.  Polarized starlight has been shown to provide
a viable diagnostic of the orientation and structure of the local
interstellar magnetic field over both large and small spatial scales
\citep[e.g.][]{MathewsonFord:1970,Piirola:1977,Tinbergen:1982,PereyraMagalhaes:2007vela,planetpol:2010,Frisch:2012ismf2,BerdyuginPiirola:2014s1,Wiktorowicz:2015vesta,Frisch:2015ismf3,Frisch:2015fil,Cotton:2016brightS,MarshallCotton:2016hotdustlism,Cotton:2017FGK}.
Stars within 40 pc show interstellar linear polarizations of $\sim
0.0004\%$/pc or less, corresponding
\footnote{An average local linear polarization of 0.0004\% per parsec
  yields \nHI$\sim 0.1$ \cc, based on the relation \Polpc=9\ebv\ for
  percentage polarization \Polpc, color excess
  \ebv\ \citep{Serkowski:1975ebvpol,FosalbaLazarian:2002apj} and
  \NHI+\NHtwo=$5.8 \times 10^{21}$\ebv\ \cmtwo\ mag
  \citep{BohlinSavageDrake:1978}.  Ionization effects must be included
  to obtain accurate ratios \citep{Frisch:2015ismf3}.  Note that the
  polarization pseudo-vector is aligned parallel to the ISMF direction
  \citep{Lazarian:2007rev} so that these conversion formulae provide a
  lower limit to the amount of interstellar gas corresponding to the
  observed polarization strengths. \label{fn:ebv} } to mean densities
of \nHI $\sim 0.08$ \cc, with a tendency for larger polarizations in
southern compared to northern galactic regions
\citep{Piirola:1977,Tinbergen:1982,MarshallCotton:2016hotdustlism,Cotton:2017FGK}.
In this study the magnetic field orientations are explicitly assumed
to be parallel to the position angles of the linearly polarized
starlight, so that possible radiative torques would affect the
efficiency of grain alignment but not the alignment orientation
\citep{HoangLazarian:2016rat}.

Local space is sparsely filled with interstellar material so that
polarizing dust grains may be entrained in multiple magnetic fields
that are foreground to nearby stars.  For volume densities similar to
the densities of the interstellar material entering the heliosphere,
\nHI$\sim 0.2$ \cc\ and \nel$\sim0.07$ \cc, and typical column
densities \NHI$\sim 10^{17.6}$ \cmtwo, and cloud lengths are less than 1
pc (FRS11).  A kinematical model of the CLIC
as consisting of fifteen clouds predicts that less than 19\% of
nearby space is filled with gas \citep{RLIV:2008}.  Interstellar gas
absorption components detected toward individual nearby stars indicate
filling factors of 5\%--40\% \citep{FrischSlavin:2006book}.  These low
cloud filling factors suggest that the polarizing interstellar dust
grains are distributed over less than 20\% of local interstellar
space.

Heliospheric data have recently provided an entirely new perspective
of the CircumHeliospheric interstellar Medium (CHM) forming the
interstellar boundary conditions of the heliosphere.\footnote{The
  need for a new term arises from the different kinematical
  definitions of the Local Interstellar Cloud (LIC) in the literature
  \citep[FRS11,][]{RLIV,GryJenkins:2014}.  In this paper the term
  ``LIC'' will be reserved for the component by that same name in the
  15-cloud model of \citet{RLIV}, while the two velocity groups found
  by \citet{GryJenkins:2014} will be referred to as the CLIC (which
  Gry and Jenkins termed the ``LIC'' in their 2014 paper) and the
  ``Cetus Arc'' (which is the decelerated and possibly shocked gas
  identified by Gry and Jenkins). \label{fn:fnchmlic}} The CHM
establishes the interstellar boundary conditions of the heliosphere,
feeds interstellar gas and dust into the heliosphere, and absorbs the
escaping energetic neutral atoms (ENAs) that were created by charge-exchange
between the solar wind and CHM neutral gas inside of the
heliosphere.  {In situ} measurements of
interstellar \HeI\ in the inner heliosphere by Ulysses and the
Interstellar Boundary Explorer (IBEX) give an upwind direction for the
CHM velocity of \glon=3.7\deeg, \glat=15.1\deeg, and low volume
densities of \nHeI=0.0196$\pm 0.0033$
\cc\ \citep{Schwadron:2015He,WoodMuellerWitte:2017hebreeze}.  The
angle between the CHM and LIC velocity vectors is $4.3^\circ \pm
2.6^\circ$.

An unexpected tracer of the interstellar magnetic field 
embedded in the CHM was provided by the discovery of a ``ribbon'' 
of ENAs forming where the local interstellar magnetic field (ISMF) 
draping over the heliosphere is perpendicular to sightlines
\citep{McComas:2009sci,Schwadron:2009sci,Funsten:2009sci,Fuselier:2009sci}.\footnote{An 
accurate description of the formation of the IBEX Ribbon 
has been elusive, although a secondary charge-exchange mechanism
beyond the heliopause has been favored by MHD heliosphere simulations 
\citep{Zirnstein:2016ismf}.  
The secondary hypothesis is supported by the sensitivity of 
the parent population of the IBEX ribbon ENAs to variations
in solar wind fluxes due to the phase of the solar cycle
\citep{McComas:2017yr7}.} MHD simulations of the ribbon origin
predict that ribbon ENAs are born 10 AU or more upstream of the
heliopause, depending on the ENA energies \citep{Zirnstein:2016ismf}.
The center of the $\sim 73$\deeg-radius circular ENA ribbon provides
the energy-dependent orientation of the magnetic field in the CHM.  
An angle of $88^\circ \pm 3^\circ$ is found between the CHM velocity in the Local
Standard of Rest (LSR), and the interstellar magnetic field direction traced
by the IBEX ribbon, based on in situ \HeI\ data and the magnetic field
orientation of the CHM from the 1 keV ribbon, indicating the CHM
velocity and magnetic field are perpendicular
\citep{Schwadron:2014sci}.
The secondary population of heated interstellar \HeI\ atoms,
consisting partly of interstellar ions that are deflected in the
outer heliosheath, become re-neutralized and are measured in the
inner heliosphere
\citep{KubiakBzowski:2014breeze,Kubiak:2016breeze,Bzowski:2017breeze}.
The direction of this secondary neutral ``breeze'' traces the magnetic
distortion of the heliosphere.

The Sun is at the edge of the CHM in the upwind direction of the
interstellar neutral wind.  Absorption components at the CHM or LIC
velocities are not detected toward 36 Oph in the upwind direction
\citep{Wood:20036Oph}, or toward the nearest star $\alpha$ Cen,
49\deeg\ from the heliosphere nose
\citep{Lallement_etal_1995,Gayleyetal:1997}.  Voyager 1 crossed the
heliopause at 122 AU in August of 2012 and is acquiring \emph{in situ}
data on the interstellar magnetic field
\citep{Burlaga:2014ismf,BurlagaNess:2016ismf}.
\footnote{Voyager 1, launched in 1977, exited the heliosphere and
  entered the outer heliosheath at 121 AU in August 2012.  The Voyager
  1 trajectory is toward a direction that is close to the solar apex
  motions.  A disappearance of heliospheric low energy ions and sharp
  increase in cosmic ray fluxes marked the heliopause crossing
  \citep{StoneCummings:2013sci}.}  A laminar magnetic field with
Gaussian turbulence was detected in the outer heliosheath during
periods outside of solar magnetic storms, and was identified as the
interstellar magnetic field \citep{Burlaga:2014ismf}.  The gradient of
the interstellar magnetic field detected by Voyager 1 during the first
quiet time beyond the heliopause, uninterrupted by solar storms, was
found to converge on the interstellar magnetic field orientation given
by the center of the IBEX ribbon \citep{Schwadron:2015triBismf}.  The
low levels of magnetic turbulence detected in the heliosheath by
Voyager 1 beyond the heliopause at 135 AU is consistent with
Kolmogorov turbulence with an outer scale of 0.01 pc, indicating the
spacecraft and heliosphere are near the cloud edge
\citep{BurlagaFlorinskiNess:2018turb}.

We find here that interstellar dust grains interacting with the
heliosphere affect linearly polarized starlight data.  In situ dust
measurements indicate that interstellar dust interacts with and is
deflected around the heliosphere
\citep[e.g.][]{FrischGruenHoppe:2005,Mann:2010araa}.  The interstellar
gas and dust inside of the heliosphere have the same velocity vectors
\citep{Frisch:1999,KimuraMann:2003clicvel}.  Ulysses and Galileo data
show that grains with the largest charge-to-dust mass ratios are
excluded from the inner heliosphere and indicate gas-to-dust mass
ratios of 192 (+85,-57), vs. typical interstellar values of 100
\citep{Frisch:1999,Krueger:2015,Sterken:2015}.  The grains have
silicate compositions, based on the composition of the gaseous CHM and
\emph{in situ} dust measurements
\citep{SlavinFrisch:2008,Altobelli:2016sci}. Detailed interactions
between interstellar dust grains and the heliosphere depend on the
magnetic phase of the Hale solar cycle due to Lorentz forces on
charged grains in the heliosheath
\citep{Frisch:1999,Landgrafetal:1999,MannCzechowski:2004dustdefl,SlavinFrisch:2012,Sterken:2015,AlexashovKatushkinaIzmodenov:2016dust}.
The spectrum of locally polarized starlight is bluer than that of more
obscured distant stars, suggesting the presence of smaller grains
\citep{MarshallCotton:2016hotdustlism}.

Measurements by Voyagers 1 and 2 inside of the heliosphere identified
low frequency radio emissions upstream of the heliopause that are
associated with disturbances of the interstellar plasma and magnetic
field \citep{GurnettKurthetal:1993,KurthGurnett:2003pos3khz}.  The
heliopause crossing was also marked by the reappearance of low
frequency radio plasma emissions at the plasma frequency of an
interstellar electron population with density 0.08
\cc\ \citep{Gurnett:2013sci}.

Here we use a statistical approach for defining structure in
the ISMF traced by the polarization data.  Earlier efforts
to obtain the orientation of the very local interstellar magnetic field 
from linear polarization data found an ISMF direction consistent with
the IBEX ribbon magnetic field direction, after
omitting a data subset that appears to define a filament of polarizing
dust grains around the heliosphere
\citep{Frisch:2015fil,Frisch:2015ismf3}.  The coincidence of the
magnetic field traced by the dust filament and the direction of the ``warm
breeze'' of secondary interstellar helium discovered by IBEX is
confirmed in this paper.

Achieving our goal of mapping the interstellar magnetic field in the
very low column density of the local interstellar medium is possible
only because of recent advances in precision polarimetry that give
accuracies of parts-per-million
\citep{Magalhaes:1996iagpol,WiktorowiczMatthews:2008polish2,planetpol:2010,Piirola:2014spie,Bailey:2015hippi,BaileyCotton:2017minihippi}.
In this paper we combine older less precise 20th century data
\citep{Heiles:2000} with high precision polarization measurements made
in the 21st century in order to probe the geometrical properties of
the interstellar magnetic field within 15 pc.  These results are built
on combining polarization position angles based on methods introduced
in \citet{Frisch:2016method2}, and on a classic least-squares-fit
analysis.  The analysis in this paper is restricted to stars within
$\sim 15$ pc, where interstellar column densities are low.  

 Data used here are summarized in Section (\S) \ref{sec:data}.
 Polarization measurements of over 270 nearby stars made with the
 DIPOL-2 instrument \citep{Piirola:2014spie} have enabled this survey
 and are being published separately.The methodology for obtaining
 magnetic structure by counting the number of stars that predict
 magnetic fields at each location according to statistical criteria is
 explained in \S \ref{sec:swath}.  Probability constraints permit
 sorting the magnetic structure maps according to whether polarization
 position angles trace magnetic field lines that are either
 quasi-parallel or quasi-perpendicular to the plane of the sky.  A
 band of favored magnetic field directions is found (\S
 \ref{sec:ang360}).  Magnetic structure is sampled for intervals of
 $\pm 90^\circ$ (\S \ref{sec:ang90}), and $\pm 60^\circ$ and $\pm
 45^\circ$ (\S \ref{sec:ang60ang45}).  A least-squares analysis was
 performed on the best data describing the polarization band, yielding
 a dipole component at the ecliptic poles (\S \ref{sec:heiles57}).  A
 previously discovered filament of polarized interstellar dust grains
 \citep{Frisch:2015fil} is re-examined with a least-squares analysis
 (\S \ref{sec:heilesfil}), confirming a magnetic pole aligned with the
 direction of the warm breeze of secondary interstellar \HeI\ (\S
 \ref{sec:heilesfil}).  Maps of mean polarization angles yields angles
 of the magnetic field throughout the sky (\S \ref{sec:meanang}).  The
discussion (\S \ref{sec:discussion}) relates magnetic structure to supplementary
data on the outer heliosheath, including
 Voyager interstellar plasma emissions, the coincidence of the
 magnetic fields traced by the dust filament and warm breeze of secondary
 interstellar helium, as well as Loop I and southern TeV cosmic ray data.  
Concluding
 remarks are presented in \S \ref{sec:conclusion}.  Table 1 summarizes
locations quoted in this paper.  Additional
 information on the statistical distributions, the polarization band
 and supplementary data, the relation between the
 polarization band and Cassini belt, and the spatial
counting of target stars with the most detectable polarizations
is given in Appendices A--D, respectively.

\section{Polarization Data Used in Analysis \label{sec:data}}

This study utilizes linearly polarized starlight as the basis of
determining the structure of the nearest interstellar magnetic field.
Starlight that is linearly polarized while traversing a dichroic
interstellar medium created by magnetically aligned interstellar dust
grains provides a longstanding diagnostic of the interstellar magnetic
field orientation
\citep[e.g.][]{MathewsonFord:1970,Piirola:1973,Serkowski:1975ebvpol}.
Linearly polarized starlight is parallel to the interstellar magnetic
field direction in the absence of a strong local radiation source that
can perturb (or enhance) alignment \citep[see the review of
][]{Andersson:2015araa}.  Linear polarizations are derived from
measurements of the Q and U Stokes parameters, which yield
polarization position angles \PA\ (the angle between the meridian of
the coordinate system and the polarization pseudo-vector) and their
uncertainties \citep[\dPA, e.g.][]{Tinbergen:1982,NaghizadehClarke:1993stat,Plazwiktor}.

Observed orientations of the polarization position angles, \PA,
represent the projection onto the plane of the sky of the
three-dimensional linearly polarized pseudo-vector, and is regarded as
insensitive to the polarity of the interstellar magnetic field.  In
principle the observed polarization strength will decrease with the
angle between the star and the location where the magnetic field is
perpendicular to the plane of the sky. In this study we
utilize polarization position angles to map the structure of the
nearby interstellar magnetic field.  Polarization strengths are not
directly utilized in this study because of the heterogeneous
underlying data that are drawn from diverse sources with variable
sensitivity levels, some collected during the 20th century
\citep{Piirola:1977,Tinbergen:1982,Leroy:1993lism,Heiles:2000}.

The development of high sensitivity polarimeters has allowed the
precise measurements of starlight polarizations arising in the low
column density nearby interstellar material where typically
N(\HI+\HII) $ \sim 10^{17} - 10^{18.5}$
\cmtwo\ \citep{Wood_etal_2005}, corresponding to expected polarization
strengths of 0.0002\%--0.005\% compared to
mean measurement errors of parts-per-million or better.

This study includes essential new data that have been acquired for this survey of interstellar
polarizations in the local interstellar medium using the DiPol-2
polarimeter \citep{Piirola:2014spie}.  This instrument yields high
precision polarization measurements at better than $10^{-5}$ levels.
Three copies of DIPOL-2 have been built and are being used at the UH88
telescope at Mauna Kea Observatory, the T60 telescope at Haleakala
Observatory, and the 1.3 m telescope of the Greenhill Observatory
(H127), University of Tasmania, Australia.  Additional new data have
been obtained with IAGPOL \citep{Magalhaes:1996iagpol} at the Pico dos
Dias Observatory in Brazil.  These new data from the DIPOL-2 instrument
are being published separately (Berdyugin et al., in preparation; Frisch et al.,
in preparation).

Data from the literature on the polarizations of nearby stars are also used in this study
\citep{planetpol:2010,Santos:2011,Frisch:2012ismf2,Frisch:2015ismf3,Frisch:2015fil,Bailey:2015hippi,Wiktorowicz:2015vesta,MarshallCotton:2016hotdustlism,Cotton:2016brightS,Cotton:2017FGK,BaileyCotton:2017minihippi}.
Data from the literature include high precision data 
collected at Lick Observatory in California,
at the Anglo-Australian Telescope at Siding Spring Observatory in Australia,
and at the 14" telescope at the UNSW observatory at the Kensington campus in Australia. 

The analysis in this paper is based on stars with distances within 15
pc, where a star is considered to be within 15 pc if any part of the
parallax error conical section is less than 16.0 pc.  This requirement
is the only distance information utilized in this paper.
Stellar distances are available for all of the stars
included in this analysis, and are determined from Hipparcos parallax data
\citep{Perryman_etal_1997} by utilizing the parallax uncertainties to
identify the distance that divides the volume of the parallax
uncertainty conical section into two equal volumes.  The current
polarization data base consists of measurements of the polarizations
of over 760 stars within 40 pc, of which 134 stars are
utilized in this study.

Newly acquired data for this survey avoid known intrinsically
polarized stars, and binary systems where intrinsic polarization may
be present \citep[see][for a discussion of intrinsic stellar
  polarizations]{Cotton:2017FGK}, however it is likely that
intrinsically polarized stars remain in the data set utilized in this
paper.  In the present analysis it is assumed that intrinsically
polarized stars, if present, would contribute randomly oriented
polarization position angles that would not bias the results.
Polarization strengths are not used directly in this study but will
affect uncertainties on the polarization position angles 
(Appendix \ref{app:stat}).  
The merit function used to determine the probabilities, \Grot, is
based on the probability distribution for polarization position angles
given by \citet[][Appendix A]{NaghizadehClarke:1993stat}.
An upper limit of 3.5 is therefore placed on the probabilities \Grot\
that are incorporated into the calculation of the statistical significance of
individual data points.  This limit, which affects 16\% of the stars
utilized in this study, is required to minimize the influence
of possible intrinsic stellar polarizations that could mimic interstellar
polarizations and bias outcomes.

The physical distribution of the stars in the sky is unrelated to the
magnetic structure derived in this paper. For instance, compare the
distributions of stars in Appendix \ref{app:polband} with the magnetic
structure maps presented here.

\section{Mapping Magnetic Structure with Polarization Position Angles  \label{sec:swath}}

Starlight that becomes linearly polarized while traversing a dichroic
medium formed by magnetically aligned dust grains provides the basis
for determining the structure of the local interstellar magnetic
field.  Magnetic fields in regions devoid of interstellar dust will
not be sampled in this analysis.  Three approaches are used here to map
the structure of the local interstellar magnetic field using
polarization position angle data. The first approach (this section) is
based on counting the number of overlapping polarization position
angle (PPA) swaths that trace a ``true'' magnetic field towards each
location on the sky according to statistical criteria, and normalizing
that count by the number of stars within \angmax\ degrees of that
location, and then characterizing the magnetic field orientation using
patterns of those counts on the sky \citep{Frisch:2016method2}.  The approach imposes either minimum or
maximum limits on the values of \Grot\ that will be
mapped at each location, in an approach that is conceptually analogous to
binning the data according to the statistical probabilities.  
The second approach (\S \ref{sec:heiles}) is
a least-squares analysis that finds the most probable location for the
dipole component of the magnetic poles traced by the polarization
position angles of the stars tracing the polarization band feature (\S
\ref{sec:heiles57}), and the polarization filament (\S
\ref{sec:heilesfil}).  In the third approach we map the mean values of \PArot\
at each position on the sky.

The method presented here for identifying structure in the interstellar magnetic
field is based on the statistical probability that a polarization
position angle will trace a magnetic field directed toward any
location on the sky for a grid of longitudes \Lrot\ and
latitudes \Brot, and then mapping the number of data points
at each grid point that meet the statistical criterion for that map,
after normalization by the total number of data points available
for mapping at that location according to the geometrical constraints.

The statistical probability that a polarization position angle traces
a magnetic field oriented toward \Lrot, \Brot\ is denoted \Grot. 
The value for \Grot\ is derived by first calculating the angle
between the polarization orientation and a meridian that
passes through the location \Lrot, \Brot.
The statistical probability that \PArot\
traces a ``true'' magnetic field direction is then given by the probability
distribution for position angles and measurement uncertainties.  
Appendix \ref{app:stat} shows this
probability distribution for several levels of measurement
uncertainties.

Two types of statistical limits are placed on \Grot\ values that
constrain the plotted PPA values.  The selection of data that will
trace magnetic field directions that are quasi-parallel to the
sightline requires that \Grot, is large so that \PArot\ has a high
probability of tracing a magnetic field with that orientation. The
limits on the values of \Grot\ that are counted and plotted therefore
restrict the orientations of the magnetic field lines that will be
plotted.  Magnetic field orientations that are quasi-parallel
to the sightline are analogous to a dipole
field with an orientation approximately aligned with the sightline 
with respect to the location \Lrot, \Brot.
Orientations quasi-perpendicular to the sightlines are closer to the
plane of the sky.
These different magnetic field orientations are implemented with the
criteria \Grot$>$\Gmin\ for field directions quasi-parallel to the
sightline, and \Grot$<$\Gmax\ for field directions quasi-perpendicular
to the sightline. The values of \Gmin\ and \Gmax\ are selected
to yield samples of the data large enough to
cover all of the sky for the spatial constraints 
imposed on the map construction.  These maps of the number of
stars meeting the assumed statistical criteria then becomes the
diagnostic of structure in the interstellar magnetic field.  The
parameters \Lrot, \Brot\ serve both to define the grid over which the
magnetic field structure is plotted and the pole of the meridian that
must be used for calculating the rotated polarization position angles,
\PArot.  Note that polarization position angles are generally
calculated with respect to meridians of either the equatorial (\PAra)
or galactic (\PAgal) coordinate systems, however they can be rotated
into any spherical coordinate system defined by an arbitrary pole
location \Lrot, \Brot , to yield the value \PArot\ evaluated with
respect to that rotated coordinate system.

A second geometric requirement is imposed that requires stars to be
located within \angmax\ degrees of \Lrot, \Brot\ to be included in the
plots.  The number of data points that statistically qualify to be
plotted is normalized by the total number of data points that satisfy
the geometric criteria, e.g. the number of
stars meeting the statistical criterion at \Lrot,\Brot\
is divided by the total number stars within 60\deeg\ of that
location for the geometric criterion \angmax$<$60\deeg.  
Normalization compensates for the
uneven spatial distribution of stars in the polarization data set, and
has little effect on the \angmax=90\deeg\ maps but is significant for
the higher resolution maps at \angmax=60\deeg\ and 45\deeg\ (\S
\ref{sec:ang60ang45}).

Mapping magnetic structure is implemented by rotating the coordinate
system through all possible poles for \Lrot, \Brot, evaluating
\PArot\ at each location for stars satisfying the geometric
constraint.  In \S \S \ref{sec:ang360}, \ref{sec:ang90}, and
\ref{sec:ang60ang45} data are counted and plotted that meet the statistical
requirement on \PArot\ after normalizing by the number of stars that
meet the geometric constraint.  Values of \PArot\ are small when they
trace true interstellar magnetic field orientations that are nearly
parallel to a meridian of the spherical coordinate system defined by a
pole at \Lrot,\Brot.

Maps built on the statistical requirement \Grot$>$\Gmin\ are
unaffected by data points with low \Pol/\dPol\ for reasonable values
of \Gmin. Maps built on \Grot$<$\Gmax\ would not display correct
magnetic structure for stars that lack foreground interstellar dust
grains.  Therefore, maps built using the condition \Grot$<$\Gmin\ also
require that \Pol/\dPol $>1.9$ in order to avoid confusing magnetic
structure with stars devoid of foreground dust grains.

An example of a probability swath of possible magnetic pole locations
is shown in Figure \ref{fig:dipol}
for the star HD 104304, located $12.9\pm 1.0$ pc away at
\glon,\glat=283\deeg,50\deeg.  
The HD 104304 data were obtained with the DIPOL-2 polarimeter at the T60 telescope at Haleakala.
HD 104304 has a polarization position angle of \PAra=$59.4 ^\circ\pm 4.9 ^\circ$,
which corresponds to a position angle in galactic coordinates of 
\PAgal=$41.5^\circ \pm 4.9^\circ$.
Figure \ref{fig:dipol} and other maps in this paper utilize an Aitoff projection.

Figure \ref{fig:gfact} shows an example of maps of the probability
distribution \Grot, based on the polarization position angles of stars
within 40 pc that trace a magnetic field orientation toward
$\ell=36.7^\circ,~b=57.0^\circ$, corresponding to the 
the weighted mean center of the IBEX ribbon energy bands
\citep[][Table1 ]{Funsten:2013}.  This distribution shows polarization
data for stars within 40 pc that have high, as well as low,
probabilities for tracing the IBEX ISMF direction, indicating that
multiple magnetic field orientations are found.  Appendix
\ref{app:stat} also shows an additional example of the
statistical distributions for the polarization position
angles of stars tracing a field direction at the location \glon=135\deeg\ and
\glat=5\deeg.

The use of polarization position angle uncertainty swaths introduces
several intrinsic biases to the output maps.  Measurements with larger
values of \Pol/\dPol\ will have larger values of \Grot\ for some
locations but will also tend to be counted at fewer grid locations
because the angular extent of the uncertainty swath is smaller.  In
contrast, large numbers of data points with low \Pol/\dPol, and
therefore large angular uncertainty swaths, will enlarge the number of
grid points where the individual data points are counted (according to
requirement \Grot$>$\Gmin) and therefore spatially blur the magnetic
structure.

The extended polarization data base for stars within 40 pc is used to
display the data that trace the IBEX interstellar
magnetic field orientation according to the statistical criteria.  
Figure \ref{fig:ibexswath} gives an example of statistically combining
polarization position angle probability swaths that produce the
magnetic structure obtained in the direction of the center of the IBEX
ribbon.  The plotted probability swaths are those intersecting in the
region outlined by the cyan-colored polygon in Figure \ref{fig:ibexswath}.  
No angular restrictions are placed on the locations of the stars counted 
at each value of \Lrot, \Brot\ in these two figures, where \angmax=360\deeg, 
so that normalization is not needed.  Rotated polarization position angles, 
\PArot, have probabilities \Grot$>$\Gmin\ larger than \Gmin=0.6 (left), or
\Gmin=0.9 (right), for tracing a magnetic field orientation inside 
the box.  The color coding is dynamically generated for each map so
that the small increase in \Gmin\ between maps modifies the color
scheme and tends to give the appearance of more tightly defined
spatial regions with higher probability data points.  
The color-bars in Figures \ref{fig:ibexswath} and \ref{fig:ang360} give the
total number of stars counted into each location according to the
statistical criteria.  The polarization
position angles of the contributing data points are plotted using
black symbols, and the variations in uncertainties for the diverse
data sample is shown by the variations in the angular uncertainties
representing \dPA.

Comparisons between Figures \ref{fig:ibexswath} right and left show
that the lower probability data points trace more extended regions
defined by weaker statistical constraints.  Note that the color-bar
nomenclature in Figure \ref{fig:ibexswath} and other maps may include
the annotations ``n0'', ``n1', ``n2'' or ``n3'.  The value ``n0''
indicates that probabilities less than a \Gmax\ are plotted.  Values
``n1', ``n2'' or ``n3'' indicate, respectively, that \Gmin\ displayed
on the figure is n=1, n=2, or n=3 times an assumed baseline
probability (\Gmin/n).  The values of the map constraints \Gmax\ and
\Gmin\ are arbitrarily selected to provide maps with enough counts to
be useful for color coding, and are shown on the figures.  Smaller
\Gmin\ values produce more extended high-count regions, while larger
\Gmin\ values sharpen the magnetic structure, have lower counts, and
may lead to inadequate sampling of spatial structure.  The practical 
impacts of varying \Gmin\ and \Gmax\ is to change the color-coding 
in the figure and the visual impact of the color-coded magnetic 
pattern (Figure \ref{fig:ibexswath}).

Identifying locations with small or large counts of data that meet the
statistical criterion for tracing true magnetic field directions
provides a diagnostic of magnetic structure that can be applied over
all spatial scales if an adequate sample of data is available.
Considerations when constructing maps from polarization data are that
small values of \Pol/\dPol\ may indicate the lack of foreground dust,
the absence of a magnetic field, and/or a strongly depolarizing
foreground screen.  Polarization data alone do not distinguish between
these possibilities.  This approach is free from prior assumptions
about the physical configuration or origin of the magnetic fields that
align the dust grains aside from the assumption that position angles
are parallel to the intervening interstellar magnetic field
orientation.

For any location \Lrot, \Brot, the number of polarizations that trace
a ``true'' magnetic field orientation tends to be less than the number
of polarizations that do not trace a magnetic field orientation, so
that generally more data are included in the maps constructed with the
criterion \Grot$<$\Gmax\ than for maps constructed for
\Grot$>$\Gmin\ (Figure \ref{fig:gfact}).

The construction of maps with enough counts to be useful requires
angular smoothing of the data.  Smoothing over large angles masks
small scale structure but highlights large-scale features.  Smoothing
is accomplished by allowing all stars within angle \angmax\ (degrees)
of \Lrot, \Brot\ to contribute to the counts.  Maps displaying angular
smoothing over hemispheres (\angmax=90\deeg, \S \ref{sec:ang90}) and
smaller scales (\angmax=60\deeg, \angmax=45\deeg,\S
\ref{sec:ang60ang45}) are shown.  Values of \Gmin\ and \Gmax\ are
selected so that a sufficient number of stars are mapped to provide
useful color coding of magnetic structure and are responsive to the
numbers of counted points that increase with larger angular sampling
intervals.

\subsection{Globally Smoothed Magnetic Structure and Polarization Band (\angmax=360\deeg) \label{sec:ang360}}

The probability that a polarization position angle traces a true
magnetic field orientation is the parameter used in
this study to map the structure of the ISMF within 15 pc.  The first
stage of the analysis is to assume that the interstellar magnetic
field is uniform in the nearest 15 pc, using \angmax=360\deeg, so that
stars throughout the sky are included in the mapping of overlapping
position angle swaths at each location.  Rotated position angle,
\PArot, values are selected to satisfy the criterion \Grot$>$\Gmin=1.0
(Figure \ref{fig:ang360}, top) or \Grot$>$\Gmin=1.5 (Figure
\ref{fig:ang360}, bottom), as the criterion for tracing a true
magnetic field oriented toward each \Lrot,\Brot\ grid point.  An
implicit property of the constraint \angmax=360\deeg\ is that the
resulting magnetic pattern will have mirror symmetry since each
polarization position angle traces two locations in opposite
directions on the sky.  The labels on the color-bars in Figures
\ref{fig:ang360} shows counts of the number of stars with
\Grot$>$\Gmin=1.0 (top), or \Grot$>$\Gmin=1.5 (bottom) of Figure
\ref{fig:ang360}.  The probability distributions in Appendix
\ref{app:stat} suggest that \Grot$>$\Gmin=1.0 will test values of
\absPArot\ that are typically smaller than $\sim 20^\circ$, although
the limiting angle depends on \Pol/\dPol\ for each data point.  The
color-bar in Figure \ref{fig:ang360} is labeled with the counts of
numbers of overlapping position angle swaths at each grid locations.
Color coding levels vary between the maps.  Polarization position
angles of the stars contributing to the \angmax=360\deeg\ maps are
plotted with black symbols with triangular extensions showing the
uncertainties $\pm$\dPA\ projected onto the figures, and the maps are
shown centered on the galactic center and anti-center.  

The magnetic pattern in Figure \ref{fig:ang360} (\angmax=360)
is dominated by a
prominent band where the magnetic field orientation is restricted to
being quasi-parallel to the sightline by the statistical requirements
\mbox{\Grot$>$\Gmin=1.0}, top, and \mbox{\Grot$>$\Gmin=1.5}, bottom.
A great circle that provides a good approximation to the curvature of
the polarization band feature has an axis toward galactic coordinates
\glon=214\deeg, \glat=67\deeg\ and is tilted by 23\deeg\ with respect
to the plane of the Galaxy (see Table 1 for values in equatorial
and ecliptic coordinate systems).  The equator of the great circle 
tracing the polarization band is plotted with a cyan-colored line on 
most figures.

The polarization band is defined by overlapping polarization position
angle swaths and the axis of the great circle that corresponds to the
polarization band configuration does \emph{not} correspond to a
magnetic pole direction (Appendix \ref{app:polband}).  Instead, the
polarization band overlaps the direction of the warm breeze flowing
into the heliosphere (\S \ref{sec:fil}), and tends to separate the
port and starboard
\footnote{A nautical analogy has been used to describe locations in the
heliosphere defined by ecliptic coordinates, and that analogy is
adopted here.  The ``bow'' of the heliosphere is the nose direction
defined by the upwind direction of interstellar neutral \HeI\ gas
flowing into the heliosphere \citep{Schwadron:2015He}.  The
port/starboard sides of the heliosphere indicate the directions of
increasing versus decreasing ecliptic longitudes compared to the central
heliosphere nose direction, and the ``up'' direction refers to the
north ecliptic pole.
\label{fn:nautical}}
sides of the heliosphere at lower
latitudes (see the nose-centered ecliptic projections of Figures \ref{fig:ang45}, \ref{fig:parot2}).

The band is patchy and shows a gap of about 15\deeg\ on the
galactic-west of the heliosphere nose starting at \glon$\sim
243^\circ$ and extending south to \glon$\sim 228^\circ$ (Figure
\ref{fig:ang360}). Such a gap is produced, for the mapping
procedure used here, only if fewer magnetic field-lines
overlap inside of the gap in comparison to adjacent regions (since the
number of overlapping position angle swaths traces the
number of overlapping magnetic field lines according to 
probability constraints).  In the absence of prior knowledge about the true
magnetic field direction that is being traced, it is not possible to 
distinguish between a true gap and 
unknown biases in the data underlying the analysis.  The physical locations
of the underlying stellar data set does not cause the gap (Appendix \ref{app:starnum}).
This gap is located close to the upwind direction
of the flow of the CLIC through the LSR, and is labeled ``W'' for
``wind'' on the figures.  The CLIC flows through the LSR away from the
direction \glon=335\deeg, \glat=$-$7\deeg, with a velocity $-17$
\kms\ (FRS11).  Maps based on smaller \angmax\ values further
constrain the gap (\S \ref{sec:ang60ang45}).  Most locations with high
counts of statistically qualifying position angles in the
\angmax=360\deeg\ maps are located in the polarization band feature
(Figure \ref{fig:ang360}).

\subsection{Hemispheric Smoothing of Magnetic Structure (\angmax=90\deeg)  \label{sec:ang90} }

Mapping counts of overlapping PPA swaths using \angmax=90\deeg\ is
equivalent to counting polarization position angles of stars located
in the same hemisphere as the grid point \Lrot,\Brot.  For smoothing
over subsections of the sky two criteria are used selecting stars that
are counted into the figures, the statistical criterion and the
angular criterion.  The numbers of data points within \angmax\ degrees
of a grid point (\Lrot, \Brot) is more likely to vary across the maps
for smaller smoothing angles because of the inhomogeneous distribution
of the target stars on the sky.  This possible differences in the
number of data points found within \angmax\ of the \Lrot, \Brot\ grid
locations can be compensated for by normalizing the number of
statistically qualifying data points at each grid point with the total
number of stars within \angmax\ of that grid point.  The color-bar
labeling in Figure \ref{fig:ang90} therefore is based on a scale of
0.0--1.0 (as are the later maps that are plotted with a linear color
scale).  The polarization band feature becomes most apparent in the
first and second galactic quadrants (\glon=0\deeg -- 180\deeg) and is
patchy.  
The polarization band feature is asymmetric between Galactic quadrants
I and II (\glon=0\deeg -- 180\deeg) versus III and IV (\glon=180\deeg
-- 360\deeg) in the \angmax=90\deeg\ map, and for this set of 
constraints is not evident for galactic \glon$>$180\deeg,
or equivalently the ecliptic south of the heliosphere nose.

Polarizations included in the angmax=90\deeg\ map have a relatively high
probability, \Grot$>1.0$, of tracing an ISMF near the sightline but
only 27\% of the geometrically qualifying stars satisfy this
statistical criteria as shown in Figure \ref{fig:ang90}. The
probability distributions in Appendix \ref{app:stat} suggest that
probabilities larger than 1.0 typically correspond to
\absPArot$<12^\circ$, and therefore trace magnetic fields oriented
within $\sim 12^\circ$ of the plotted locations, although this value
varies somewhat with \Pol/\dPol\ of the measurement.  Based on this
argument it appears the polarization band is formed of values of
\absPArot\ that are near the radial sightline.  

The gap seen to the right of the heliosphere nose in the \angmax=360\deeg, 
\Grot$>1.00$ galactic projection enlarges to 
over 60\deeg\ wide in the \angmax=90\deeg\ map.  The gap
is evident as large reddish regions with few stars tracing magnetic
field orientations parallel to sightlines.  This gap borders
the upwind direction of the CLIC flow through the LSR (Table 1),
marked by a "W' on figures.  Figure \ref{fig:ang90}, lower right.  For
the ecliptic projection centered on the heliosphere nose, the gap also
coincides with the southern ecliptic latitudes below the heliosphere
nose.  This southern region overlaps the regions of highest magnetic
pressures for heliosheath plasma according to IBEX ENA data
\citep[][Table 1]{Schwadron:2014sep2,McComas:2017yr7}.

\subsection{Smoothing Magnetic Structure over Angular Scales of 60\deeg\ and 45\deeg \label{sec:ang60ang45}}

Different magnetic field orientations can be selected through 
the probability constraints. For the data utilized here, the number of data points 
found to have small probabilities for tracing a magnetic field at 
a value of \Lrot,\Brot\ is typically greater than the number of data points
with large probabilities for tracing a magnetic field at the
location (Figure \ref{fig:gfact} and Appendix \ref{app:stat}).
Smaller spatial sampling intervals can be used when larger numbers of stars qualify statistically for plotting.

Maps are constructed by binning probability distributions using two limits,
 those with
\Grot\ larger or smaller than the limiting value.
Maps constructed with the criterion \Grot$<0.63$ (\S \ref{sec:ang60ang45ang63})
will count larger numbers of stars with magnetic fields favoring an
field orientation near the plane of the sky than maps based on 
the criterion \Grot$<0.1$ (\S\ref{sec:ang60ang45}
because of the peaked probability distribution for \Pol/\dPol $>1.9$ (Appendix
\ref{app:stat}).  Maps built with conditions \Grot$<$\Gmax\ are also restricted to
data where \Pol/\dPol$>1.9$ in order to avoid counting regions where no dust grains are present.
Relative numbers of stars that have values \Grot\ larger (smaller) than the probability cutoffs \Gmin (\Gmax),
then become a diagnostic of the typical value of \absPArot\ in each sightline.
The goal is to obtain the best angular
resolution that maximizes the statistical significance of the derived
magnetic structure and also avoids spatial gaps not sampled by
the available data.  Smaller values of \angmax\ reduce the number of geometrically
qualifying data points at each location.  
Generally, the numbers of data
points traced at each location using the criteria \Grot$<$\Gmax\ tends
to be larger than the numbers of data points with \Grot$>$\Gmin\ at
the same location if \Gmin=\Gmax. The criteria \Grot$<$\Gmax\ 
therefore allows more spatial detail in the plotted magnetic structure.  
The limit \Grot$<$\Gmax\ emphasizes magnetic field directions with
statistically low probabilities of tracing magnetic fields aligned
with \Lrot, \Brot\ and there emphasizes field
field orientations quasi-parallel to the plane of the sky. Fewer
data points are counted generally with the limit
\Grot$>$\Gmin\ (Appendix \ref{app:stat}).
Smaller angular smoothing intervals enable comparisons between
magnetic structure and interstellar and heliospheric phenomena that
are sensitive to magnetic structure.

\subsubsection{Emphasizing Fields Quasi-Parallel to Plane of Sky \label{sec:ang60ang45ang63}}

Figure \ref{fig:ang60norm} shows the importance of normalization on the
plotted magnetic structure and varies the data display to show the
influence of the color scale on perceived magnetic structure.  The unnormalized
top left figure in Figure \ref{fig:ang60norm} has extended reddish
regions with star counts of 10 or below. Variations in the total
number of data points available at each location for the 
geometric criteria set by \angmax\ indicate that the more useful number of
normalized statistically qualifying data points (top right, Figure
\ref{fig:ang60norm}) should be used.  The top left figure counts about eight stars
in the heliosphere nose region that statistically qualify to be counted as for the condition \Grot$<0.63$,
which corresponds to a field orientation with respect to the sightline
of \PArot\ larger than $>19$\deeg\
(Appendix \ref{app:stat}, the exact angle limit depends on \Pol/\dPol).
The top right figure shows that roughly 60\% of the geometrically
qualifying data points have \Grot$<$0.63, and the remaining 40\%
of the data trace an ISMF in the heliosphere nose region  
with \PArot$<19$\deeg, and field lines quasi-parallel to sightlines.
Normalization of the number of
stars meeting the statistical constraint by the total number of stars
meeting the geometric constraint therefore yields a more accurate description of
magnetic structure by avoiding map patterns dominated by the distribution
of the stars.  Counts of stars in the data base that are within 60\deeg\ of each 
position \Lrot, \Brot, for data with \Pol/\dPol$>1.9$, are shown in Appendix \ref{app:starnum}.
There is no relation between the derived magnetic structure in this paper
and the spatial distribution of the stars.  

Color contrasts differ if a logarithmic color scale (to
base 10) is used, as seen in the middle and lower images of Figure
\ref{fig:ang60norm}.  These figures also display supplementary data
for phenomena that are sensitive to local interstellar or heliosphere magnetic field structure,
overlaid on galactic, ecliptic and equatorial projections.  The
normalized figures displayed with a logarithmic color scale 
give different color contrasts that emphasize different
physical structures in the data.

Loop I is plotted with concentric lines in Figure
\ref{fig:ang60norm}.  An elongated green polarization
filamentary feature is present on the east side of the southern part of Loop I,
extending between \glon,\glat$\sim 30^\circ,0^\circ$ to \glon,\glat$\sim 0^\circ,-35^\circ$ 
(east of the heliosphere nose for the plot in ecliptic coordinates).
This polarization filament extends to within
15 pc of the heliosphere and could alternatively be associated
with an extension of the Loop I superbubble to the solar location.
Interior to Loop I magnetic field orientations that are quasi-parallel to the
sightlines are avoided (Figure \ref{fig:ang60norm},
middle left). 

An alternative picture places this same green filament on the
east side of the region of highest heliosheath pressure found by IBEX
\citep{McComasSchwadron:2014V12pressure}.  The green circle in the
lower right panel in Figure \ref{fig:ang60norm} is centered south of
the ecliptic nose at ecliptic coordinates \elon=249\deeg,
\elat=--20\deeg, with a radius of 47\deeg, corresponding to the
central regions of highest heliosheath pressures.  The interior of
Loop I and the region of highest heliosheath pressure can not be
distinguished geometrically in the plane of the sky, most likely
because Loop I has expanded to the solar location, dominates the CLIC
configuration, and dominates the CLIC and LIC kinematics \citep[see
  reviews FRS11,][]{FrischDwarkadas:2018}.

Maps are displayed in galactic,
equatorial, and ecliptic coordinate systems to allow comparison with
other data sets.  \footnote{Figures in galactic coordinates are
  centered on the galactic center and anti-center.  Figures in
  equatorial coordinates are centered on RA=0\deeg\ and RA=180\deeg.
  Figures in ecliptic coordinates are centered on \elon=0\deeg\ and
  heliosphere nose (\elon=255.5\deeg, based on the interstellar wind
  upwind direction, Table 1).}

Several external data sets that are sensitive to local magnetic
phenomena are plotted in Figures \ref{fig:ang60norm}--\ref{fig:ang45}
and discussed in Section \ref{sec:discussion}.  These data include the
contours that outline the Loop I configuration defined by polarization
data \citep{Santos:2011}, low frequency radio emissions measured by by Voyager 1 and
Voyager 2 that originate as plasma emissions in the interstellar gas
upwind of the heliopause
\citep[][]{KurthGurnett:2003pos3khz,Gurnett:2013sci}, the
polarizations of a dust filament around the heliosphere
\citep{Frisch:2015fil}, and the ICECUBE59 cosmic ray point sources.

\subsubsection{Emphasizing Fields Quasi-Parallel to Sightlines \label{sec:ang60ang45p1}}

Broad regions are found in nearby space where magnetic fields tend to be
oriented closer to the sightlines than to the plane of the sky.
A consistent picture is obtained for the polarization band magnetic structure in
the first two Galactic quadrants, \glon=0\deeg--180\deeg, based on
the maps with lower limits imposed on \Gmin, \Grot$>$\Gmin\ (Figures
\ref{fig:ang360}, \ref{fig:ang90}).  The polarization band is a
semi-continuous feature in these maps.  The more restrictive
constraint of \Grot$<$\Gmin=0.1 in Figure \ref{fig:ang60} (\angmax=60\deeg)
and Figure \ref{fig:ang45} (\angmax=45\deeg) selects out the stars for
plotting that have \PArot$>37^\circ$ so the position angles are not aligned
with the sightline.  The result is a patchy magnetic structure
that discriminates between data points on the shoulders of the 
probability distributions (Appendix \ref{app:stat}) along the great 
circle of the polarization band.  In Figure
\ref{fig:ang60} the polarization band great circle is visible as
reddish regions (small counts of \Grot$<$0.1 and \PArot$>$37\deeg data points) 
in Galactic intervals 0\deeg--90\deeg, and as the
border of greenish regions in galactic intervals 180\deeg--270\deeg.
The difference between maps made with \Grot$<$\Gmax=0.63 (Figure
\ref{fig:ang60norm}) and maps made with \Grot$<$\Gmax=0.1 (Figures
\ref{fig:ang60}, \ref{fig:ang45}) is that the \Grot$<$\Gmax=0.1 constraint
counts stars with \PArot$\sim 19^\circ--37^\circ $ that
are not included in the \Grot$<$\Gmax=0.63 counts (again with details sensitive to values of\Pol/\dPol).
Regions where there are few
or no stars with \Grot$<0.1$ (reddish regions of low counts in Figure
\ref{fig:ang60}) will be highlighting sightlines where the ISMF makes
angles typically less than $\sim 37^\circ$ with respect to the
sightline.  For maps constructed with \Grot$<0.63$ the 
reddish regions 
correspond typically to stars with \PArot\ within $\sim 19^\circ$ of the sightline.  
These general characteristics of the statistical
distributions suggest that the differences between mapped values in Figures
\ref{fig:ang60norm} and \ref{fig:ang60} may be attributed to magnetic
field orientations with respect to the sightline of
\PArot$\sim$19\deeg--37\deeg, which are counted in Figure
\ref{fig:ang60norm}, but not counted for \Gmax=0.1 (Figure \ref{fig:ang60})
Evidently the magnetic
field along the polarization band has a tendency to rotate roughly
20\deeg--40\deeg\ away from the sightline in parts of the second
Galactic quadrant.  This result is consistent with the mean polarization
position angles plotted in \S \ref{sec:meanang}.

\section{Least-squares Analysis of the Best-fitting Magnetic Pole \label{sec:heiles}}

Mapping overlapping polarization position angle swaths gives
qualitative but not quantitative results on the direction of the
magnetic field.  An important symmetry of the polarization band data
will be given by the direction of the dipole component of the magnetic
field.  The best-fitting magnetic pole to a set of position angle data
can be found by performing a least-squares fit to sine(\PA).
We find below that the
dominant dipole component of the magnetic fields for two well-defined
subsets of these polarization data, the polarization band (\S
\ref{sec:heiles57}) and the filament stars (\S \ref{sec:heilesfil}),
reflect the geometry of the heliosphere rather than the local
interstellar gas.

A two-parameter least-squares fit is performed on position angles for
subsets of the data in order to determine the most probable location
for the magnetic pole sampled by these data.  The dipole component of
the magnetic field, toward \glonP,\glatP, can be found by minimizing
the sum of the weighted values of sin(\PA) for the stars.  Data points
are weighted using values of \Pol/\dPol, rather than values of
(\Pol/\dPol)$^2$, in order to minimize biases introduced by the
heterogeneous underlying data sample where \Pol/\dPol\ varies
systematically between sets of data.  In addition, \Pol/\dPol\ is
capped at \Pol/\dPol$<$6 to minimize biases from possible unrecognized
intrinsic stellar polarizations.

The expected value for the sine of each polarization position angle
\PAi\ is zero so the chi-squared variance (\vari) of each measurement
\emph{i} is sin(\PAi) for N data points.  The least-squares estimate
of the sky position of the magnetic field will be at the location
\glonP, \glatP\ given by the minimum of the variance of the data
points:
\begin{equation} \label{eqn:heiles}
S_{\ell,b} = \frac{N}{N-2} {\Sigma_\mathrm{i} } 
\frac{  w_\mathrm{i} (sin(PA_\mathrm{i,obs}) -sin(PA _\mathrm{i,\ell,b}))^2 }
{  w_\mathrm{i} }. 
\end{equation}
The location \glonP, \glatP\ of the minimum in \var\ is determined by
constructing a grid on the sky of 360x180 points (the \Lrot,
\Brot\ grid) and evaluating \var\ at each location on the grid.
Contours of \chisq\ are evaluated and plotted for the polarization
band stars (\S \ref{sec:heiles57}) and the filament stars (\S
\ref{sec:heilesfil}), using a red contour to indicate standard
uncertainties that are 68.3\% probable where \chisqeq\ 2.30 (Figures
\ref{fig:heiles57} and \ref{fig:heilesfil}).

\subsection{Quantitative Analysis of the Polarization Band \label{sec:heiles57}}

The polarization band feature shown in Figures \ref{fig:ang360} and
\ref{fig:ang90} arises from a locus of points in the sky where the
interstellar magnetic field has a relatively high probability of being
parallel to the sightlines.  
A subset of data that best traces the polarization band feature is
selected to include stars with probabilities \Grot$>1.5$ for tracing
an interstellar magnetic field orientation that is within 10\deeg\ of
the polarization band (i.e. as given by the equator of a sphere with a
pole located at the axis of the polarization band, \S
\ref{sec:ang360}, Table 1).  Stars that trace the filament of
polarizing dust grains around the heliosphere \citep{Frisch:2015fil}
are excluded from this subset since those polarization position angles
are expected to have a different origin than that of the polarization
band.  Fifty-seven non-filament stars are found to have probabilities
\Grot$> 1.5$ for tracing a magnetic field pole within 10\deeg\ of the
polarization band equator.  Note that the filament stars were not
excluded when the polarization band feature was first identified in
Figure \ref{fig:ang360}, but are excluded here where the origin of the
polarization band is tested.  The data for these 57 stars include
polarization data collected during the 20th century (25 stars) and
data collected in the 21st century.  The diversity of data sources indicates that
instrumental biases are unlikely to dominate the properties of the
polarization band. The polarization band defined by these 57 stars then
provides the basis for a least-squares analysis to determine the
best-fitting magnetic field orientation to the polarization band.

A two-parameter least-squares fit was performed to the polarization
position angles of these 57 stars to determine the most probable
values of the interstellar magnetic pole sampled by
these data.  The location of the minimum in
\var\ (equation \ref{eqn:heiles}) is at ecliptic coordinates 
$\ell, \beta = 247^\circ,~82.2^\circ (\pm
4.5^\circ)$, for the \chisqeq =2.3 fit (red
contours, Figure \ref{fig:heiles57}), 
indicating that the best-fitting magnetic pole
to the 57-star polarization band is within 7.8\deeg\ of
the ecliptic poles (Table 1 gives values for galactic and equatorial coordinate
systems).  Figure \ref{fig:heiles57} shows the
\chisq\ contours of likelihood that a magnetic pole is found at each
location for the 57 stars, in a linear (left) and a stereographic
projection that is centered on the north ecliptic pole (right).  The
equator of the polarization band feature, with respect to the axis of
the polarization band (Table 1), is based on a larger sample of stars
that includes the filament stars and is 17.6\deeg\ away from the
ecliptic poles at the position of closest approach.  The polar
location of the best-fitting magnetic pole to the 57-star sample
of the polarization band suggests strongly that the dust grains are 
aligned with respect to the ISMF that is distorted by interactions 
with the heliosphere.

\subsection{Quantitative Fit to Filament Stars \label{sec:heilesfil}} 

A quantitative least-squares fit was performed to the polarization
position angles of the 13 filament stars originally identified in
\citet{Frisch:2015ismf3}.  The best-fitting magnetic field orientation
found from the least-squares analysis of these filament stars is
toward \glon $ = 5.9^\circ \pm 8.2^\circ$, \glat$ = 17.0^\circ \pm
3.5^\circ$, a result consistent with the earlier value
\citep{Frisch:2015fil}.  Table 1 gives the filament magnetic pole in
ecliptic coordinates.  Figure \ref{fig:heilesfil} shows the
\chisq\ contours for the filament stars in ecliptic coordinates,
centered on the ecliptic nose (left) and in a stereographic projection
centered on the north ecliptic pole (right).  The red contour shows
the \chisqeq\ 68.3\% probability contour.

\section{Mapping Magnetic Field Mean Position Angles}  \label{sec:meanang}

The methodology described in \S \ref{sec:swath} is analogous to mapping values of \PArot\
according to preset probability bins.  An alternate display of 
magnetic field structure relies on plotting mean values of \PArot\ at each
location \Lrot, \Brot\ without imposing constraints on the probability \Grot.
Direct mapping of \PArot\
is useful only if the mapped data are required to have measurable polarization signal
so data are also required to have \Pol/\dPol$>1.9$ for inclusion in figures mapping 
\PArot.  Figures \ref{fig:parot1} and \ref{fig:parot2} show the
mean values of \PArot\ for each location for a sampling radius of $\pm$45\deeg.
Maps are displayed in linear (left columns) and logarithmic (base 10, right columns)
color scales.  Figure \ref{fig:parot1}, top right, where mean position angles
are plotted on a logarithmic color scale, shows that the heliosphere nose
region and the polarization band in the first galactic quadrant are dominated
by \PArot$\sim 25^\circ - 30^\circ$ or less.   The high-pressure region south of the ecliptic
nose is seen in Figure \ref{fig:parot2} to be dominated by mean polarization position
angles of \PArot$>$50\deeg.  The large values for \PArot\ in this
region are expected if these polarizations are influenced by the draping of the
interstellar magnetic field over the heliosphere, since this is the region
of maximum magnetic pressure according to models of the
heliosphere \citep{PogorelovFrisch:2011} and where the IBEX ENA ribbon forms \citep{Zirnstein:2016ismf}. 
The direction of the interstellar magnetic field traced by the IBEX ribbon,
marked by the ``B'', is in a region where \PArot\ varies spatially and tends to be
\PArot$>$50\deeg.  However the smoothing radius for these figures is $\pm 45^\circ$ so
the plotted mean \PArot\ should not be expected to aligned directly with the
magnetic field direction if the field orientation spatially varies as expected during the
interaction with the heliosphere.  The heliosphere nose region clearly stands out
as a region where the magnetic field is quasi-parallel to sightlines.  

\section{Discussion}  \label{sec:discussion}

The patterns of the interstellar magnetic field found for stars within 15 pc show strong
geometric symmetries related to the interaction between dust-bearing
interstellar magnetic fields and the heliosphere, and extended patterns
of field directions with angles larger or smaller than $\sim 19^\circ
- 37^\circ$ with respect to sightlines (depending on probability
selection criteria).  These extended patterns may trace either purely
interstellar field directions or the interaction between the
interstellar magnetic field and the heliosphere.  

\subsection{Polarization Band \label{sec:band} }
The most unexpected feature in this study is the
polarization band created by the overlapping position angle swaths
(Figures \ref{fig:dipol}, \ref{fig:ang360}).  The band was identified
in an unbiased plot of all data that had probabilities larger than
\Gmin=1.5 for tracing an ISMF direction toward the \Lrot,\Brot\ locations.
Selection of data for mapping with the statistical condition \Grot$>$\Gmin=1.5 
restricts plotted data to tracing magnetic field directions that are
nearly parallel to the sightlines and only stars with relatively large
values of \Pol/\dPol~ would qualify for plotting under this condition.

The band is tilted by 23\deeg\ with respect to the galactic plane and has a geometry that
follows a great circle with an axis at \glon=214\deeg, \glat=67\deeg.
The band was discovered and plotted using the statistical constraint
\Grot$>1.5$ that counts values of \PArot\ close to sightlines.  The
best visual rendition of the polarization band is obtained for the
logarithmic color scale for \Grot$<$\Gmax=0.63 and \angmax=60\deeg\ in
Figure \ref{fig:ang60norm}, where the statistical constraint
corresponds to the counting of position angles with \PArot\ larger
than about 19\deeg\ with respect to sightlines \Lrot, \Brot.
Color-coding in this map plots non-compliant data (\PArot\ less than
about 19\deeg) as reddish regions.  Figure \ref{fig:ang60norm} is
dominated by regions with high counts of data that comply with this
constraint (blue-green regions, \PArot$>19^\circ$), including green
filamentary structures where roughly one-third of the data do not
comply with the statistical constraint and have \PArot\ values
oriented quasi-parallel to sightlines.  The band is most prominent in
the galactic interval $\ell=0^\circ - 180^\circ$, corresponding to
regions in the ecliptic north of the heliosphere nose.  

In Figure \ref{fig:ang60norm} the polarization band is visible as a feature with
lower fractions of statistically compliant data for all intervals
along the great circle except for a gap centered in the third galactic
quadrant (or equivalently south of the ecliptic nose).  The origin of
the gap is ambiguous because it corresponds spatially to both galactic
features and heliospheric features.  In a galactic context, the
polarization band gap is centered on the center of the Loop I
superbubble, which models and data indicate has expanded to the solar
vicinity \citep{FrischDwarkadas:2018}.  In the heliospheric context
the band gap is centered on the regions south of the heliosphere nose
where IBEX ENA data show that plasma pressures are highest
\citep{McComasSchwadron:2014V12pressure}, and MHD heliosphere models
find that both plasma and magnetic pressures are highest, with the
larger magnetic pressures centered somewhat south of the maximum
thermal pressure \citep{PogorelovFrisch:2011}.

Mean polarization position angles are determined using a smoothing
radius of \angmax=45\deeg\ in Figures \ref{fig:parot1} and
\ref{fig:parot2}.  The nose-centered ecliptic projection shows 
that the ISMF has angles larger than 40\deeg\ with respect to the
sightline in the region of the IBEX ENA pressure bulge.  This same
region overlaps the location of the IBEX ribbon
\citep{McComas:2009sci} that is predicted to originate with magnetic
field lines draping over the heliosphere so that they are
perpendicular to sightlines \citep{Schwadron:2009sci}.  Throughout the
ecliptic north section of the band, the ISMF is within about
30\deeg\ of the sightline.  

The band symmetry is not likely to be related to galactic phenomena.
The small offset of the Sun above the galactic plane \citep[$\sim
  15^\circ$,][]{Cohen:1995sungalplane} should not influence small
scale structure in the nearby interstellar material or magnetic field.
The inter-arm interstellar magnetic field in the solar neighborhood of
the Galaxy toward \glon=$82.8^\circ \pm 4.1^\circ$
\citep{Heiles:1996curve} has a direction that differs significantly
from the interstellar field orientation shaping the heliosphere based
on the IBEX ribbon center (Table 1).

A possible explanation for the low inclination of the polarization
band feature with respect to the galactic plane is 
that the solar rotational and the ecliptic poles make
small angles with respect to the galactic plane.  The poles of the ecliptic are
tilted by 30\deeg\ with respect to the galactic plane and the northern
pole of the solar rotational axis is located at \glon=94\deeg,
\glat=23\deeg.  The Cassini belt of 5.2--55 keV ENAs also follows the
configuration of the galactic plane but does not align with the
polarization band (Appendix \ref{app:cass}).

The dipole component of the magnetic field traced by the
57 stars that trace the polarization band (Table 1) is within
$7.8^\circ \pm 4.5^\circ$ of the ecliptic poles.  The polarization
band great circle passes within $5.0^\circ \pm 1.4^\circ$ of the
heliosphere nose, within 0.9\deeg\ of the warm breeze direction, and
within 17.6\deeg\ of the ecliptic poles at the locations of
closest approach.  The geometric relation between the
polarization band symmetries and heliosphere features suggest that the
polarization band is related to the interaction of
the heliosphere with dust-bearing interstellar magnetic field lines.  

The polarization band divides the upwind heliosphere between the port and
starboard sides of the heliosphere, and is disrupted in the regions of
high plasma pressure south of the heliosphere nose.  Magnetic
turbulence in these high plasma pressure regions may disrupt the
magnetic field south of the heliosphere nose so the polarization band
is no longer a distinct magnetic feature.  An alternate possibility is
that the field lines are quasi-parallel to the heliopause in the gap
and therefore lack a component that is parallel to sightlines for this region.
This is a configuration that MHD heliosphere models predict
\citep[e.g.,][]{Zirnstein:2016ismf}

The magnetic pole of the polarization band from least-squares fitting
(\S \ref{sec:heiles57}) is located 96\deeg\ away from the central
portion of the region of highest ion pressures found by IBEX.  
The polarization band also has a gap that coincides with this high-pressure region.
IBEX skymaps of ENA distributions allow direct comparison between
the heliosheath regions and the magnetic field traced by
the polarizing dust grains.  ENA maps can be separated into 
two principal components, a globally distributed ENA flux and the
excess ENA fluxes of the IBEX ribbon
\citep{Schwadron:2014sep2}. Globally distributed ENAs arise in
the inner heliosheath regions.\footnote{Outflowing ENAs created by the first
charge-exchange between the solar wind and interstellar \nHI\
can not be measured by IBEX. A second charge exchange
with heliosheath plasma casts ENAs back toward the IBEX detectors for in situ measurements.}
An extended asymmetrical region of maximum plasma pressure was found in
the globally distributed ENAs \citep{McComasSchwadron:2014V12pressure}.  
The high pressure region is centered near \elon=255\deeg,
\elat=--14\deeg, south of the heliosphere nose (Figure
\ref{fig:ang60norm}).  MHD models of heliosheath pressures predict
a maximum for magnetic pressure south of the heliosphere nose and
south of maximum thermal pressure region
\citep{PogorelovFrisch:2011}.   

These geometrical properties suggest the polarization
band is a heliospheric feature, but interstellar origins are 
not ruled out.  We have searched for
possible interstellar features that spatially coincide with the axis of
the polarization band great circle, located at \glon=214\deeg,
\glat=67\deeg.  The first test was to search for a correspondence
between the location of the axis of the polarization band and either
the LSR velocity vectors \citep{FrischSchwadron:2014icns}, or the
heliocentric velocity vectors of clouds in the 15-cloud model
\citep{RLIV:2008}.  If the band arose from an interaction between an
interstellar cloud and the heliosphere, then the heliocentric velocity
(measured in the inertial frame of the Sun) would be relevant, while
if the band is a purely interstellar phenomena the LSR velocity would
be relevant.  However, neither the heliocentric nor LSR velocities of
the 15 clouds match the direction of the axis of the polarization
band.

The next possibility is to search for a geometric coincidence between
the polarization band axis and the location of one of the 15
clouds. This comparison is more successful and the band axis is found
to be directed close to the Leo cloud as defined in the 15-cloud
model.  Both polarization data and hydrogen column density data are
available for the star $\alpha$ Leo (HD 87901, 24 pc).  Two
interstellar components are in front of $\alpha$ Leo.
\citet[][GJ17]{GryJenkins:2017leo} assign one of these components to
the LIC. \citet{RLIV:2008} place this same Leo component in a different
cloud but allow the LIC as a possible assignment.  The total
interstellar hydrogen column density toward $\alpha$ Leo is
\mbox{\NHI+\NHII= $2.83~(+1.18,-0.69) \times 10^{18}$ \cmtwo}.  The
polarization of $\alpha$ Leo has also been measured, with
\Pol=36.7 ppm although a
contribution from rotational flattening of the star is possible
\citep{planetpol:2010,MarshallCotton:2016hotdustlism,Cotton:2016brightS,Cotton:2017natregulus,Cotton:2017natregulus}.
\footnote{The polarization of $\alpha$ Leo has not been utilized for the figures
in this paper because the star is at 24 pc, however we tested the
direction of the polarization position angle of $\alpha$ Leo and find
that it has a probability \Grot$>$1.5 of tracing an interstellar
magnetic field direction that is within 15\deeg\ of the polarization band great circle.}

Utilizing the standard relations between hydrogen column densities and
polarization strengths(footnote \ref{fn:ebv}), the polarization
strength of $\alpha$ Leo predicts column densities of \NHI+\NHII= $2.4
\times 10^{18})$ \cmtwo\ that are close to the measured values.  It is
therefore plausible that the LIC cloud, which contains roughly 70\% of
interstellar gas in the $\alpha$ Leo sightline for the Gry Jenkins
model, is associated with an interstellar disturbance that creates the
polarization band.  

A puzzling and probably coincidental geometry is that the polarization
band great circle passes through the direction of the solar apex
motion,
\footnote{The heliosphere is
  shaped by the ISMF, the Doppler combination of the interstellar wind velocity through
  the LSR and the motion of the Sun toward the ''solar apex direction''.
  The solar apex motion consists of a solar LSR velocity of
$18.1 \pm 0.9$ \kms\ toward \glon=$47.8^\circ  \pm 3.0^\circ $  \glat=$23.7^\circ  \pm 2.1^\circ$
\citep{Frisch:2015ismf3}.  \label{fn:apex}} 
to within $1.4^\circ \pm 3.6^\circ$. The motion
of the Sun through the LSR toward the apex of solar motion could
affect the polarization band configuration if it forms where the
interactions between the solar and interstellar magnetic fields
\citep[as first modeled by][]{Yu:1974} are distorted by the solar apex
motion.

\subsection{Magnetic Field Direction and IBEX Ribbon \label{sec:ribbon}}

The mean magnetic field directions displayed in Figures \ref{fig:parot1}
and \ref{fig:parot2} provide an opportunity to directly compare the
magnetic structure of the IBEX ribbon and the interstellar field
lines traced by the polarization data.  Two ribbon regions are notable in the
IBEX 2.2 keV map showing seven years of data
\citep[Figure 23 in ][]{McComas:2017yr7}.  The first ribbon segment is 
located at ecliptic coordinates \elon=210 \deeg -- 240\deeg, \elat$\sim -30$\deeg.
The second ribbon segment is located at \elon=0\deeg--315\deeg, \elat=30\deeg -- 60\deeg.
Mean polarization position angles are shown for these
regions in the nose-centered ecliptic projection in Figure \ref{fig:parot2}, upper left.
Both of these two IBEX ribbon regions correspond to locations where the mean polarization position angles are
inclined to the sightlines by over $\sim 60^\circ$.  Future higher spatial resolution 
studies of the magnetic field may provide confirmation that the IBEX ribbon is found
in sightlines where the magnetic field is perpendicular to the sightlines.

\subsection{Polarization Band and \BV\ Plane Symmetry of the Heliosphere \label{sec:ns}}

The 40\deeg\ angle between the heliocentric CHM velocity
of interstellar neutrals
(\Vchm) and interstellar magnetic field direction (\Bchm) leads to
differences between the propagation of interstellar neutrals and charged
particles through the outer heliosheath due to Lorentz forces.  A warm
breeze of secondary interstellar \HeI\ atoms, originating mainly with interstellar
helium ions displaced by Lorentz forces,\footnote{Interstellar
  neutrals are ionized by charge-exchange with plasma in the outer
  heliosheath regions to form a new secondary ion population.  Both
  the pristine interstellar ions and the secondary ions will
  charge-exchange with the interstellar neutrals to produce
  ``secondary neutrals''.  Secondary neutrals have small angular
  offsets along the \BV\ with respect to the directions of the primary
  neutrals, which is caused by the actions of the Lorentz force on the
  parent ions \citep{Bzowski:2017breeze}.\label{fnwarmbreeze}} was
discovered by IBEX and confirmed by Ulysses
\citep{KubiakBzowski:2014breeze,Schwadron:2015He,Schwadron:2016oxy,Kubiak:2016breeze,Bzowski:2017breeze,WoodMuellerWitte:2017hebreeze}.
The spatial offsets between the
directions of primary and secondary neutrals are aligned with the \BV\ plane
\citep[Figure \ref{fig:ns}, adapted from][]{Schwadron:2015triBismf,SchwadronMcComas:2017bv},
The magnetic field direction in Figure \ref{fig:ns} (star) is given by MHD
models of the IBEX ribbon \citep{Zirnstein:2016ismf}.  
Primary \HeI\ and \OI\ populations define the inflow direction of
undeflected interstellar neutral.  Secondary \HeI\ neutrals (the warm
breeze) are shifted along the \BV\ plane due to the Lorentz deflection 
of parent ions \citep{Kubiak:2016breeze,Bzowski:2017breeze,WoodMuellerWitte:2017hebreeze}.
The \HI\ population is a mix of of primary and secondary atoms and is shifted along
the \BV\ plane \citep{Lallement:2010soho}.  

The great circle of the polarization band (red line) passes through
the heliosphere nose (making an angle of $5.0^\circ \pm 6.7^\circ$ with
respect to the upwind \HeI\ directions).  The polarization
band also overlaps the direction of the warm breeze of secondary \HeI\ atoms and
the filament ISMF locations. 

The polarization band is canted toward a different direction 
than the B-V plane but appears to have a similar slope as the 
locus of the energy-dependent centers of the IBEX ribbon 
\citep[][five black crosses in Figure \ref{fig:ns}]{Funsten:2013}.  
The lowest energy ribbon data point \citep[at 0.7 keV, see][]{Schwadron:2015triBismf}
deviates slightly from the alignment of the ribbon points (Figure \ref{fig:ns}).  

The dipole component of the magnetic field of the filament of
polarized dust grains (\S \ref{sec:heilesfil}) is not plotted in
Figure \ref{fig:ns} but it coincides with the \HeI\ warm breeze
direction (the angle between the two directions is $5.2^\circ \pm
11.8^\circ$) so that the dust filament also traces the B-V plane.  The
\BV\ plane appears to have an influence on the polarization band
because they cross near the nose and warm breeze directions, but they
trace different Lorentz forces because their slopes differ.

The deflection of interstellar dust grains by heliosheath magnetic
fields is well established both observationally and theoretically,
including a prediction of dust plumes around the heliosheath
\citep{Frisch:1999,Landgraf:2000,Landgrafetal:2000,MannCzechowski:2004dustdefl,Mann:2010araa,SlavinFrisch:2012,Krueger:2015,Sterken:2015,AlexashovKatushkinaIzmodenov:2016dust}.

\subsection{Filament of Polarizing Dust \label{sec:fil}}

The filament polarizations \citep[\S
  \ref{sec:heilesfil},][]{Frisch:2015ismf3,Frisch:2015fil} and the
warm breeze of deflected interstellar neutrals
\citep{KubiakBzowski:2014breeze,Bzowski:2015isn,Kubiak:2016breeze,Schwadron:2016oxy,WoodMuellerWitte:2017hebreeze,Bzowski:2017breeze}
trace the same interstellar upwind direction at the heliosphere nose
(Table 1).  The alignment of these two directions suggests that the
parent interstellar ions of the secondary neutral atoms, and the
charged interstellar dust grains, are trapped in the same interstellar
magnetic field lines that are interacting with the heliosphere.  Both
populations would be guided through the outer heliosheath by
mass-independent interactions with the draped interstellar magnetic
field.

Warm breeze atoms survive to the inner heliosheath where they are
measured by IBEX.  Interstellar dust grains are also measured in the
inner heliosphere so that it is possible some filament dust grains
reach the inner heliosphere, where they would comprise the lower end
of the interstellar dust mass spectrum detected in situ, and have been
predicted by models of the entry and propagation of interstellar dust
through the heliosphere
\citep[][]{Frisch:1999,Landgraf_2000,MannCzechowski:2004dustdefl,SlavinFrisch:2008,SlavinFrisch:2012,Sterken:2013filter,Ma:2013fluffyinism,Krueger:2015}.

The filament polarizations 
\footnote{Evidence for a filament of dust around the heliosphere was
  originally presented by \citet{Frisch:2015fil}, who identified
  sixteen stars, 6--33 pc, with polarization position angles that were
  best fit with an interstellar field orientation at
  \glon,\glat=357\deeg,17\deeg\ ($\pm 11$\deeg). } possibly arise in
the grains trapped in the laminar interstellar magnetic field observed
in the outer heliosheath where Voyager 1 is approaching the interstellar medium
\citep{BurlagaFlorinskiNess:2018turb}, 30\deeg\ north of the
heliosphere nose and in a sightline adjacent to the polarization band.
Weak Kolmogorov turbulence observed by Voyager 1 is consistent
with an outer scale corresponding to a cloud boundary at 0.01 pc.
These low levels of magnetic turbulence in the outer heliosheath
would minimize the disruption of grain alignment. 

MHD models of dust propagation in the heliosphere
predict deflected dust grains in these same locations
\citep{SlavinFrisch:2012}.  Filament star polarizations are plotted in
Figures \ref{fig:ang60} and \ref{fig:ang45}.  Five filament stars in
the ecliptic-north of the heliosphere nose are also located toward the
polarization band.  Eight stars below the heliosphere nose (in
ecliptic projection) have polarizations that veer northwards of the
polarization band.

% current V1 approx \glon$\sim$31.5\deeg, \glat$\sim$30.0\deeg
% nose gL,gB = 3.6594404       15.047772
% ang Voyager nose = sphdist(3.7,15.0,31.5,30.,/deg)= 29.613801

Interstellar magnetic
field lines interacting with the heliosphere have different possible
effects on dust grain polarizations.  Twisted magnetic field lines and
unorganized depolarization screens would have different effects on
linearly polarized starlight.  Thick foreground screens of randomly
oriented dust grains will act to depolarize light. A twisted magnetic
field in a thin low density region will also twist polarization
position angles without disrupting the attachment of the grains to the
magnetic field lines because of the long collisional timescales
required for the grain to sweep up enough gas mass to disrupt the
grain alignment. The low levels of magnetic turbulence found in the
outer heliosheath by Voyager 1 \citep{BurlagaFlorinskiNess:2018turb}
would minimize disruption of grain alignment. 

Polarization strengths can arise from the photon path through
interstellar clouds before reaching the heliosphere, but grains must
stay tightly coupled to the ISMF interacting with the heliosphere to
explain the polarization filament.  The polarization strengths for
nearby stars tend to be comparable to strengths expected from standard
ratios between interstellar \NHI, \ebv, polarization strengths (footnote
1), however neither interstellar ions nor radiative torques
are included in these standard relations \citep{Frisch:2015ismf3}. 

\subsection{Polarizations and the Voyager Low-frequency Plasma Emissions \label{sec:voyager}}

During the years 1992--1994 the plasma wave instruments on the Voyager
1 and 2 spacecraft detected 12 sources of low-frequency radio
emissions formed beyond the heliopause
\citep[see][]{KurthGurnett:2003pos3khz}. Triangulation of source
directions gave the locations of the emitting regions, which were
found to be at 113--139 AU for the primary solutions, placing the
events beyond the heliopause in the upwind regions of the heliosphere.
The plasma oscillation frequency is consistent with electron plasma
oscillations that originate from interstellar gas with density 0.08
\cc\ \citep[][in agreement with models of the interstellar electron
  density at the heliosphere boundary, Slavin \& Frisch
  2008]{Gurnett:2013sci}.  The emissions arise from Langmuir waves
initiated by the effects of the impact of plasma from a solar storm on
the outer heliosheath region.  A magnetic field direction that is not
in the plane of the sky permits the propagation of Langmuir waves
upstream of the heliopause \citep{MitchellCairnsetal:2004}.
\citet{KurthGurnett:2003pos3khz} pointed out that the source regions
of the low-frequency emissions were roughly aligned along the galactic
plane.  (The locations of the \three\ kHz emissions are shown on the figures
with asterisks).  Most of these
emission events (75\%) are located close is to the equator of the
polarization band, and on the port side of the heliosphere nose.  The
emissions on the starboard side of the heliosphere are located above
the polarization band by ten degrees or more.  The low frequency
plasma emissions \emph{avoid} tracing interstellar magnetic field
directions near the plane of the sky for the \Grot$<0.63$ statistical
criteria of Figures \ref{fig:ang60} and \ref{fig:ang45}.  The distance
interval over which the radio emissions were detected
\citep{KurthGurnett:2003pos3khz} is comparable to the distances
predicted by 3D MHD heliosphere models of interstellar grains
deflected around the heliosphere \citep{SlavinFrisch:2012}.  The
geometric coincidence between the polarization band feature and most
of the low frequency plasma emission events suggest they are affected
by magnetic fields that are located in the same region of the outer
heliosphere beyond the heliopause.

The heliosphere nose region clearly stands out as a region where values
of \PArot\ are very small so that the interstellar magnetic field is quasi-parallel
to the sightlines.

\subsection{Impact of Loop I on Local Interstellar Magnetic Field \label{sec:loopI}}

The encroachment of an expanding superbubble on the solar
neighborhood, as indicated by both models and data describing Loop I
\citep[][FD18]{FrischDwarkadas:2018}, should affect the ISMF within 15 pc.
The interstellar magnetic field configuration in
the tangential regions of the Loop I shell is well determined from
polarized starlight that indicate the Loop I perimeter is a
magnetically defined feature \citep{Santos:2011}, which approaches to within 40 pc of the
Sun near $\ell \sim 25^\circ$ if it is spherically symmetric (FD18].  The concentric
circular lines in Figures \ref{fig:ang60} and \ref{fig:ang45} outline
the X-ray feature in the sky that has been proposed 
as an interaction ring between Loop I and the Local Bubble
\citep{Egger:1995}.  The smallest angular sampling used here is $\pm
45^\circ$ (for \Grot$<0.10$, Figure \ref{fig:ang45}) and shows that
nearby regions interior to the innermost shell contour have preferred
interstellar field orientations that are dominated by magnetic field
angles larger than $\sim 37^\circ$ with respect to the sightlines
(dark blue).  The \Grot$<$\Gmax=0.63 
(Figure \ref{fig:ang60norm}) indicates that the dominant blue regions
inside of Loop I shell are within 40\deeg\ of the plane of the sky.
The interior of Loop I and the location of highest heliosheath plasma
pressures extend over similar regions of the sky.

Figure \ref{fig:ang60norm} shows maps of magnetic structure for
\Grot$<0.63$ plotted with a logarithmic color scale.  Those maps show
that an elongated organized green polarization filament is found
within 15 pc that corresponds to the eastern side of the southern part
of Loop I, south of the galactic plane for figures in galactic
coordinates.  Equivalently this filament is east of the heliosphere
nose for plots in ecliptic coordinates.  The alignment of the local
filament with the Loop I superbubble shell suggests strongly that Loop
I has expanded to the solar location.  Previous work has also shown
that polarization strengths rise steadily with distance in this region
\citep{planetpol:2010}.

\subsection{Magnetic Structure and Local Interstellar Medium \label{sec:lic}}

The role that magnetic fields play in the configuration of local
interstellar clouds is not known.  Striations or filaments that
align with the direction of the interstellar magnetic field are common
in high-latitude diffuse interstellar regions
\citep{ClarkPeek:2016fibfilstriation}.  Interstellar gas within 15
parsecs fills less than 20\% of space, and is asymmetrically
distributed around the heliosphere so that filamentary structures are
allowed by the data.  The closest interstellar cloud is the Local
Interstellar Cloud (LIC, see footnote \ref{fn:fnchmlic}), which is
observed mainly towards stars in the galactic anti-center hemisphere
(Figures \ref{fig:ang60}, \ref{fig:ang45}).  The LSR velocities of the
gas and dust flowing through the heliosphere yield an upwind direction
in the direction of the center of the Loop I superbubble
\citep[FRS11,][]{Schwadron:2014sci}.  The angle between the
heliocentric velocities of the CHM 
\citep{Schwadron:2015He}) and the LIC \citep{RLIV:2008}
is $4.3^\circ \pm 2.6^\circ$.  The absence of
kinematical components at the LIC velocity toward the nearest star
$\alpha$ Cen and toward 36 Oph, 7 pc beyond the heliosphere nose
\citep{Wood:20036Oph}, has several possible explanations, including
that the Sun is at the edge of the LIC for the nose direction of the
heliosphere, that the CHM is not part of the LIC, or that rapid speed
of the LIC through the LSR ($17.2 \pm 1.9$ \kms) and the long
collisional time scales (millions of years) decouple LIC neutrals 
from the LIC ions, which couple tightly to the magnetic field.

Note that the LIC border extends to the location of the IBEX
interstellar magnetic field location, shown by the ``B'', on maps
with the highest spatial resolution of 45\deeg\ (Figure
\ref{fig:ang45}) and other maps.\footnote{The LIC configuration
  \citep{RLIV:2008} was defined before the IBEX ribbon was discovered
  \citep{McComas:2009sci} so they are not biased by knowledge of the
  ribbon ISMF direction.}  A transition to magnetic field orientations
that are dominated primarily by angles within $\sim 37^\circ $ of the
radial sightlines (low counts of data with \Grot$<$0.1) is found over
most of the interior of the LIC above galactic latitudes of --60\deeg.
Figure \ref{fig:ang45} shows that the LIC does not extend to
sightlines in front of the interior of the Loop I shell, which is
dominated by high counts of data meeting the statistical condition
\Grot$<0.10$ (green/blue regions).  Results based on maps with
\Grot$<0.63$ are consistent with those based on \Grot$<0.10$.

If the CHM belongs to the LIC then IBEX data show that the LIC and
the interstellar magnetic field 
are perpendicular (\S \ref{sec:intro}).  The interstellar magnetic field
in the upwind direction of the LIC LSR motion should therefore be
dominated by magnetic field directions that are quasi-parallel to the
plane of the sky since the apex direction is 31\deeg\
from the ISMF defined by the IBEX ribbon.  These results on the magnetic field orientation
support such a configuration.  IBEX data show that the interstellar
magnetic field and LSR velocity vector of the CHM are perpendicular,
with an upwind direction toward the center of the Loop I superbubble.
LSR velocities of interstellar absorption lines in nearby stars
indicate an upwind direction of the CLIC toward the Loop I superbubble
(FRS11).  The favored local ISMF direction toward the interior of the
Loop I superbubble shell is quasi-parallel to the plane of the sky
(Figures \ref{fig:ang60norm}--\ref{fig:ang45}), which is consistent
with the perpendicular relation between velocity and magnetic field
direction directly observed by IBEX and favored for LIC and CLIC by
the field directions displayed here.

\subsection{Magnetic Field in BICEP2 Region}
The spatial region searched by the BICEP2 experiment for evidence of
B-mode polarizations of the cosmic microwave background signal
\citep{bicep2:2014}, is outlined with cyan-colored contours in Figures
\ref{fig:ang60} and \ref{fig:ang45}.  The nearest 15 pc of the BICEP2
region contains a high percentage of polarization position angles with
angles larger than 37\deeg\ with respect to the sightlines, according
to the statistical criteria \Grot$<0.1$.  Such a magnetic
configuration would favor the detection of synchrotron emissions
should they arise in this local region and would inhibit the
detectability of Faraday rotations of background sources.

\subsection{Small-Scale Structure in Cosmic Ray Data \label{sec:gcr}}

IceCube IC59 maps of cosmic rays in the TeV--PeV energy range at
declinations less than --30\deeg\ identified regions showing
excesses of cosmic rays over small angular scales, 10\deeg\ or
less \citep{Abbassi:2012icecubeapj,Aartsen_icecube:2016}.  Figures
\ref{fig:ang60}, \ref{fig:ang45} show that the IC59 events arrived
from directions that are outside of the LIC and where
magnetic field directions with angles larger than $\sim$37\deeg\ with
respect to sightlines dominate.  The lack of IC59 sources inside of
the LIC suggests that the LIC impedes the propagation of galactic
cosmic rays to the heliosphere.  It is not possible to make a
clear-cut statement about the magnetic field orientation in the LIC using
Figure \ref{fig:ang45}, with an angular resolution of $\pm 45^\circ$,
since a range of field directions are found inside of the cloud.
Global properties of the magnetic field orientation outside of the LIC show
extended regions where the magnetic field orientation is
quasi-parallel to the plane of the sky.  If the IceCube IC59 tiny
structures originate in the interstellar medium then possible reconnection
between magnetic fields may allow the inward-directing cosmic rays.  The
IC59 source angular widths of 10\deeg\ or smaller are not resolved by
the maps in this paper.

The structure of the magnetic field surrounding the heliosphere
impacts the propagation of PeV cosmic rays into the heliosphere.  The
interstellar magnetic field around the heliosphere will be frozen into
the partially ionized CHM (\nel$\sim 0.08$ \cc, magnetic field
strengths $\sim 3$ \microG, FRS11).  The 15-cloud model lacks
interstellar absorption components at the LIC velocity towards nearby
stars over most of the galactic center hemisphere.  If the CHM is part
of the LIC this indicates a distance to the LIC edge, 0.1 pc, that is
comparable to the gyroradius of 2.5 PeV cosmic rays. The magnetic
structure traced by the local polarization data will influence
trajectories of cosmic rays with energies up to 250 PeV or more.

\section{Summary and Conclusions  \label{sec:conclusion}  }

We have used new data from eight observatories in both hemispheres and
data in the literature on linearly polarized starlight to map the
structure of the local interstellar magnetic field within 15 pc.
Magnetic structure is evaluated two ways, by evaluating sightlines
where multiple stars predict the magnetic fields with the same
orientation using statistical considerations, and with a
least-squares-fit that finds the most probable magnetic field
orientation for subsets of the data.  Magnetic features have been
discovered that can be attributed to the local interstellar medium and
to the interaction between interstellar dust and the heliosphere.
Magnetic structure maps are presented in three coordinate systems,
galactic, ecliptic, and equatorial, to facilitate comparison with
other sets of data.

The initial expectation for this study was that the interstellar polarizations
of nearby star would reveal the structure of the ISMF in the interstellar medium
within $\sim 15$ pc.  This expectation proved to be incorrect and instead the 
most recognizable structures in these data are related to the
heliosphere.

The primary results of this study are:

(1) A prominent polarization feature is found that consists of a
polarization band (PB) that extends throughout much of the sky and displays
magnetic field orientations that are quasi-perpendicular to the plane
of the sky.  The PB is more prominent in the first two Galactic
quadrants and has a gap in the fourth galactic quadrant, corresponding
to south of the
ecliptic nose.  The PB configuration can be described with a great
circle that has an axis toward \glon=214\deeg, \glat=67\deeg, and is
tilted by 23\deeg\ with respect to the galactic plane.  A least
squares fit to the best data tracing the polarization band yields the
dipole component of the magnetic field aligned with \elon= $247.0^\circ,$
\elat=$82.2^\circ~(\pm 4.5^\circ) $, which is close to the ecliptic
poles.  The PB also passes close to the heliosphere nose,
to the dipole component of a magnetic field traced by a dusty
filament, and to the 
direction of the warm breeze of secondary interstellar neutrals
discovered by IBEX.  Plots of the mean magnetic field direction along the
polarization band show values within $\sim 30^\circ$ of the sightlines
in the ecliptic north of the heliosphere nose.

(2) Seventy-five percent of the low-frequency kHz radio emissions
discovered beyond the heliopause by the Voyager spacecraft follow the
alignment of the polarization band.  By coincidence the polarization
band is close to the direction of Voyager I, and close to the
direction of the solar apex motion.

(3) The geometrical relation between the polarization band and
structure of the heliosphere strongly suggests that the polarization
band represents a magnetic symmetry of the heliosphere that is traced
by interstellar dust grains entrained in the ISMF interacting with and
draping over the heliosphere.  The PB separates the flow of dust
between the port and starboard heliosphere flanks.  Similar results
for dust interacting with the heliosphere have been predicted by
several theoretical studies.

(4)  The nose region of the heliosphere stands out clearly as a region
where the magnetic field direction, as traced by rotated
polarization position angles, is quasi-parallel to sightlines.
Mean polarization position angles in this region are within
$\sim 30^\circ$ of sightlines

(5) Two regions selected from the IBEX ribbon configuration
show magnetic field orientations that are inclined by more than
$\sim 60^\circ$ with respect to sightlines.

(6) The magnetic structure of the LIC is not yet clear, but the
boundaries of the LIC tend to correspond to regions where the ISMF is
quasi-parallel to the plane of the sky, i.e. oriented more than
37\deeg\ from the sightlines.  Maps with smaller sampling angles than
the smallest values of $\pm 45^\circ$ in this paper are needed for
more detailed comparisons between the LIC and magnetic structure.

(7) Filamentary magnetic structures that overlap the Loop I shell are
found within 15 pc of the Sun.  The interior of Loop I is dominated by 
magnetic field
directions with angles larger than 37\deeg\ with respect to the
sightlines.  The Loop I interior also coincides with the region of
highest heliosheath pressures identified by IBEX.

(8) Both the heliosphere nose region and a second region near
\glon$\sim 280^\circ$, \glat$\sim 15^\circ$ stand out on the magnetic
structure maps as regions that display predominantly interstellar
magnetic field directions that are within about 19\deeg\ of the
sightlines.

(9) The Ice Cube small scale cosmic ray sources (IC59) are located
outside of the LIC and are found in locations where the dominant ISMF
directions tend to have angles greater than 37\deeg\ with respect to
the sightline.  These cosmic ray sources also tend to be found in the
region of highest heliosheath pressure.  Maps with angular smoothing
smaller than the smallest resolution, $\pm 45^\circ$, used in this
study are needed to identify local magnetic field directions more
precisely.

(10) The axis of the polarization band is directed toward the Leo
cloud.  A possibility for an interstellar origin of the polarization
band is that the band geometry corresponds to the geometry of the
kinematically defined Cetus Ripple found by
\citet{GryJenkins:2014clic}.  However, the LSR velocity of the Cetus
Ripple heliocentric velocity indicates that it
flows away from the Aurigae cloud and is unrelated to the LIC.

(11) The polarization band does not appear to be related to the Cassini ENA
belt, which has a tilt angle of about 30\deeg\ with respect to the
galactic plane and does not follow a great circle.

The polarization band discovered here establishes
that the interaction between the heliosphere
and the interstellar medium creates an ordered magnetic structure that
may cause a weak foreground contamination of astrophysical
data that are sensitive to a magnetized and/or dusty plasma.

We have shown that our statistical filtering of the magnetic field
directions traced by multiple polarization position angles is also
capable of yielding quantitative information about the orientations of
the local interstellar magnetic field.  We have not yet mapped the
extent of the ISMF that shapes the heliosphere.  Sorting interstellar
from the unexpected heliospheric contributions to the observed
polarizations of nearby stars is required before the interstellar
component of these weak polarizations are fully understood.  Future
studies will expand the general picture of ISMF orientations given in
this paper and provide quantitative on the orientations of magnetic
field directions on the sky as traced by polarization position angles.

\acknowledgements PCF, DJM, and NAS are grateful to NASA for
supporting this work through the Interstellar Boundary Explorer
mission as a part of the NASA Explorer Program (NNG17FC93C,
NNX17AB04G).  PCF is grateful for funding from HST-GO-14084.002-A.
AB and VP are grateful to the Institute for Astronomy,
University of Hawaii, for the observing time allocated on the UH88 and
T60 telescopes.  The University of Tasmania Greenhill Observatory has
been supported by Australian Research Council grant LE110100055, the
UTAS Foundation and by the continuing support of Dr. David Warren.
AMM is grateful for support from FAPESP (grant no. 2010/19694-4) and
CNP (Research Grant). DBS, TF and NLR are grateful for support from
CAPES (MSc and PhD scholarships). The authors thank the Director and
staff of the Australian Astronomical Observatory for their support
during the observing runs with HIPPI on the AAT. We also wish to thank
Daniela Opitz and Gesa Gruening for their assistance in making the AAT
observations in 2016 February/March and 2016 June, respectively.  The
work of NAS is also supported partially by NASA SR\&T Grant
NNG06GD55G, and was also supported by the Sun-2-Ice (NSF grant number
AGS1135432) project.

\appendix
\section{Statistical Probability Distribution \label{app:stat}}
The merit function used to determine the probabilities that a
polarization position angle measurement traces a magnetic field in a
``true'' direction is from \citet{NaghizadehClarke:1993stat}:
\noindent \newline
\begin{equation}  \label{eqn:gfact}
G_\mathrm{n}(\theta_{\rm{obs}} ;~\theta_{\rm{o}},P_{\rm{o}} ) ~ =
~\frac{1}{\sqrt{\pi}}   ~ \{ \frac{1}{\sqrt{\pi}} + \eta_{\rm{o}}  ~
\rm{exp} (\eta^2_{\rm{o}} ) ~
[1 + \rm{erf}(\eta_{\rm{o}} )]\}
\rm{exp}({-\frac{P^2_{\rm{o}}} {2}} )
\end{equation}
The observed position angle is $\theta_{\rm{obs}}$, the ``true''
position angle is $\theta_{\rm{o}}$, and $P_\mathrm{o}
=\frac{P_\mathrm{true}}{\sigma}$, mean error $\sigma$=\emph{dP},
$\eta_\mathrm{o}~=~\frac{P_\mathrm{o}}{\sqrt{2}}~\mathrm{cos}~[2(\theta_\mathrm{obs}-\theta_\mathrm{o})]$,
and the Gaussian error function $ erf (Z) ~ = ~ \frac{2}{\sqrt{\pi}}
\int_0^Z {exp}({-t^2})~dt$.  Polarizations and position angles are
derived from the Stokes parameters \emph{Q} and \emph{U}.  The
probability distribution function is non-Gaussian because of the
bivariate nature of the position angle and the fact that polarizations
are always positive while the underlying Stokes parameters can be
positive or negative.  The probability distribution function reverts
to a Gaussian at large probabilities ($>6$).

Theoretical probability distributions (eqn. A1) for \Pol/\dPol=1.9 and
\Pol/\dPol=2.3 are shown in Figure \ref{fig:gfact1p9}, left, together
with the rotated polarization position angles, \PArot, for stars
located within 60\deeg\ of the heliosphere nose (Table 1)
corresponding to the statistical conditions used to construct Figure
\ref{fig:ang60}.  The relative absence of data points with
probabilities \Grot$>0.63$ for tracing an ISMF direction toward the
nose of the heliosphere (Figure \ref{fig:gfact1p9} may indicate the
presence of a depolarizing screen covering the heliosphere nose
region.  Figure \ref{fig:gfact1p9} also suggests that the statistical
conditions \Grot$>$\Gmin=1.0 and \Grot$>$1.5 used for Figure
\ref{fig:ang360} to define directions near the magnetic field
orientation (in green) yield values of \PArot\ that are within
10\deeg--20\deeg\ of the sightlines.  The detailed angle between the
magnetic field orientation and the sightline depends on the
\Pol/\dPol\ of individual stars and will be studied in more detail in
a future study.  The statistical conditions \Grot$<$\Gmax=0.1 used in
Figures \ref{fig:ang60}--\ref{fig:ang45} corresponds roughly to
rotated polarization position angles \PArot\ at each grid point that
have angles that are larger than 37\deeg\ with respect to the
sightlines.

Examples of the probability distributions for \PArot\ are shown for
sightlines toward the nose of the heliosphere and the LSR upwind
direction for the CLIC in Figure \ref{fig:gfact1p9}.  These
distributions only include stars within 15 pc.

\section{Display of Supplementary Data \label{app:polband}}

The polarization band was identified using the entire data set subject
to the probability constraints, inclusive of the filament stars that
trace interstellar dust grains interacting with the heliosphere
\citep{Frisch:2015fil}.  In an effort to understand the origin of the
polarization band, the filament stars were omitted from the
least-squares fitting process in \S \ref{sec:heiles57}.  The
great circle describing the polarization band was obtained from the
entire data set used in this study, without omitting the filament
stars.  The dipole
component of the magnetic pole obtained from the least-squares fit to
the best stars tracing the polarization band (\S \ref{sec:heiles57},
Table 1) is $17.6^\circ \pm 6.6^\circ$ from the polarization
band at the location of closest approach (\glon=83.4\deeg,
\glat=15.5\deeg).  The exclusion of the filament stars from the
least-squares solution for the dipole component of the polarization
band means that the polarization band and fitted dipole component are
based on slightly different subsets of the data that will have
slightly different geometries.  There is no prior reason to predict
that the axis of the polarization band defined by the polarization
position angle swaths should correspond to the location of the dipolar
magnetic field, especially since the overlapping position angle swaths
that create the band do not intersect the polarization band at right
angles (Figure \ref{fig:polband}).  The position angle data that trace
the polarization band are selected as those stars tracing an
interstellar magnetic field falling within 10\deeg\ of the
polarization band great circle.  The polarization band is discovered
in this study and has not been predicted, so that these results can
not be compared to any formation model.

These auxiliary data that
trace structure in the heliosphere are discussed in the text and plotted on many
figures.  Here we show the auxiliary data alone, without data on magnetic structure.
Figure \ref{fig:supl} shows the auxiliary data plotted 
in galactic coordinates, and centered on the center of the galaxy.
Shown on this figure are the LIC \citep[irregular dotted line,][]{RLIV},
the two semi-concentric outlines of Loop I identified by \citet{Egger:1995}
and carefully mapped in polarization data by \citet[][semi-circular red concentric lines]{Santos:2011},
the outline of the low opacity BICEP2 region that has served as a probe of the cosmic microwave
background emission \citep[blue closed curved lines,][]{bicep2:2014}, and the IceCube IC59 measurements
of small scale structure in the TeV cosmic ray data \citep[blue squares,][]{Abbassi:2012icecubeapj,Aartsen_icecube:2016}.  The low frequency kHz emissions detected by Voyager 1 and Voyager 2 are 
plotted as blue asterisks. The three-dimensional positions of these
emissions were found to most likely be located upstream of the heliopause, according to
triangulation arguments applied to the measurements made with the two spacecraft
in different locations
\citep{GurnettKurthetal:1993,KurthGurnett:2003pos3khz}.
The red bars show polarization data that have been identified
as arising from a dusty magnetic filament that is interacting
with the heliosphere \citep{Frisch:2015ismf3,Frisch:2015fil}.  Some of the stars tracing
this magnetic filament are located beyond the 15 pc scope of this paper.
The curved cyan-colored line shows the polarization band and the green curved line shows the location of the ecliptic plane on this galactic projection.

\section{Configuration of Cassini ENA Belt vs. Polarization Band \label{app:cass}}

Cassini mapped global ENAs in the energy range 5.2--55 keV and found a
``belt'' configuration that is parallel to the galactic plane
\citep{DialynasKrimigis:2017nature}.  The belt forms beyond the
heliopause and consists of ENAs created by charge-exchange between
interstellar neutral hydrogen atoms and the inner heliosheath 28--53
keV ion population measured by Voyager 1 and Voyager
2. \citet{Krimigis:2009sci} and \citet{Dialynas:2013} discuss the belt
as a great circle, tilted by about 30\deeg\ with respect to the
galactic plane and centered near ecliptic coordinates of
\elon=190\deeg, \elat=15\deeg\ (\glon=310\deeg,
72\deeg\ in galactic coordinates).  The Cassini ENA belt is thick and
irregular, and disappears at the lowest ecliptic latitudes
\citep{DialynasKrimigis:2017nature} where the mean polarization position
angles tends to be parallel to sightlines (Figure \ref{fig:parot2}.  
The Cassini belt also roughly follows
a circle of 77\deeg\ radius, with an
axis at \glon,\glat=$110^\circ,-67^\circ$ (ecliptic
coordinates \elon= 185\deeg, \elat=--7\deeg). The axes of these
two different descriptions of the Cassini belt are consistent to
within 10\deeg.  It is clear that the
polarization band and the Cassini belt do not share the same geometry
because of their different tilts compared to the galactic
plane (23\deeg\ and 30\deeg\ respectively).

Surprisingly, the axis of the Cassini belt is within 6.2\deeg\ of the
{heliocentric velocity of the Cetus Ripple disturbance
identified by \citep{GryJenkins:2014clic}.  Heliocentric velocities
are suitable for describing physical interactions that take place in
the inertial frame of the solar system. The fact that the Cassini belt
has an axis near the Cetus Ripple heliocentric velocity vector allows
the possibility that the shock proposed by Gry and Jenkins to explain
the Cetus Ripple could also be influencing the heliosphere.

\section{Distribution of Stars within 15 pc \label{app:starnum}}

The results reported here on the orientations of the interstellar
magnetic field are not the result of the
distribution of the stars from which these conclusions are drawn.
Figure \ref{fig:starnum} shows the 
number of stars that are within 60\deeg\ of each location \Lrot, \Brot\
and that have \Pol/\dPol$>1.9$.  The figure is in equatorial coordinates.
The north/south pattern of the distribution of stars is partly a reflection
of the different times of the year for good observing conditions 
in the northern versus southern hemisphere, and is unrelated to the magnetic
structure traced by these linear polarization data. 
%%\input{figure17}
%\clearpage
%\pagebreak
\newpage
%%  TABLE 1   begin{table
\begin{deluxetable}{lcccl}
\tablecaption{Summary of Interstellar Directions  \label{tab:summary}}
\tablewidth{0pt} 
\rotate
\footnotesize{\tiny}
\tablehead{\colhead{Observable} & \colhead{Direction\tablenotemark{(A)}} & \colhead{Direction} &  \colhead{Direction} & \colhead{Ref.\tablenotemark{(B)}} \\
\colhead{} & \colhead{\glon,\glat\ (deg)} &  \colhead{$\lambda,\beta$ (deg)} & \colhead{RA,DEC (deg)} & \colhead{} }
%%%
\startdata 
Axis of polarization band & 214, 67 & 158, 17 & 167, 24 & 1 \\   %%checked

%% checked Sun Apr 29 22:12:38 CDT 2018 euler,247.0,82.2,a,b,5 & print,a,b
Polarization band dipole component & $87.9, 32.6 (\pm 4.5)$ & 247.0, 82.2 $(\pm 4.5)$ & 264.0, 59.2 ($\pm 4.5)$ & 1 \\

Filament direction & $5.9 \pm 8.2$, $ 17.0 \pm 3.5$ & $255.0 \pm 7.0$, $7.9 \pm 8.5$ & 348.4, 17.9 $(\pm 8.9)$ & 1 \\
IBEX Ribbon ISMF\tablenotemark{(C)} & $26.0 \pm 0.7$, $50.1 \pm 0.6$ & 227.3, 34.6 ($\pm 0.9$)  & 234.4, 16.3  &  2 \\
IBEX interstellar \HeI nose\tablenotemark{(D)}  &  $3.7 \pm 0.9$, $15.1 \pm 1.3$  &  
	$254.9 \pm 1.5$, $5.12 \pm 0.27$  & $254.9 \pm 1.5$, $-17.6 \pm 0.4$ &  3 \\ 
IBEX Warm \HeI\ breeze & 8.0, 21.8 $(\pm 7.8)$  & 251, 12.0 $(\pm 7.8)$ &   251.7,--10.3 $(\pm 7.8)$ &  4 \\
Heliosphere Maximum Pressure & 347, 6 &  255, --14 & 252, --36   &  5 \\

Upwind CHM HC direction\tablenotemark{(E)} & $3.7 \pm 0.9$, $15.1 \pm 1.3 $ & $255.6 \pm 1.4$, $5.1\pm 0.27$ & $254.9 \pm 1.5 $, $17.6 \pm 0.4$   & 6 \\
%%% 321.25652      -1.3364932      -17.067755

Upwind CLIC LSR direction &  $335.6 \pm 13.4$, $-7.0 \pm 9.0 $ & 260.3, -30.1 ($\pm 16.1$) & 255.9, --53.0 ($\pm 16.1$ & 7 \\
\enddata 
\tablenotetext{(A)}{Coordinates and uncertainties are given in the coordinate system degrees, 
with the exception of the uncertainties on the
ecliptic coordinates of the 57 star polarization band that represent
the radius of the error circle.  The galactic coordinates are designated \glon,\glat\ and ecliptic coordinates are designated
$\lambda,\beta$ in this paper.}
\tablenotetext{(B)}{
1.  This paper ;
2. \citet{Zirnstein:2016ismf} 
3.  \citet{Schwadron:2015He};
4.  \citet{Kubiak:2016breeze}, \citet{Bzowski:2017breeze};
5.  Central location of the region of highest-pressure plasma protons that
form the ENAs observed by IBEX \citet{Schwadron:2014sep2}.
6.  \citet{FrischSchwadron:2014icns}
7.  \citet{Frisch:2015ismf3}
}
\tablenotetext{(C)}{Based on MHD modeling of \citet{Zirnstein:2016ismf}
and a magnetic field strength $\sim 2.9 \pm 0.1 $ \microG.} 
\tablenotetext{(D)}{The upwind direction of the interstellar \HeI\ wind
is generally used for defining the heliosphere nose, although regions
of higher pressure are found south of the nose
\citep{PogorelovFrisch:2011,McComasSchwadron:2014V12pressure,McComas:2017yr7}.}
\tablenotetext{(E)}{"HC" is the heliocentric velocity.  The upwind CHM LSR direction and velocity are
\glon,\glat=321.3\deeg,1.3\deeg, --17.1 \kms.  }
%% 321.25652      -1.3364932      -17.067755}
\end{deluxetable}

	%Sun Jan 28 18:05:54 CST 2018
	%tool_euler_getunc.pro
	%Answ:  Filament stars: Lout,dL,Bout,dB =   254.96     6.92     7.89     8.52
	%Answ:  57 stars: Lout,dL,Bout,dB =  249.44    32.57    82.28     1.73
\clearpage
\pagebreak
\newpage
\begin{figure}[t!]  %% FIGURE 1
\plotone{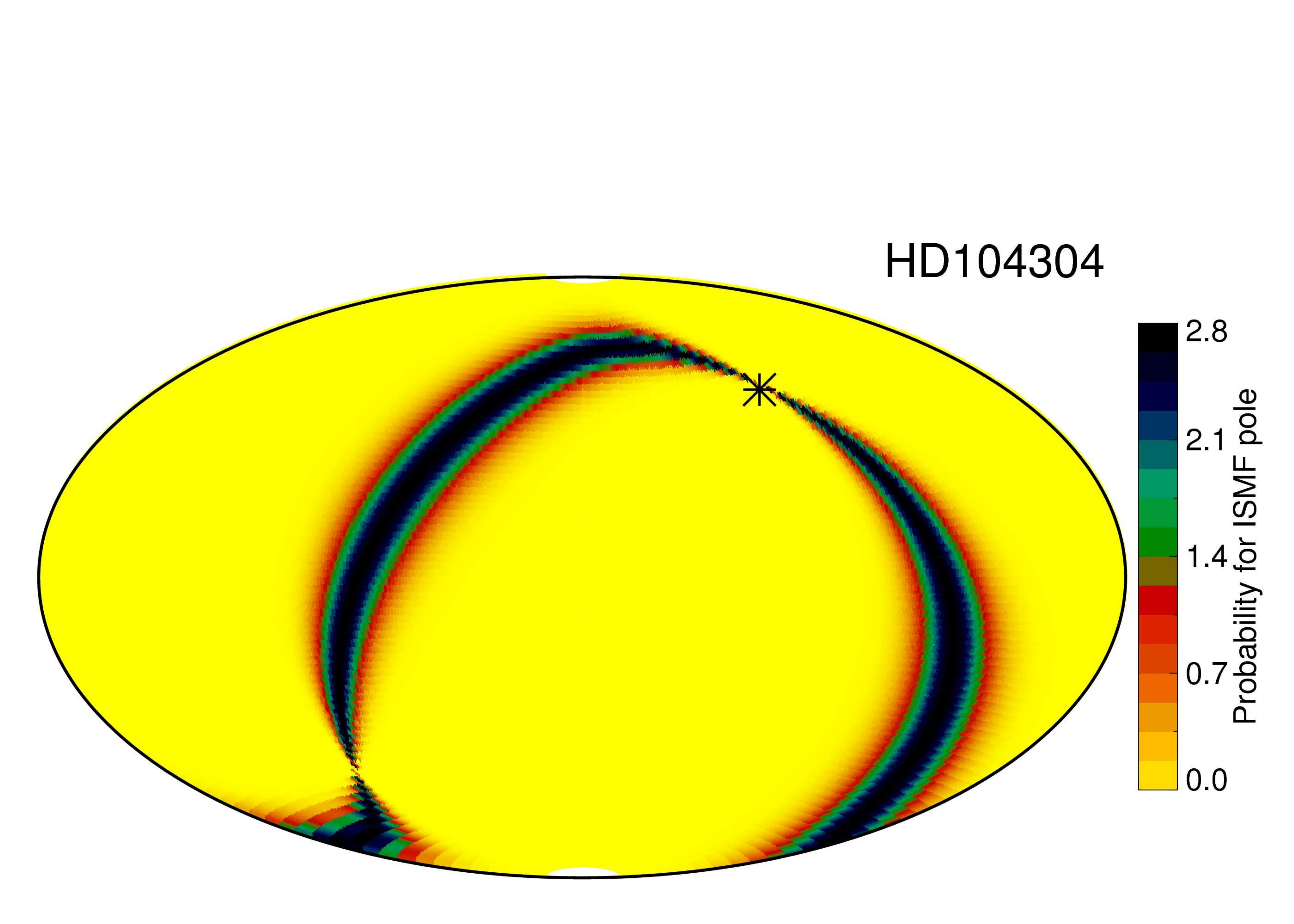}
\caption{The color scale displays values of the probabilities, \Grot,
  for measured polarization position angle, \PA, of the star HD
  104304 to trace a magnetic field located at each position on the sky,
  \Lrot,\Brot. 
The non-zero values of
  \Grot\ create the probability swath that will be sampled according to
  the probability constraints imposed on each calculation.  HD 104304
  is 13 pc away and located at \glon,\glat=283\deeg,50\deeg. 
DIPOL-2 measurements determine a polarization strengths of \Pol=$0.0033 \pm 0.0006 $ percent.
\Grot\ is zero at the position of the star because linear polarizations
are parallel to the ISMF direction and can not be measured where the
ISMF is inclined by 90\deeg\ with respect to the plane of the sky, so that
the interstellar linear polarization vanishes for a star located at the magnetic pole.
These data were acquired with DIPOL-2 mounted at
  T60 at the Haleakala Observatory.  This figure and other maps in
  this paper are displayed using an Aitoff projection.}
\label{fig:dipol}
\end{figure}

\begin{figure}[t!]   %% FIGURE 2 
\plotone{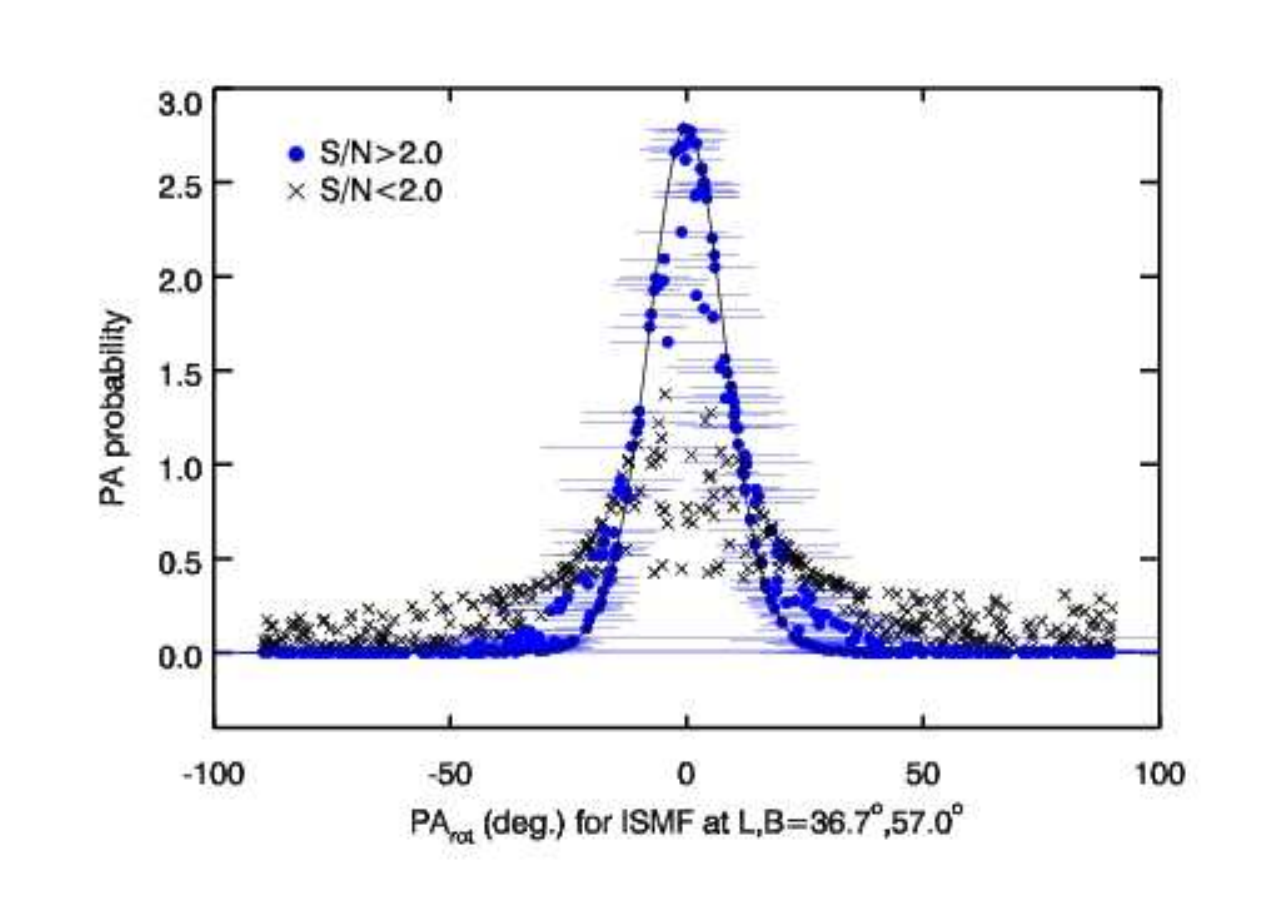}
\caption{The probabity distribution $G_\mathrm{n}(\theta_{\rm{obs}}
  ;~\theta_{\rm{o}},P_{\rm{o}} )$ (eqn. A1) is shown for an ISMF located at the direction
  \glon=36.7\deeg,\glat=56.0\deeg, for polarization data of stars within 40 pc.  This
  direction corresponds to the weighted mean value of the IBEX ribbon
  center (Table 1).  Data with \Pol/\dPol$>2.0$ ($<2.0$) are plotted
  with dots (X's), respectively.  The solid line indicates the
  underlying statistical probability distribution given by
  \citet[][and \ref{eqn:gfact}]{NaghizadehClarke:1993stat}. }
\label{fig:gfact}
\end{figure}

\begin{figure}[ht!]  % FIGURE 3 
\plottwo{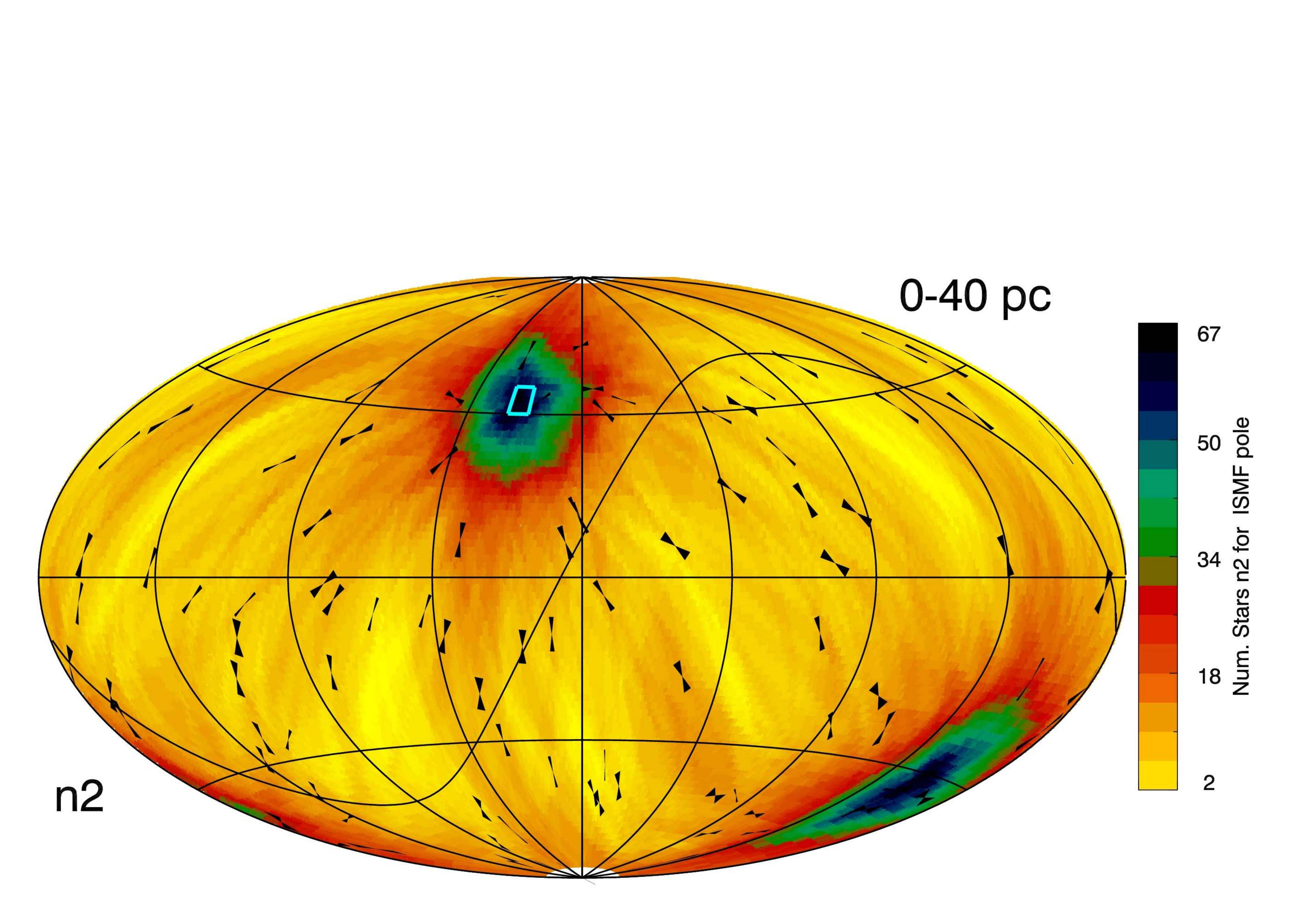}{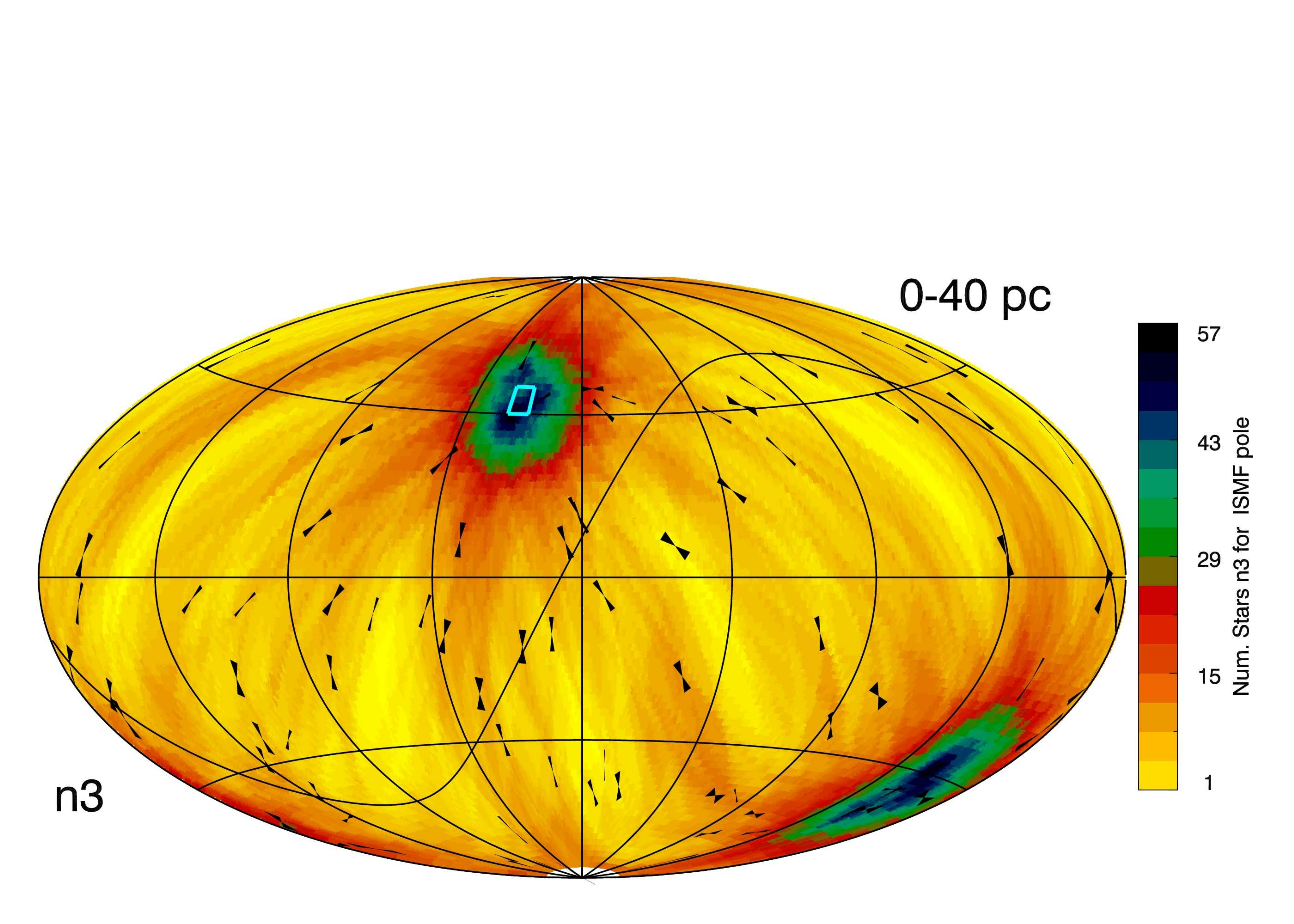}
\caption{Examples of the construction of magnetic structure maps.
  Polarization data with values of \PArot\ satisfying the
  probability criteria \Grot$>$0.6 (left) or \Grot$>$0.9 (right) for
  tracing an ISMF direction located within the cyan-colored boxes are
  counted and mapped using the color-coding shown in the
  color-bar. Maps are centered on the galactic center with longitude
  increasing toward the left. The cyan-colored box is centered on the
  interstellar magnetic field direction affecting the IBEX ribbon
  configuration, \glon=$26.0^\circ \pm 0.7^\circ$, \glat=$50.1 \pm
  0.6^\circ$, as found by the \citet{Zirnstein:2016ismf}
  ribbon-formation simulations.  Each measured polarization position
  angle creates a swath on the sky of magnetic field directions
  consistent with observed position angles \PA$\pm$\dPA\
(see Figure \ref{fig:dipol}).  The black
  symbols indicate the directions of the polarization position angles
  and their uncertainties.  The triangular uncertainties of the
  symbols illustrate the widths $\pm$\dPA\ as plotted onto the Aitoff
  projection.
\label{fig:ibexswath}}
\end{figure}

\begin{figure}[h!] % FIGURE 4 
\plottwo{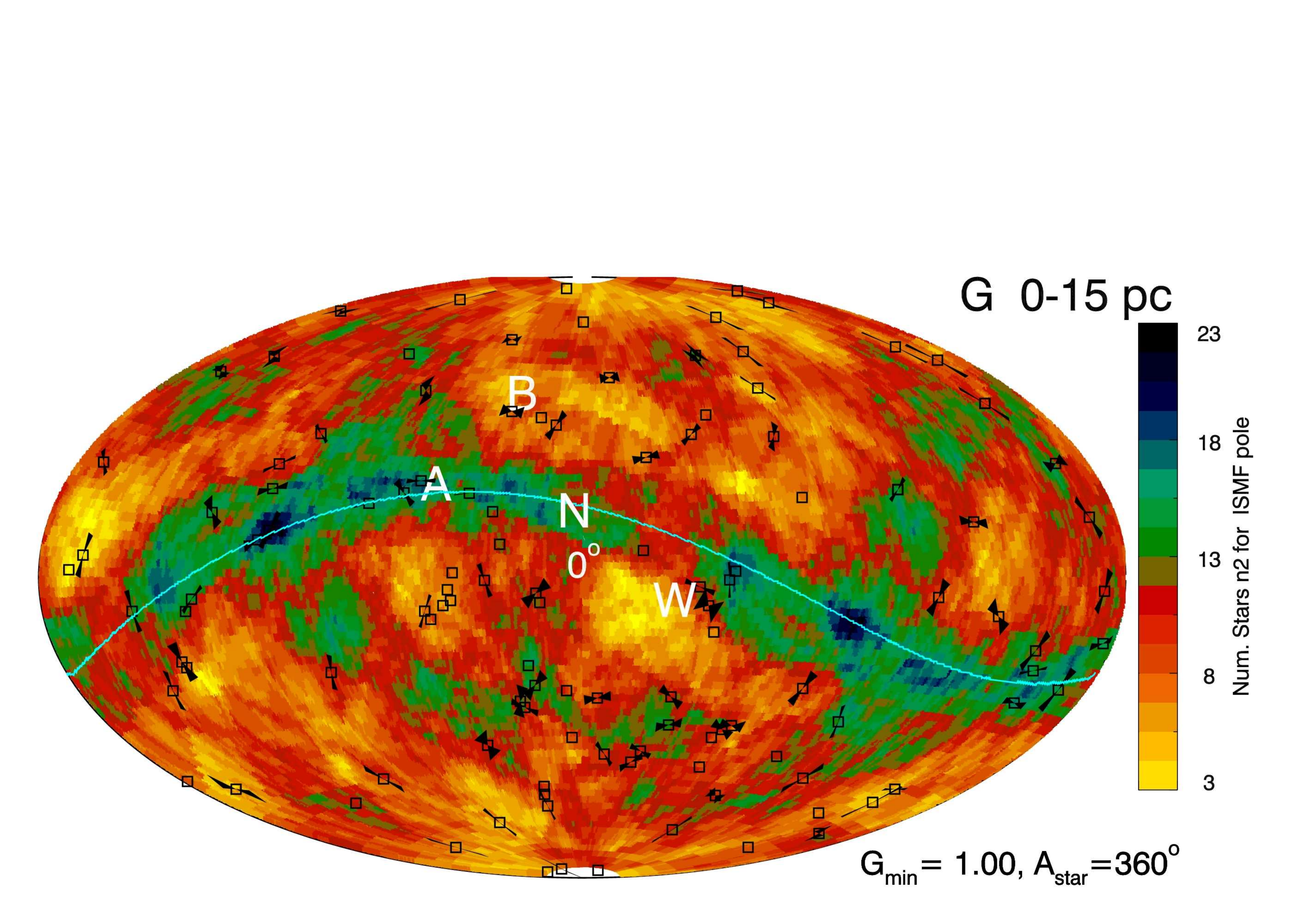}{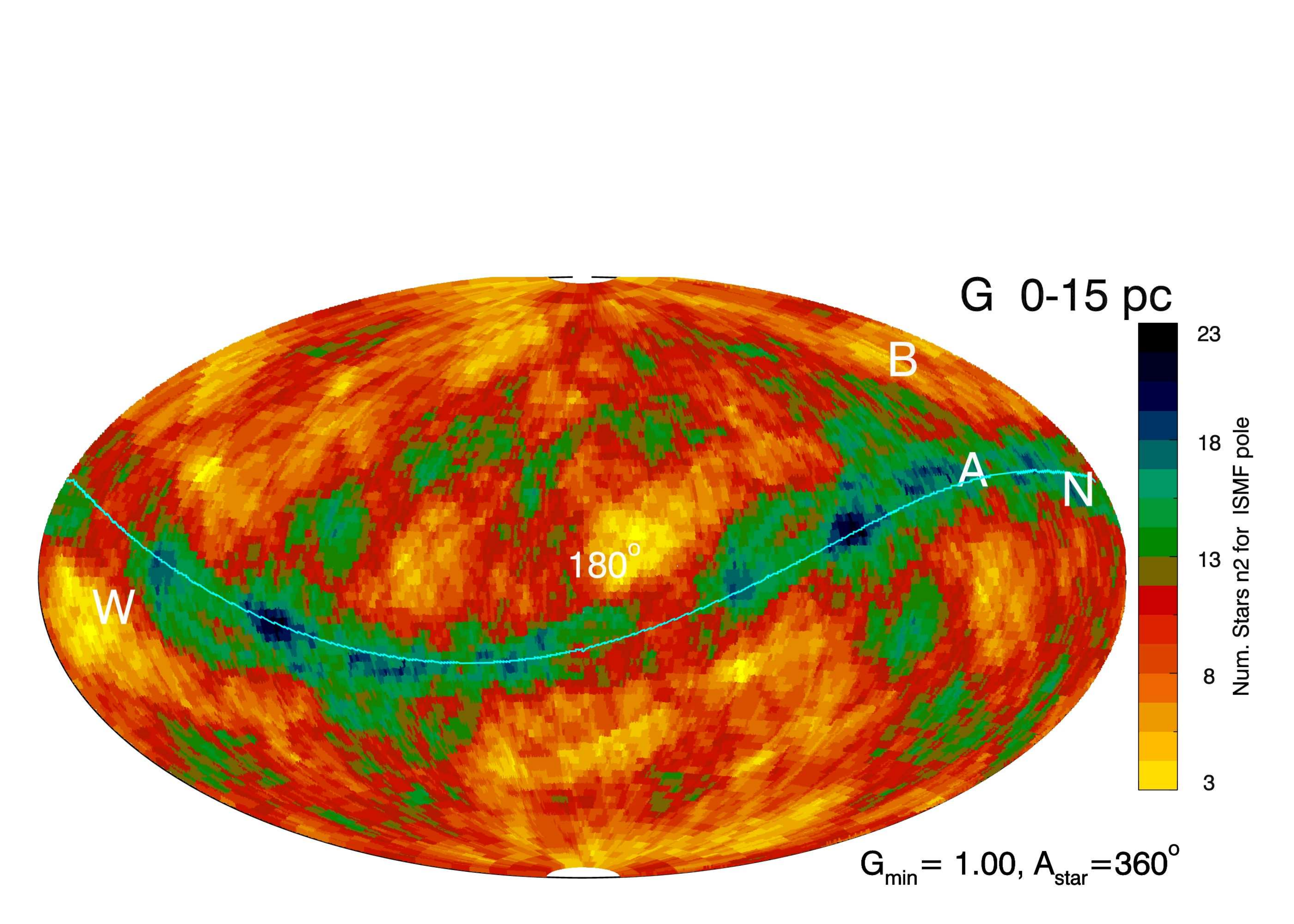}
\plottwo{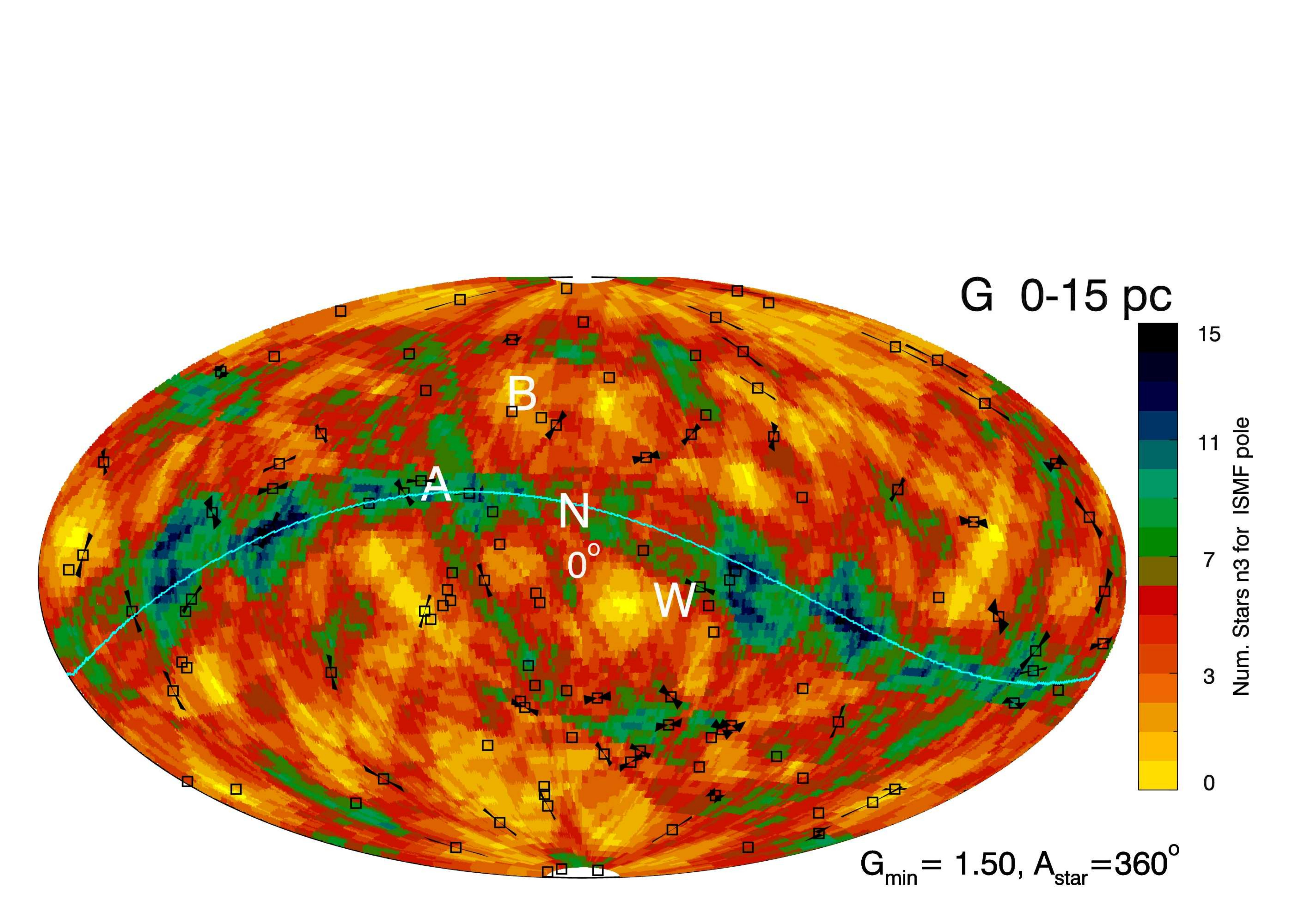}{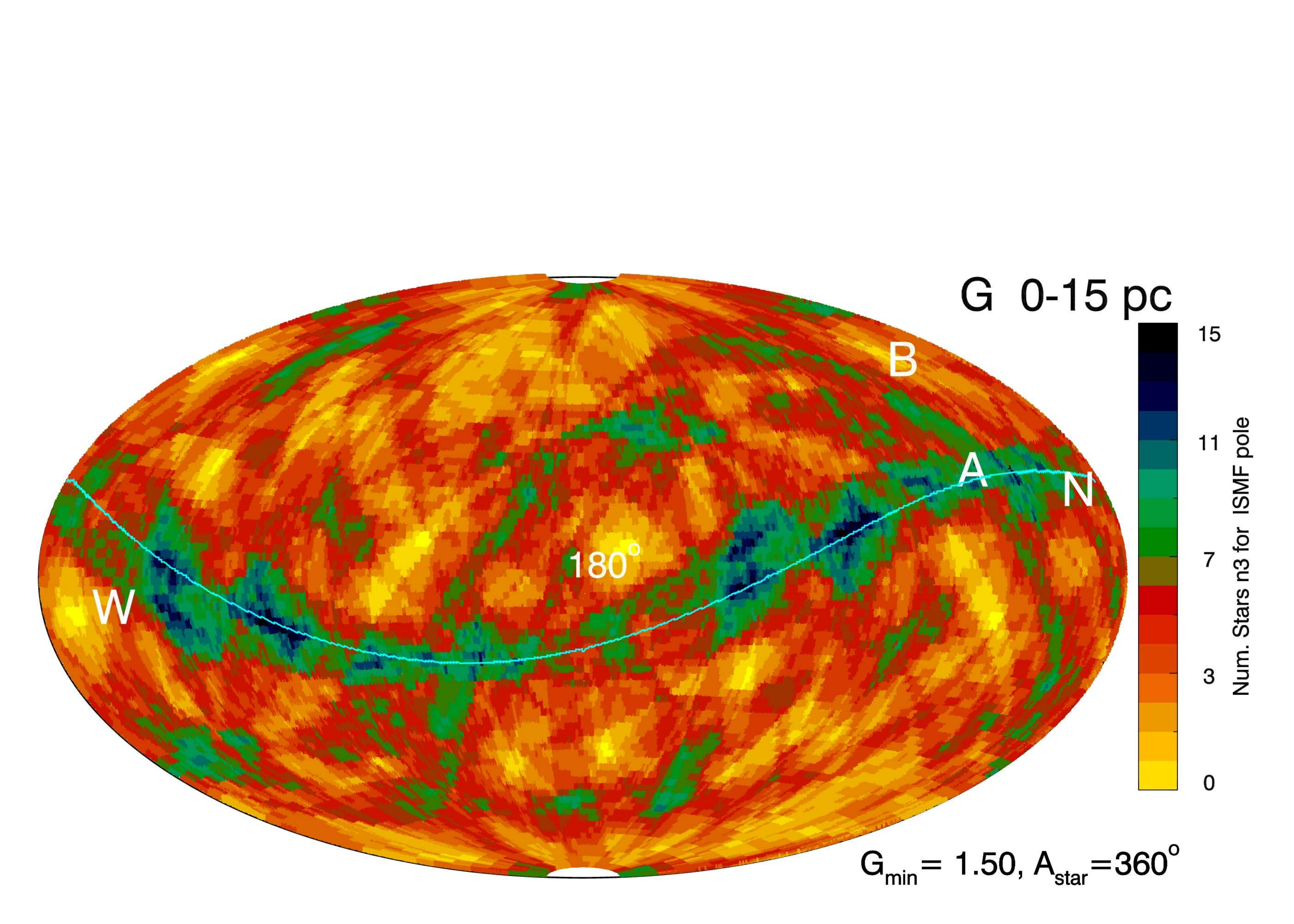}
\caption{Magnetic structure traced by polarization data within 15 pc
  are shown by counting the number of polarization position angles
  that predict a ``true'' magnetic field directed toward each position
  on the sky.  The probability at each location, \Grot\, is evaluated
  for stars throughout the sky (\angmax$= 360^\circ$), and those with
  an ISMF located at \PArot\ values that satisfy \Grot$>$\Gmin=1.0 (top) or
  \Grot$>$\Gmin=1.5 (bottom) are included in the counts plotted on the
  maps.  The cyan-colored line corresponds to the polarization band
  that has an axis toward \glon,\glat= 214\deeg, 67\deeg.  The black
  angular symbols show polarization position angles where
  \Pol/\dPol$>1.9$, while the squares show stars where
  \Pol/\dPol$<1.9$.  The letters ``N'', ``B'', ``W'', and ``A''
  indicate the directions of the heliosphere nose corresponding to the
  inflowing interstellar wind direction, the magnetic field direction
  indicated by the IBEX ribbon center, the upwind direction of the
  CLIC in the LSR, and the direction of the solar apex motion.
  Raising the minimum probability \Gmin\ used for selecting data
  included in the map decreases the total numbers of counts in the map
  and alters the color scale.  Projections are centered on the galatic
  center (left) and anti-center (right), with galactic longitude
  increasing from right to left for the Aitoff projections plotted in
  this paper. }
\label{fig:ang360}
\end{figure}

\begin{figure} % FIGURE 5 
\vspace*{-0.5in}
\plottwo{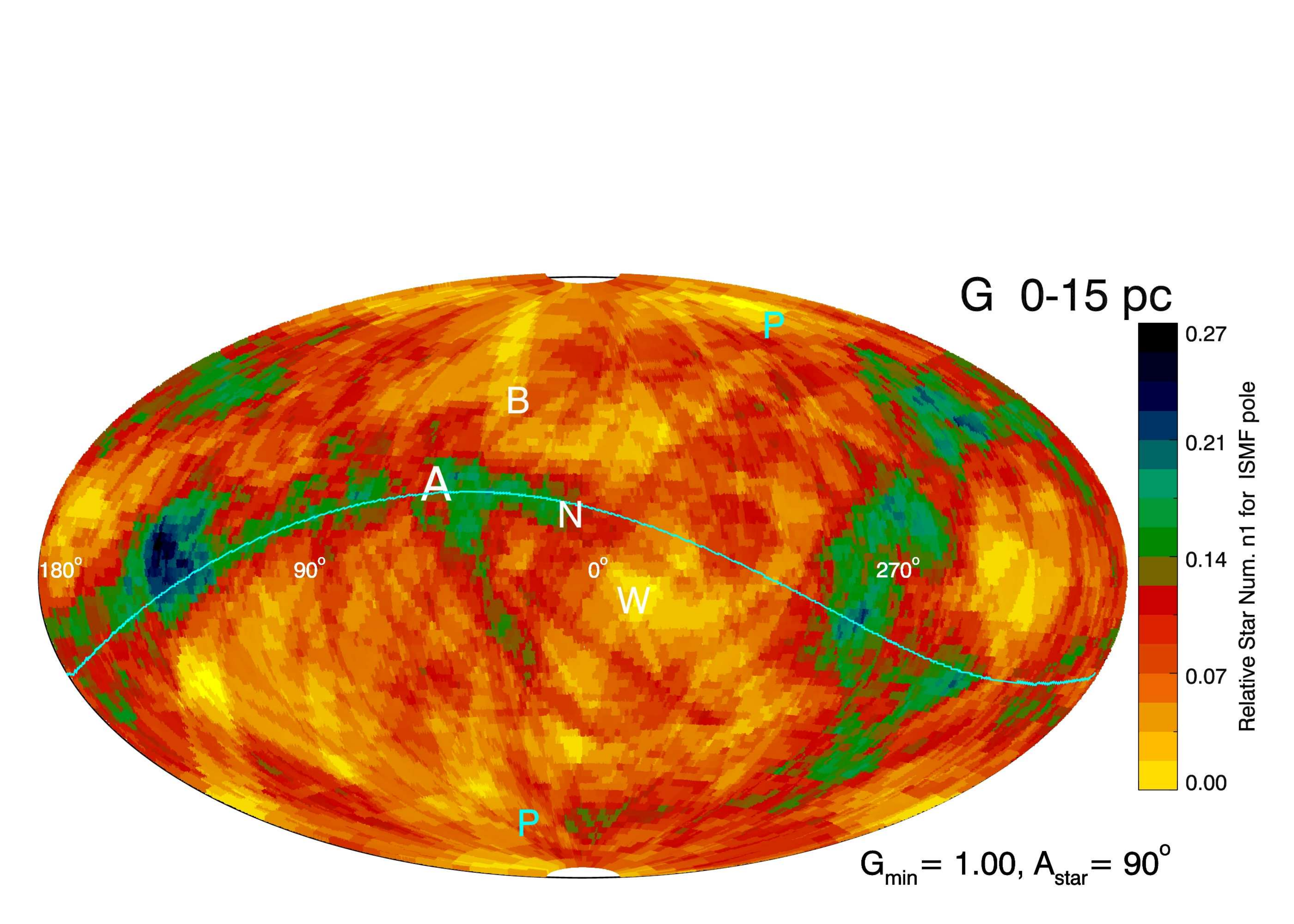}{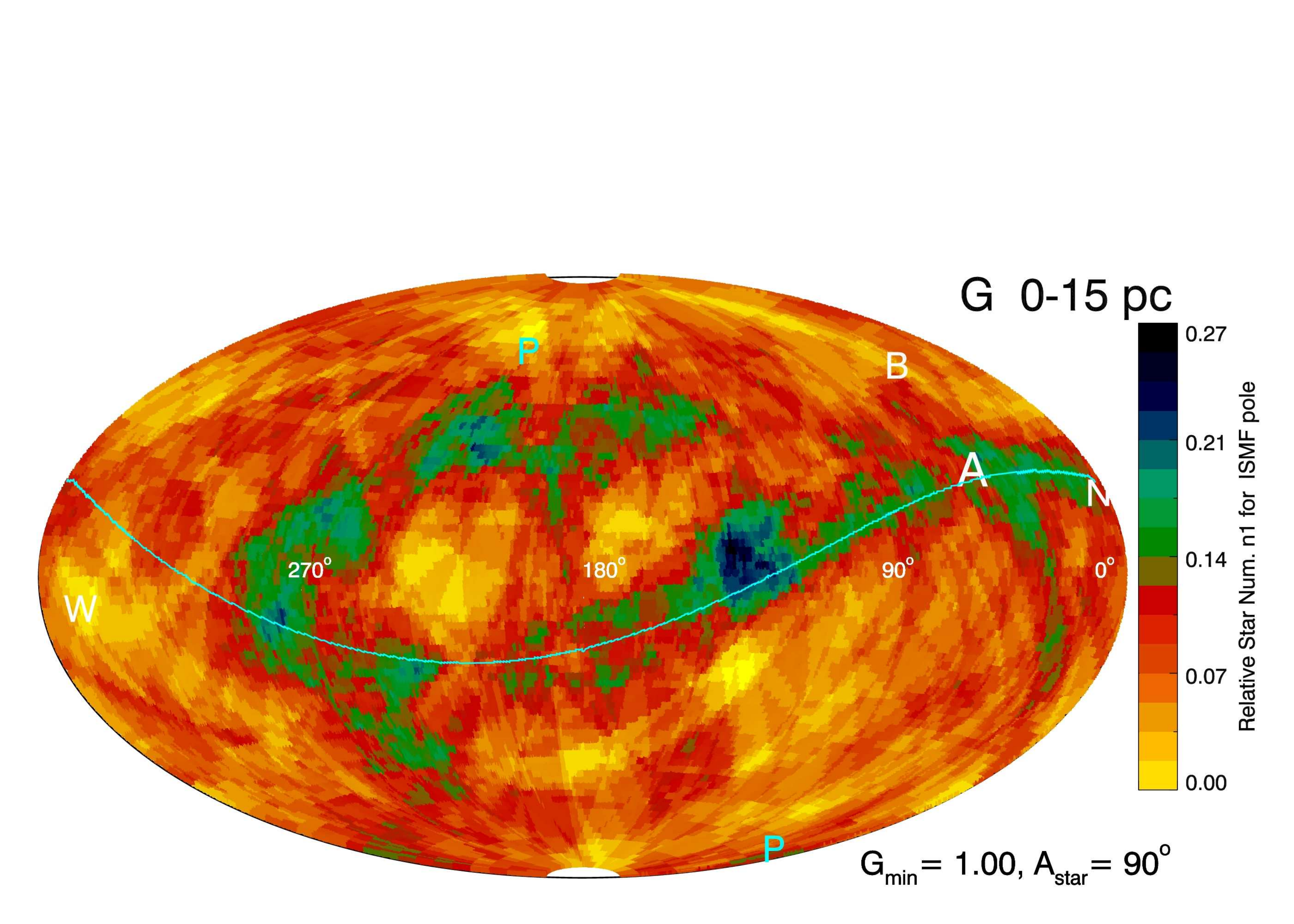}
\plottwo{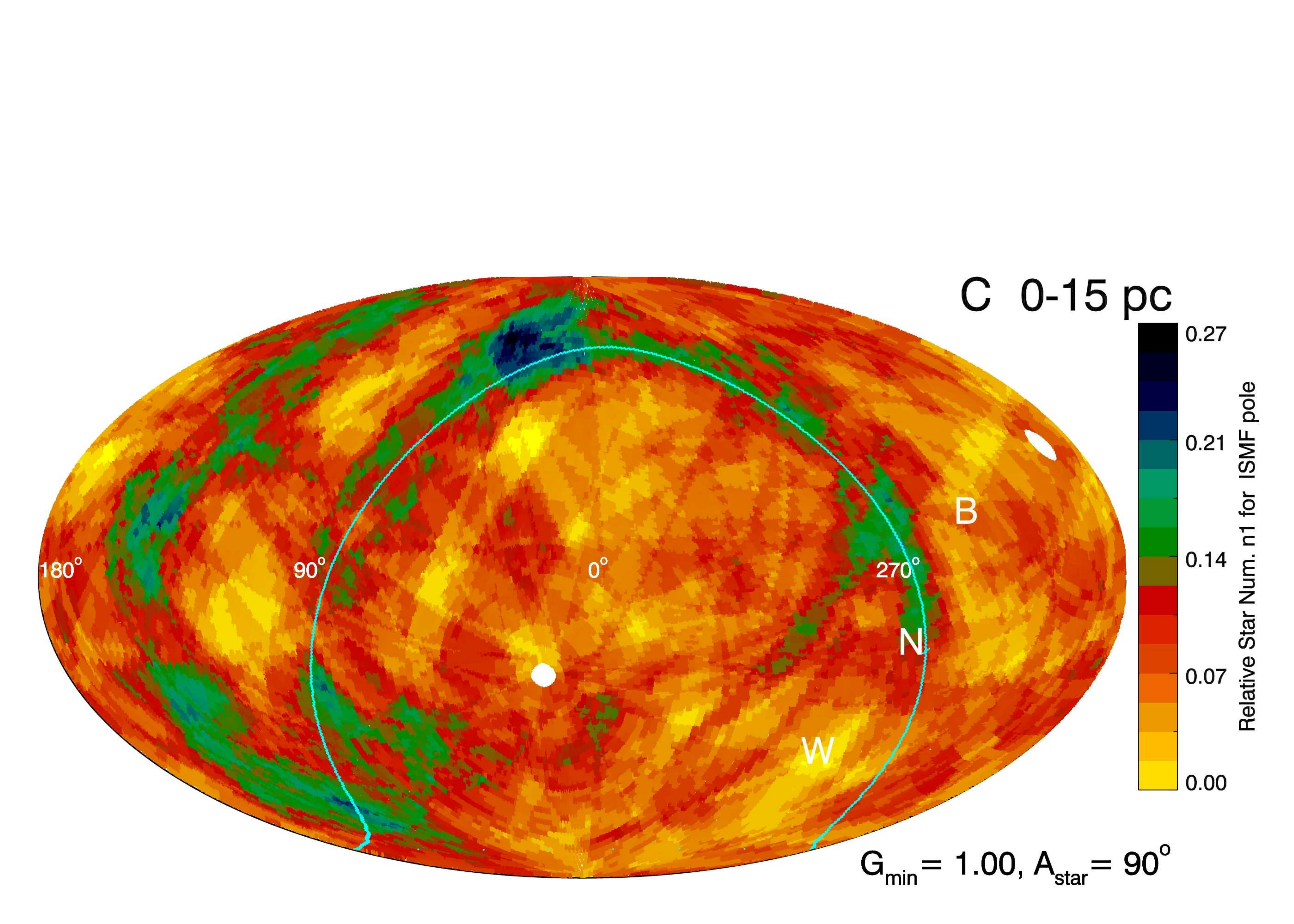}{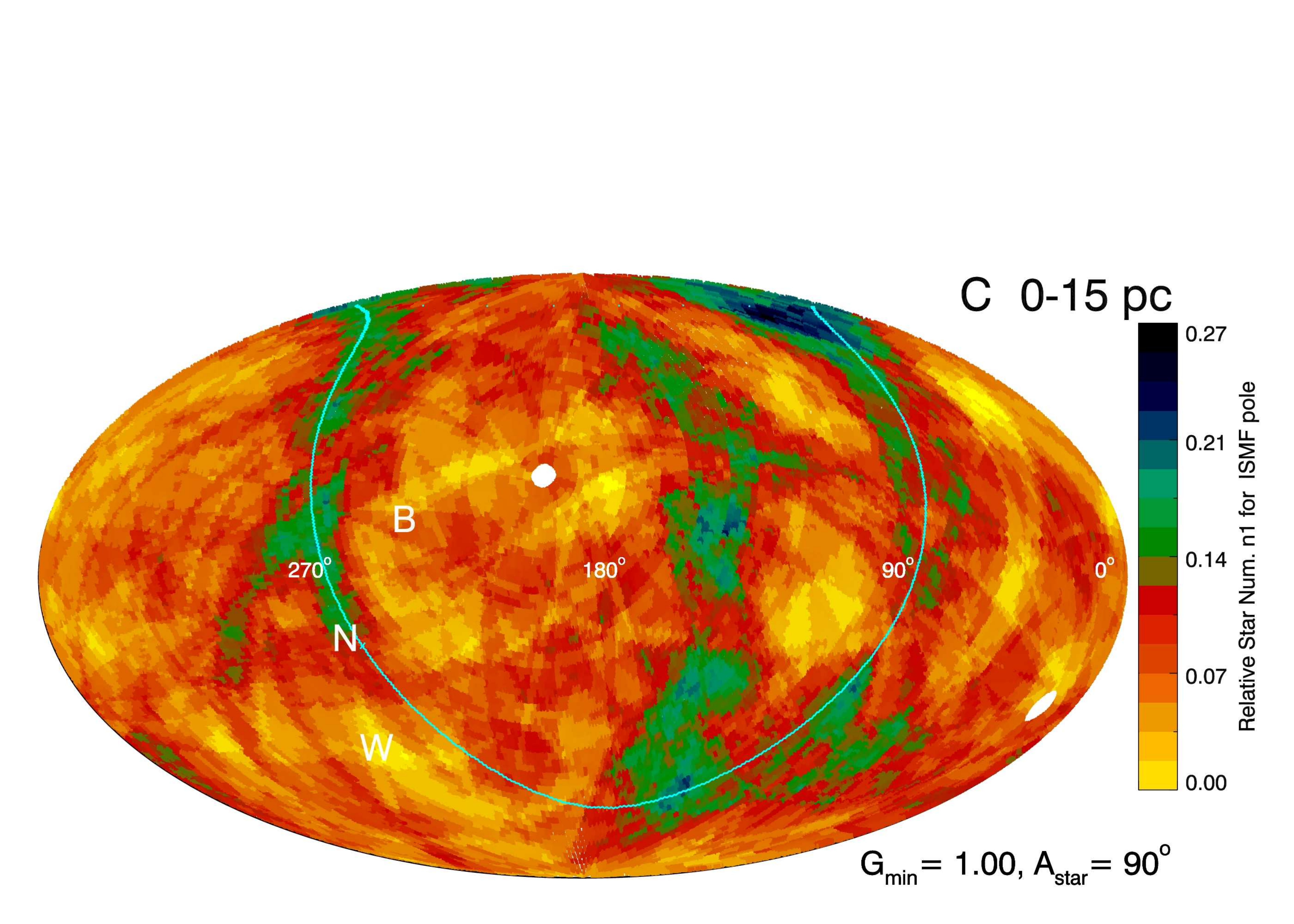}
\plottwo{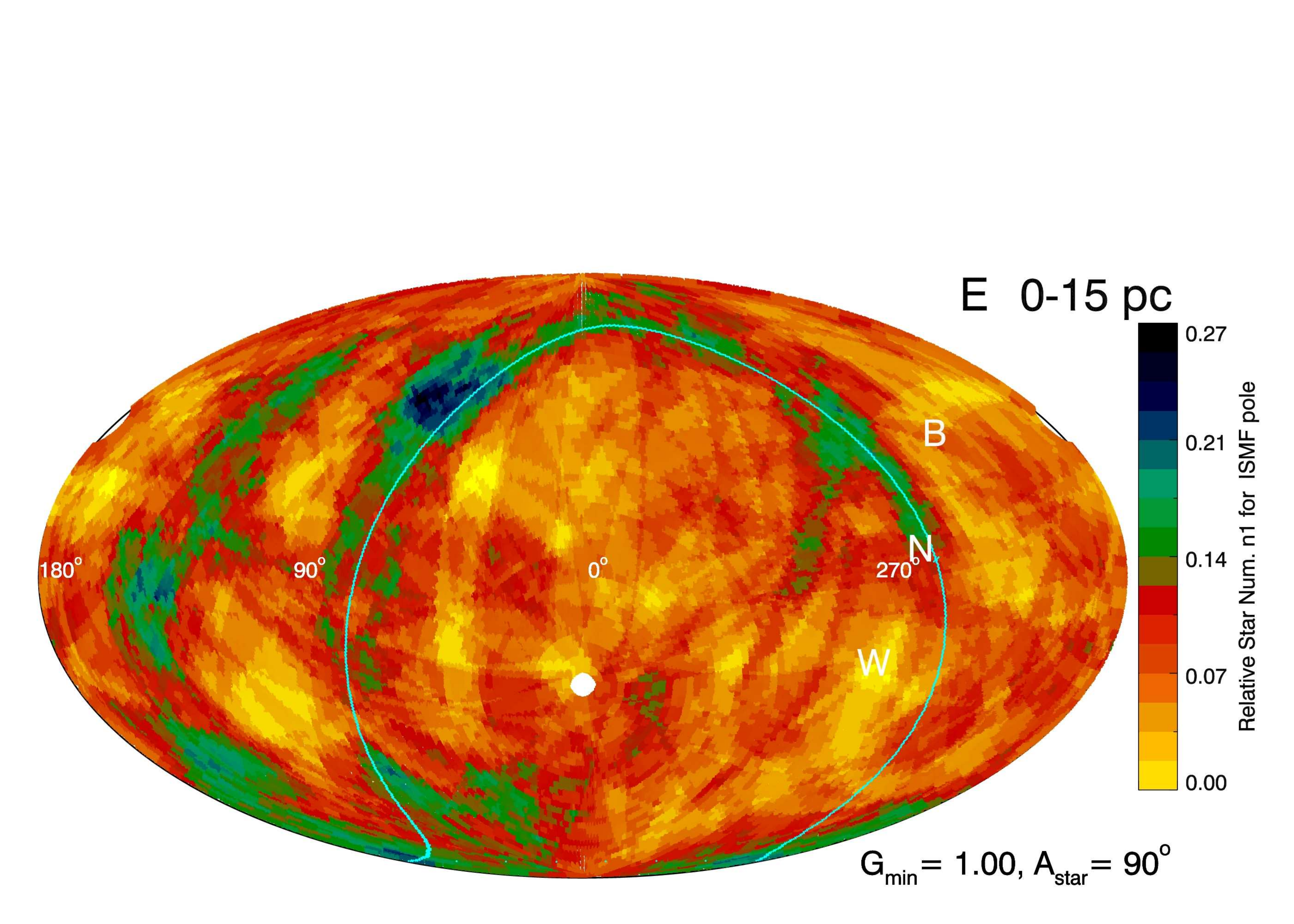}{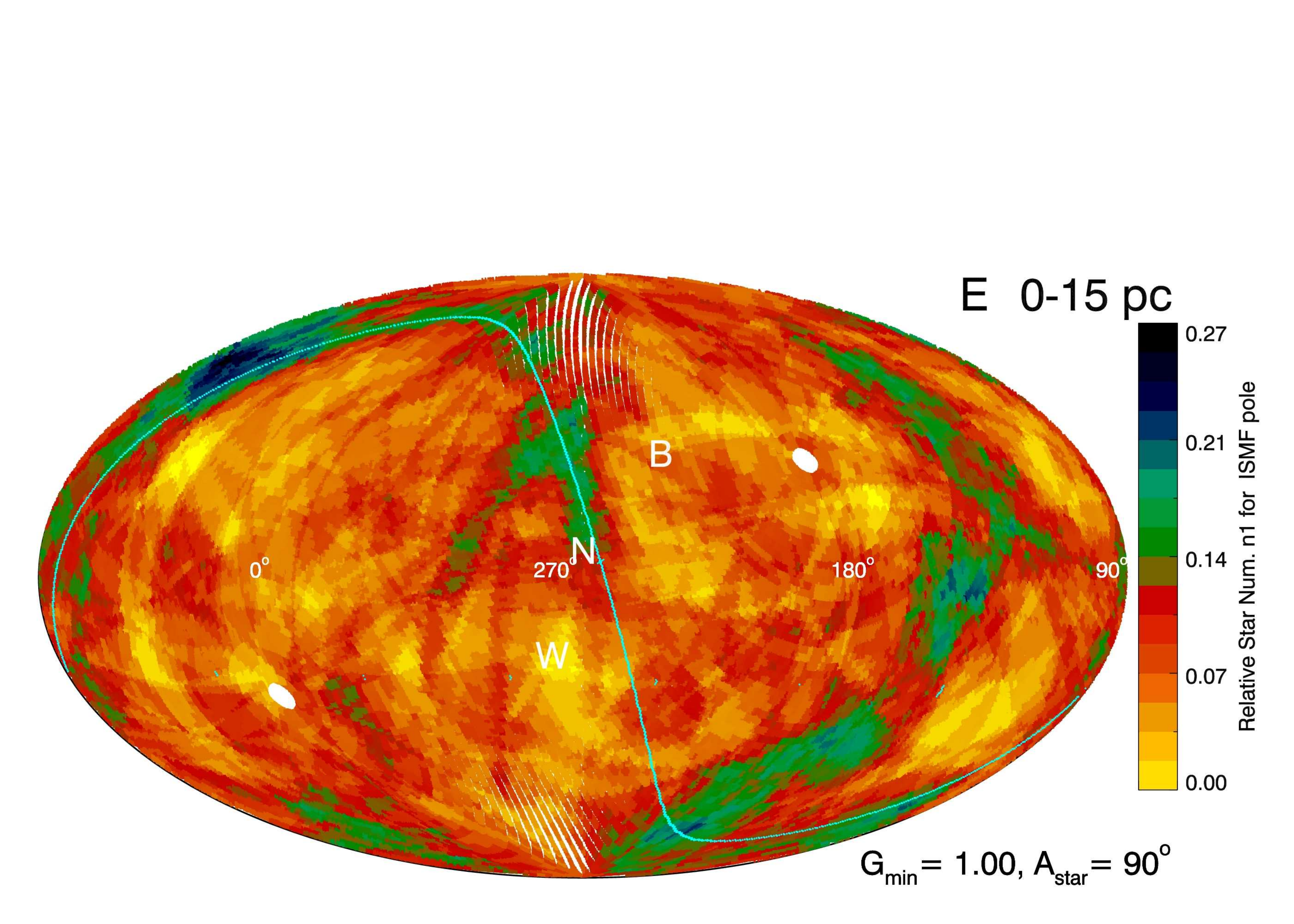}
\caption{The relative numbers of stars with probability counting criteria \Grot $>$\Gmin=1.00, 
and within 90\deeg\ of each grid point, are plotted.
  Plots are presented in galactic ('G'), equatorial ('C') and ecliptic
  ('E') coordinate systems.  Top left and right figures are centered
  on galactic center and anti-center, respectively.  Middle left and
  right figures are centered on 0\deeg\ and 180\deeg\ right ascension.
Bottom left and right figures are centered on ecliptic coordinates $\lambda,\beta =
     0^\circ,0^\circ$, left) and the heliosphere nose ($\lambda,\beta=255.6^\circ ,5.1^\circ$,
    right).  Large white dots show the galactic poles.  Other
    notations are as in Figure \ref{fig:ang360}}.
\label{fig:ang90}
\end{figure}

\begin{figure}[ht!] % FIGURE 6 
\plottwo{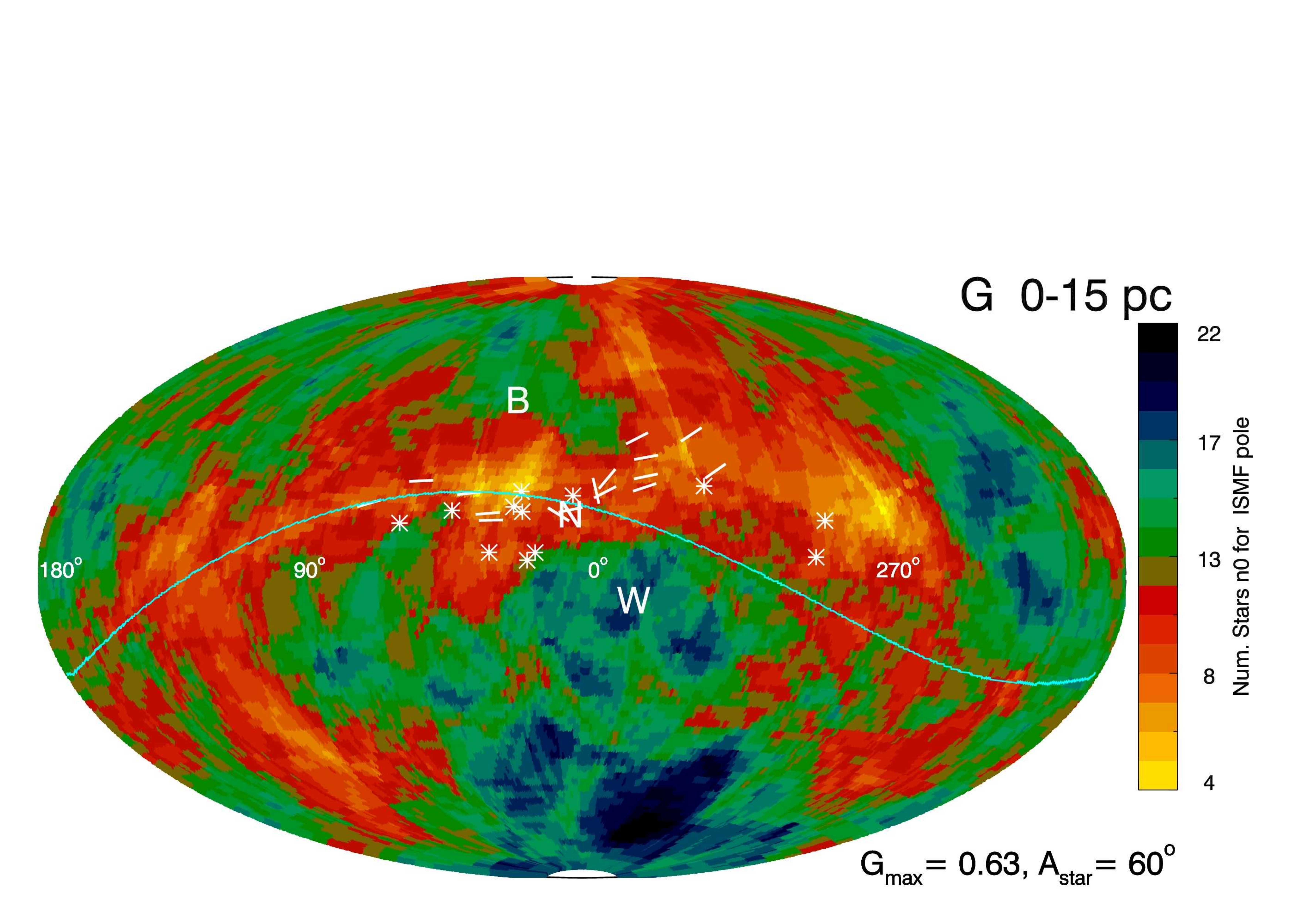}{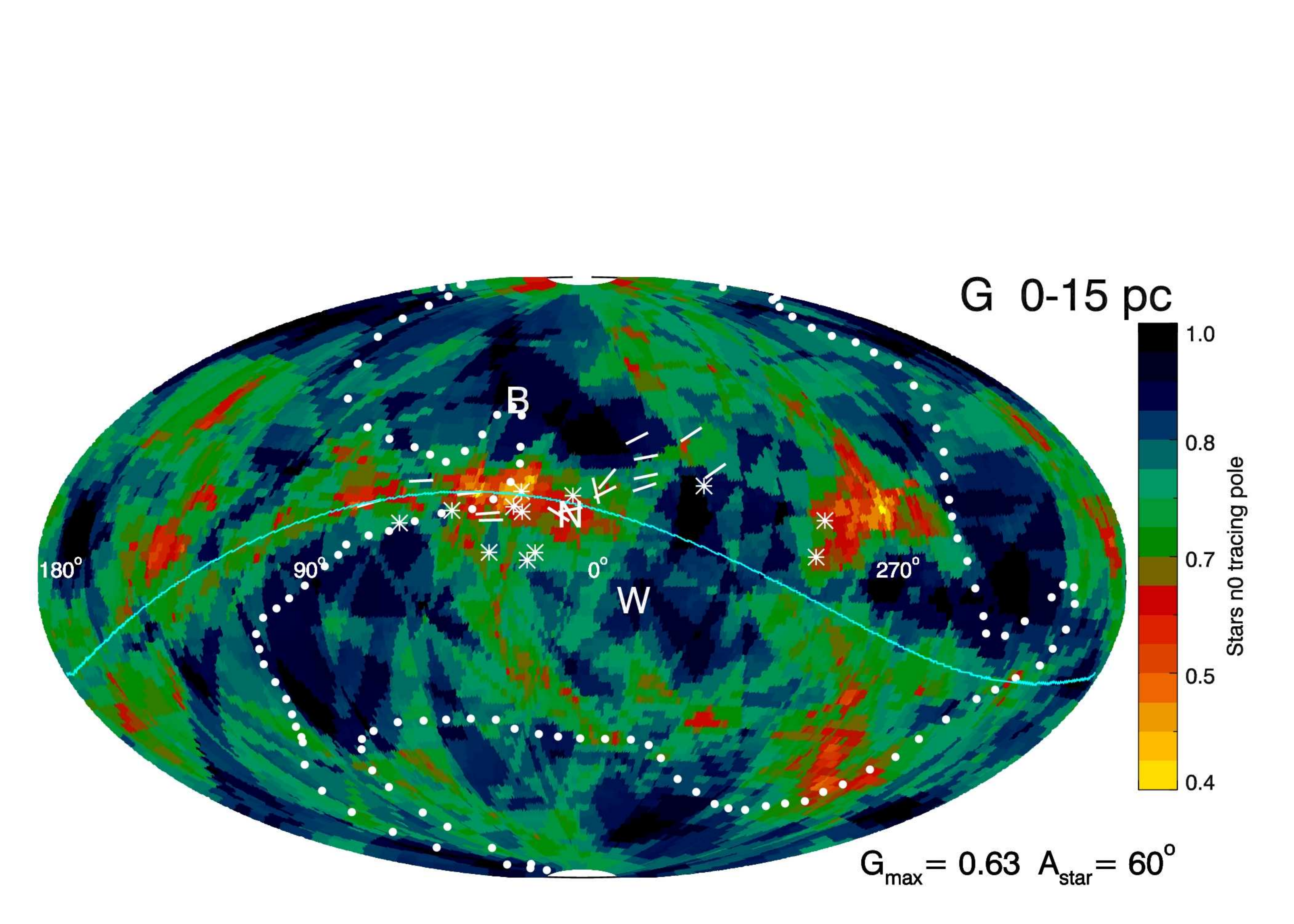}
\plottwo{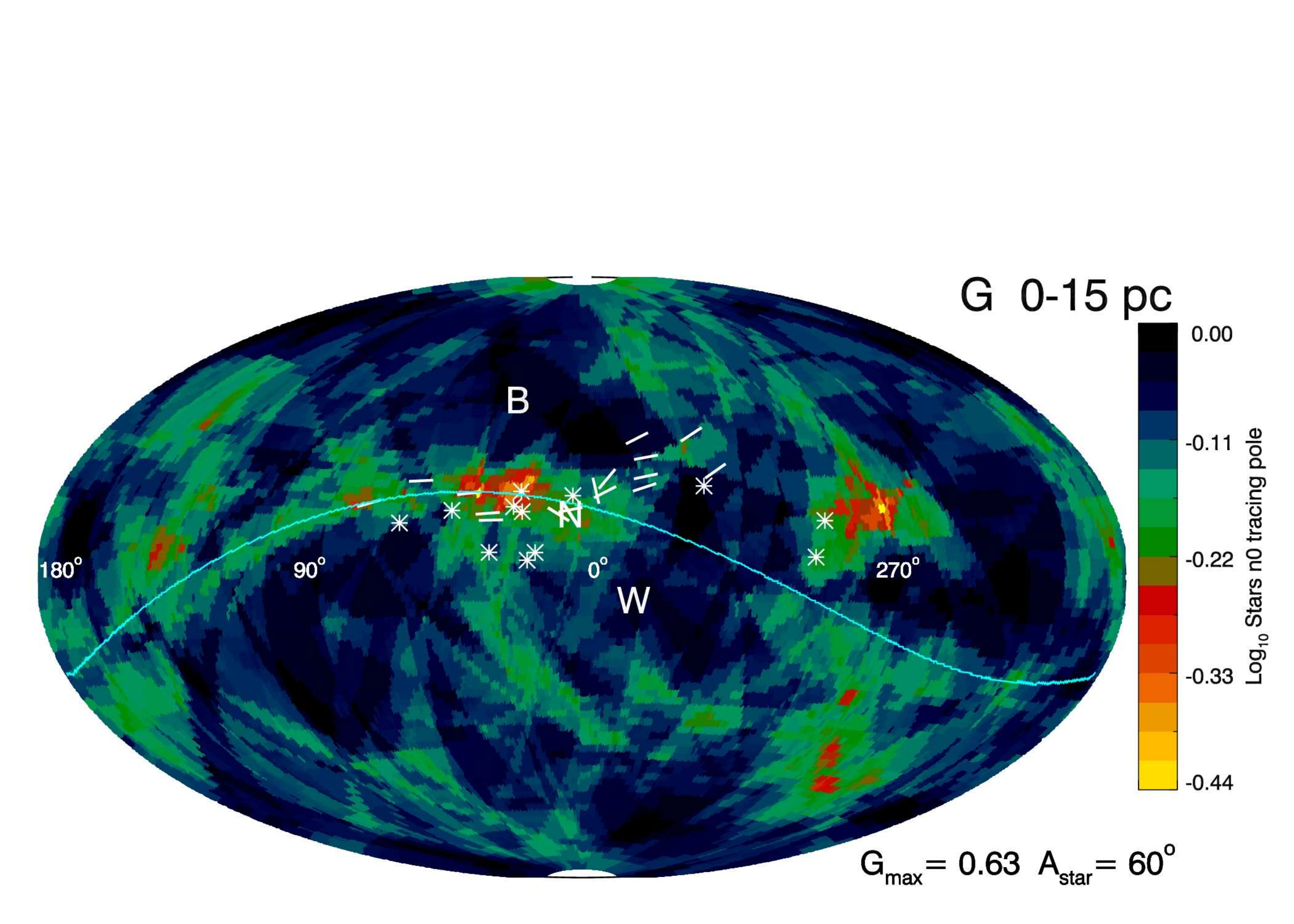}{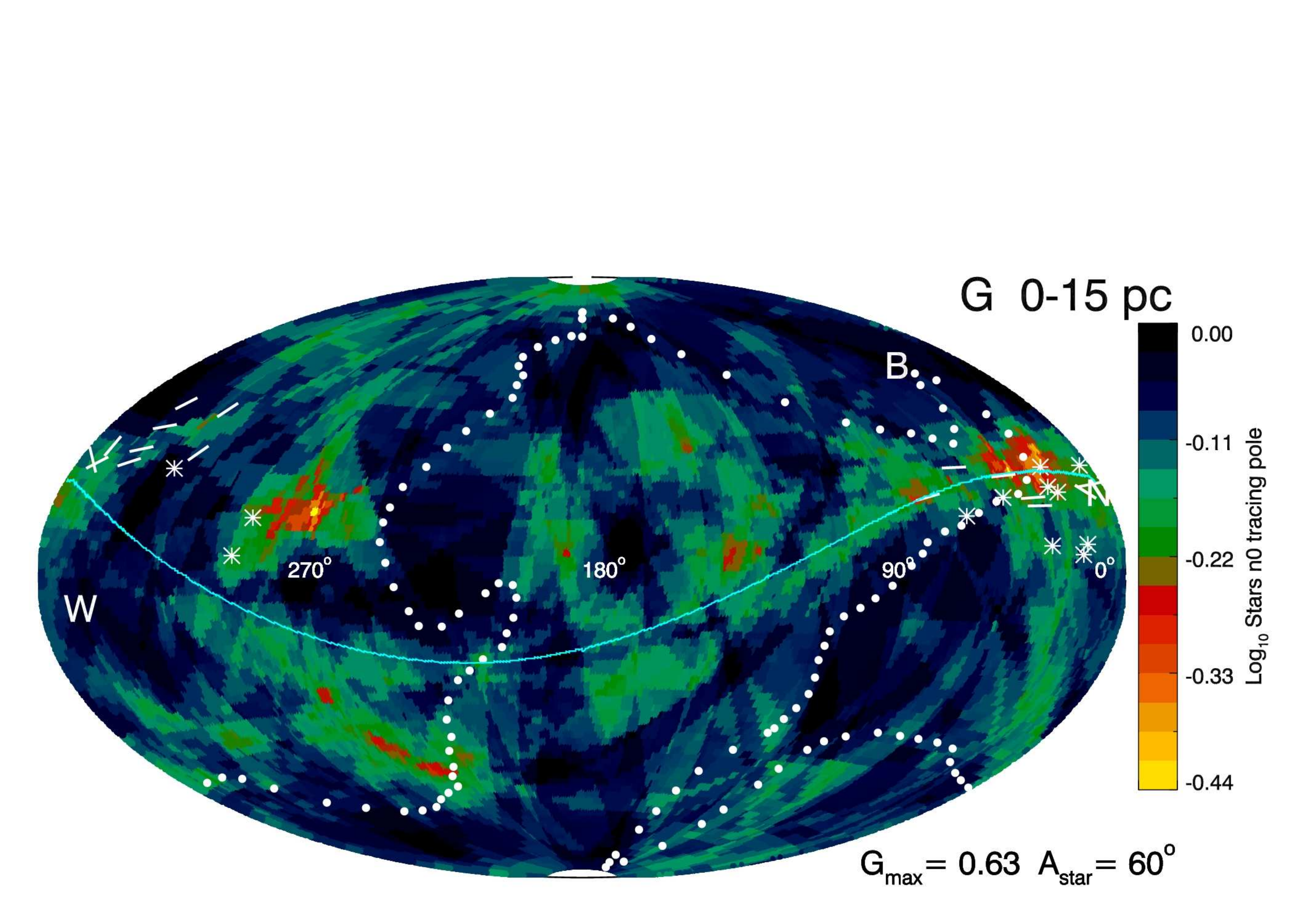}
\plottwo{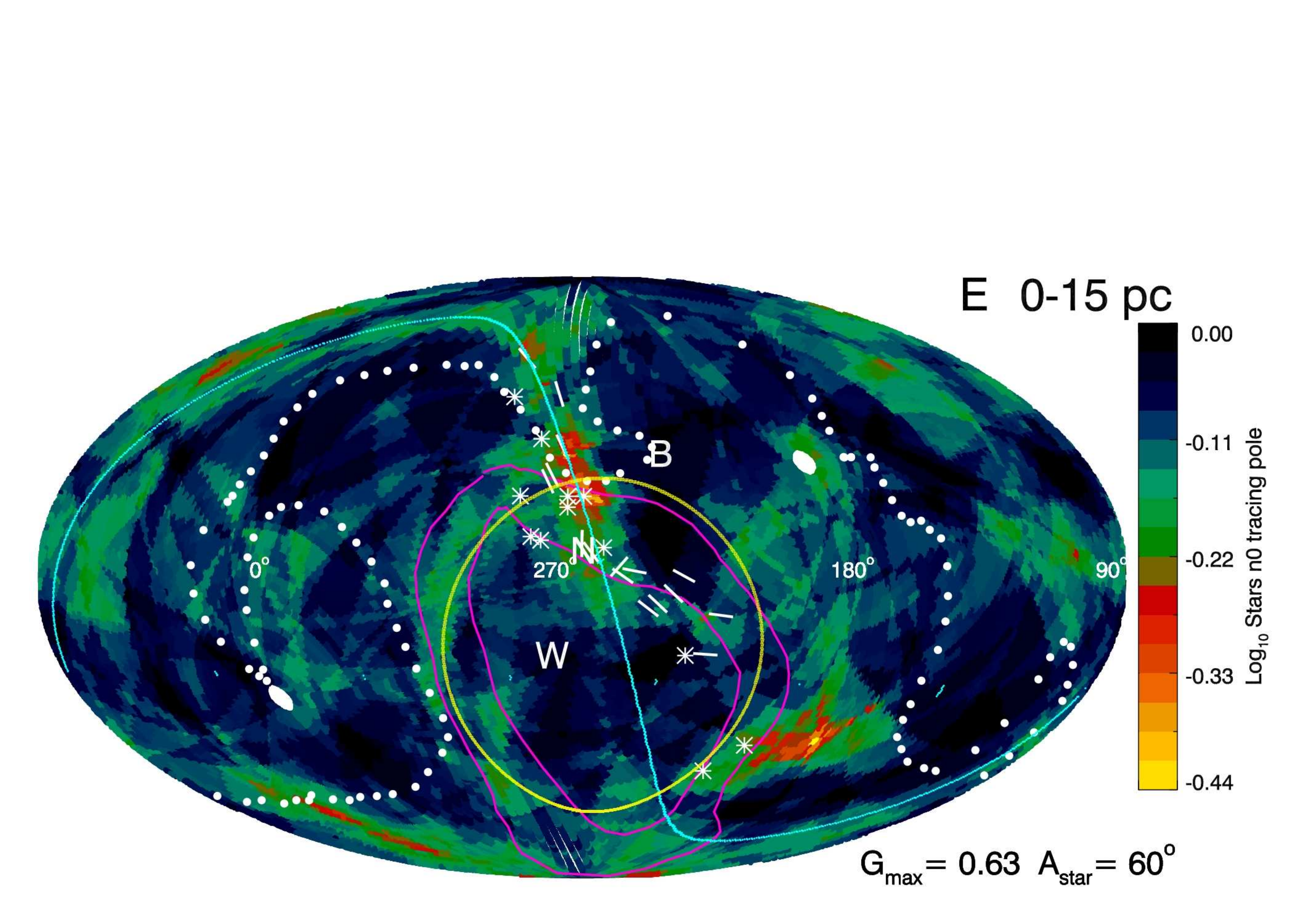}{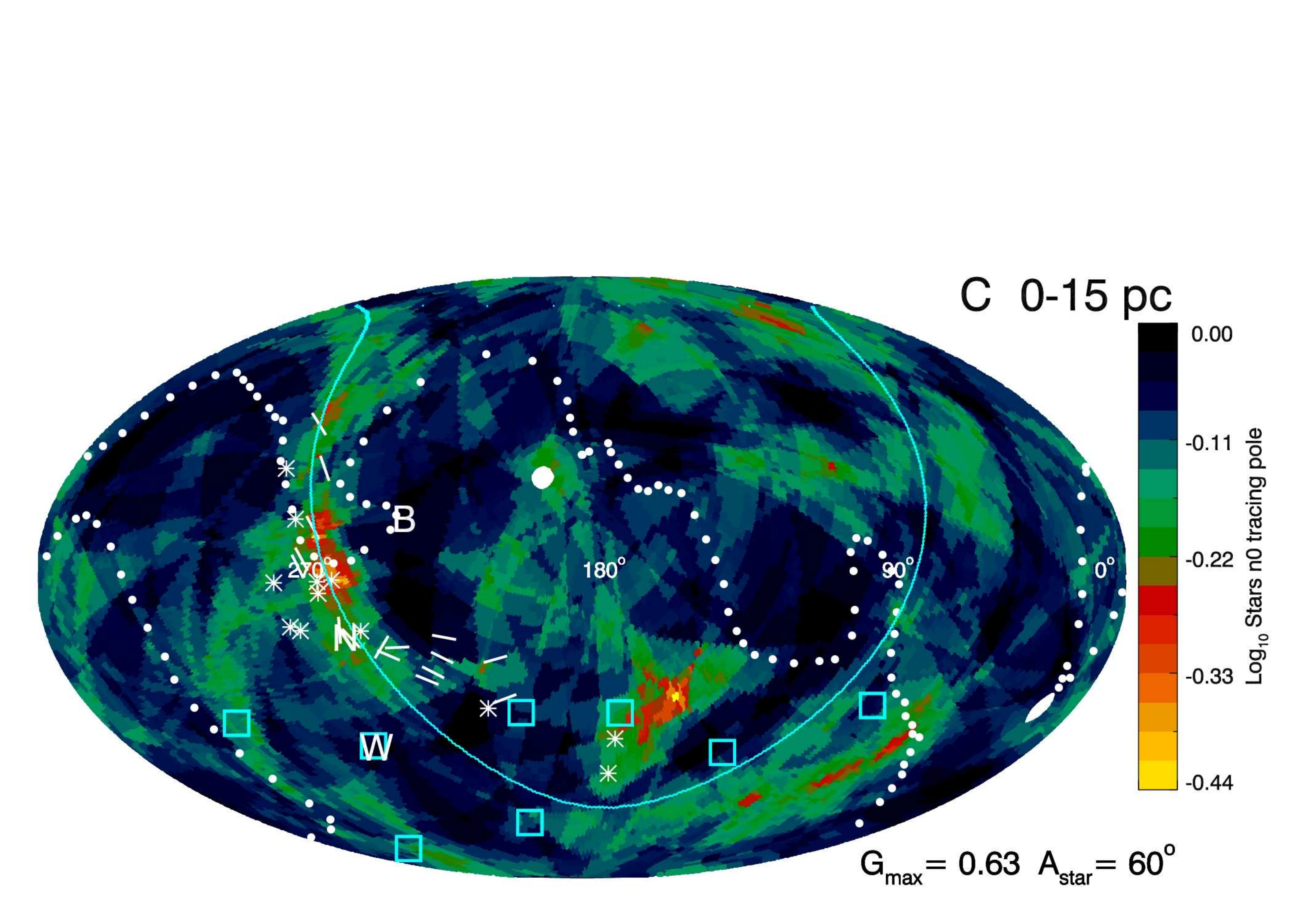}
%% \vspace*{-0.8in}
\caption{Maps of unnormalized (top left) and normalized counts
  (other figures) of data where \Grot$<$\Gmax=0.63 that are
  within 60\deeg\ of each plotted location \Lrot, \Brot, using data
with Pol/\dPol $\ge 1.9$.  Counts are
  normalized by the number of points within \angmax=60\deeg\ of each point.
  The bottom four maps display results
  with a logarithmic color scale (base 10), for projections in the
  galactic, ecliptic, and equatorial coordinate systems (`G'', ``E'', and ``C'').  
Supplementary data plotted include the
kHz emissions discovered by Voyagers 1 and 2 
  (asterisks) and the filament star polarizations (bars).  Concentric
  contours outline the Loop I shell.  The white ring in the 
  nose-centered ecliptic projection (lower left) shows the
approximate region of high plasma pressures discovered by IBEX
(centered near \elon=249\deeg, \elat=--20\deeg).
The gap in the polarization band corresponds to
  both the center of Loop I and the heliosheath high-pressure plasma region.  
Galactic cosmic ray small excess regions (IC59) are 
shown in the lower right figure (cyan-colored squares, \S \ref{sec:gcr}).  }
\label{fig:ang60norm}
\end{figure}

\begin{figure} % FIGURE 7 
\plottwo{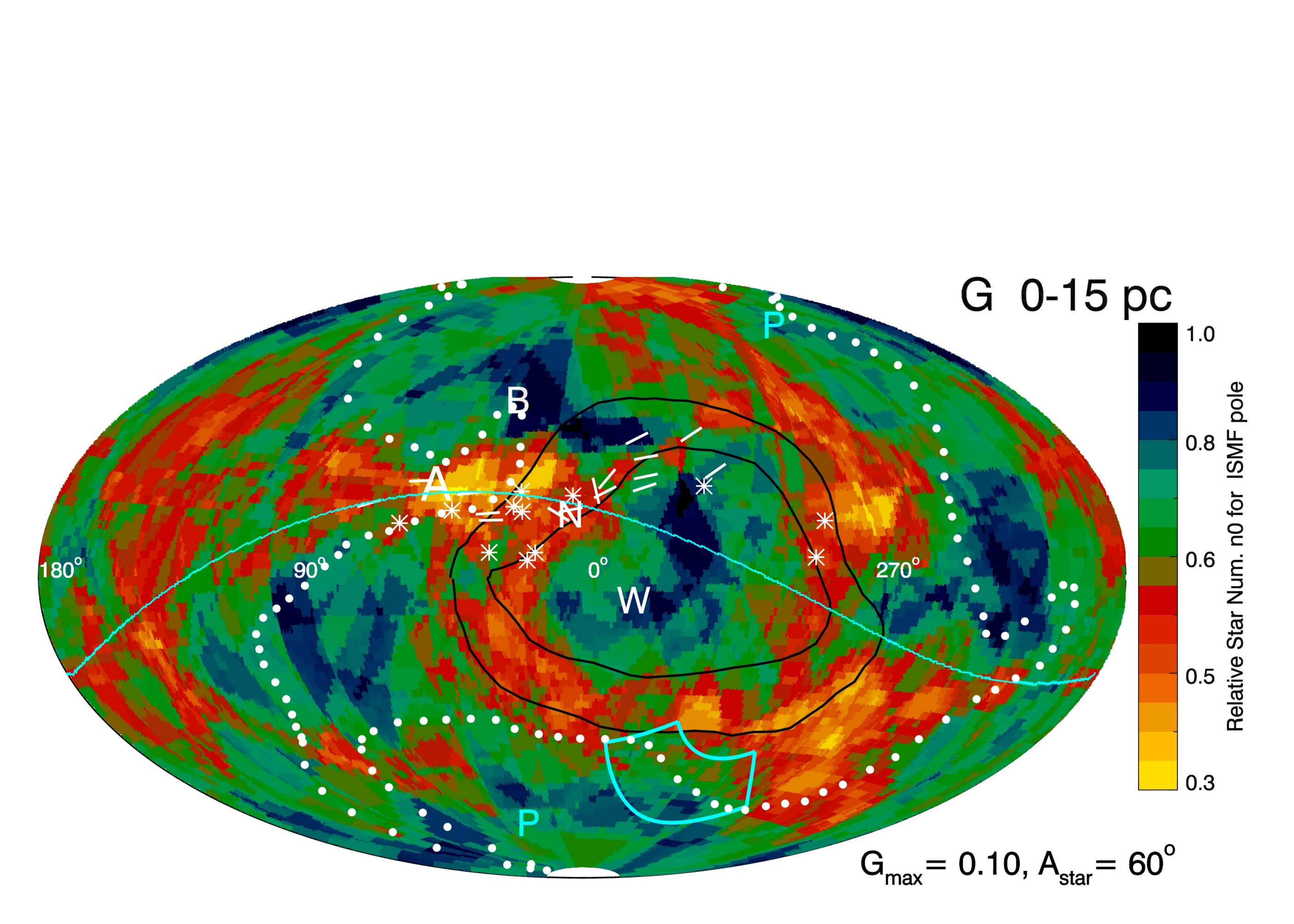}{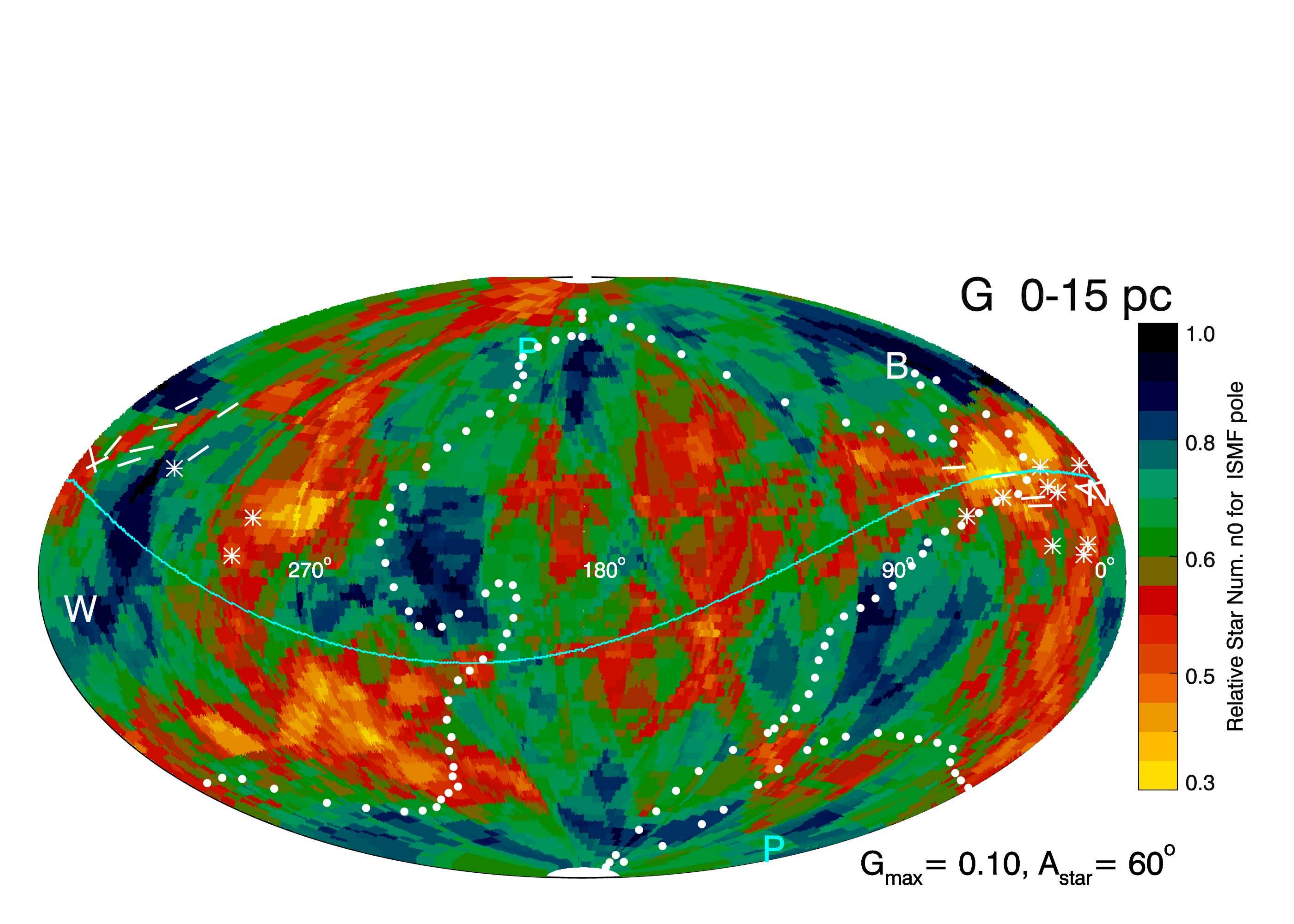}
\plottwo{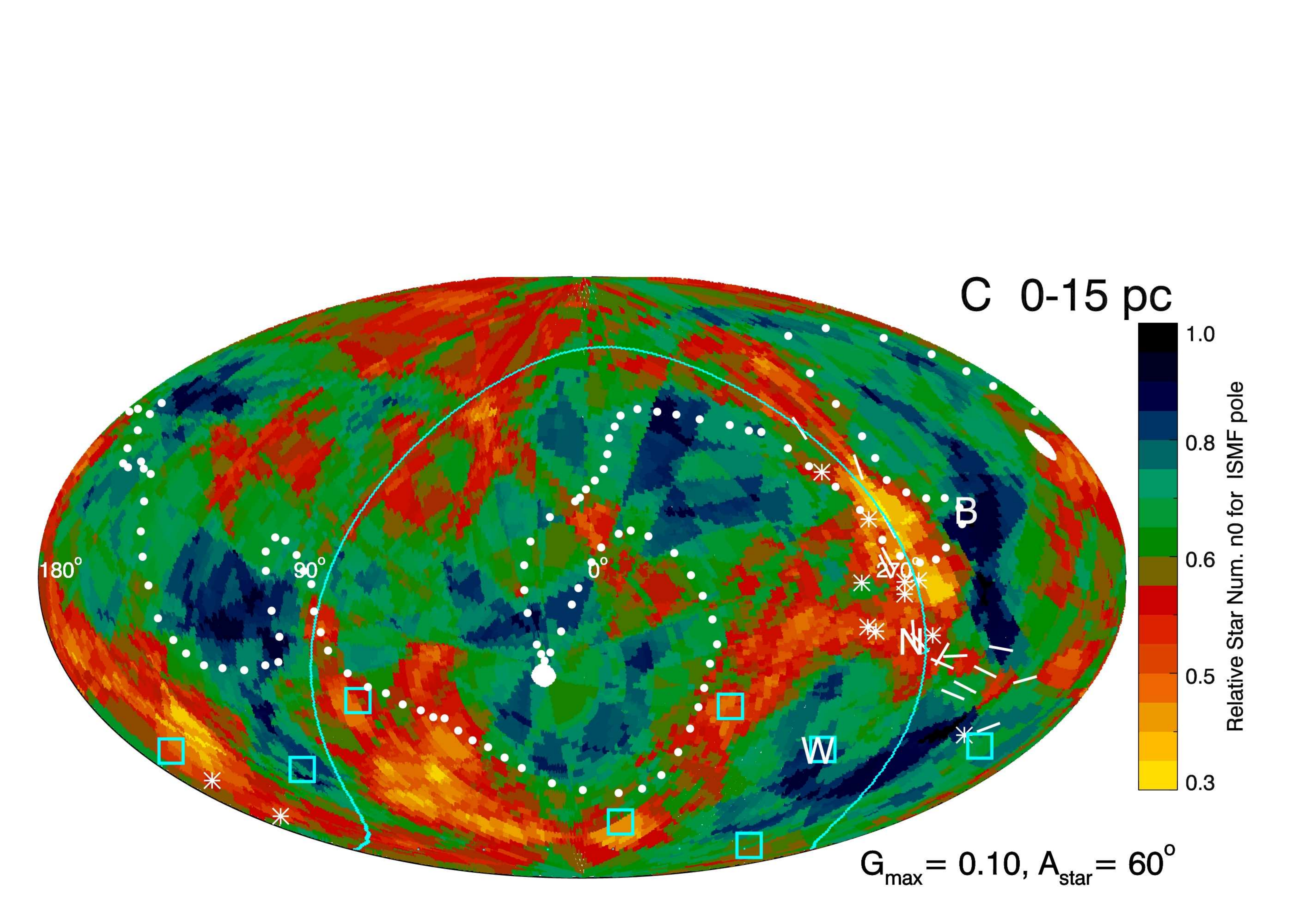}{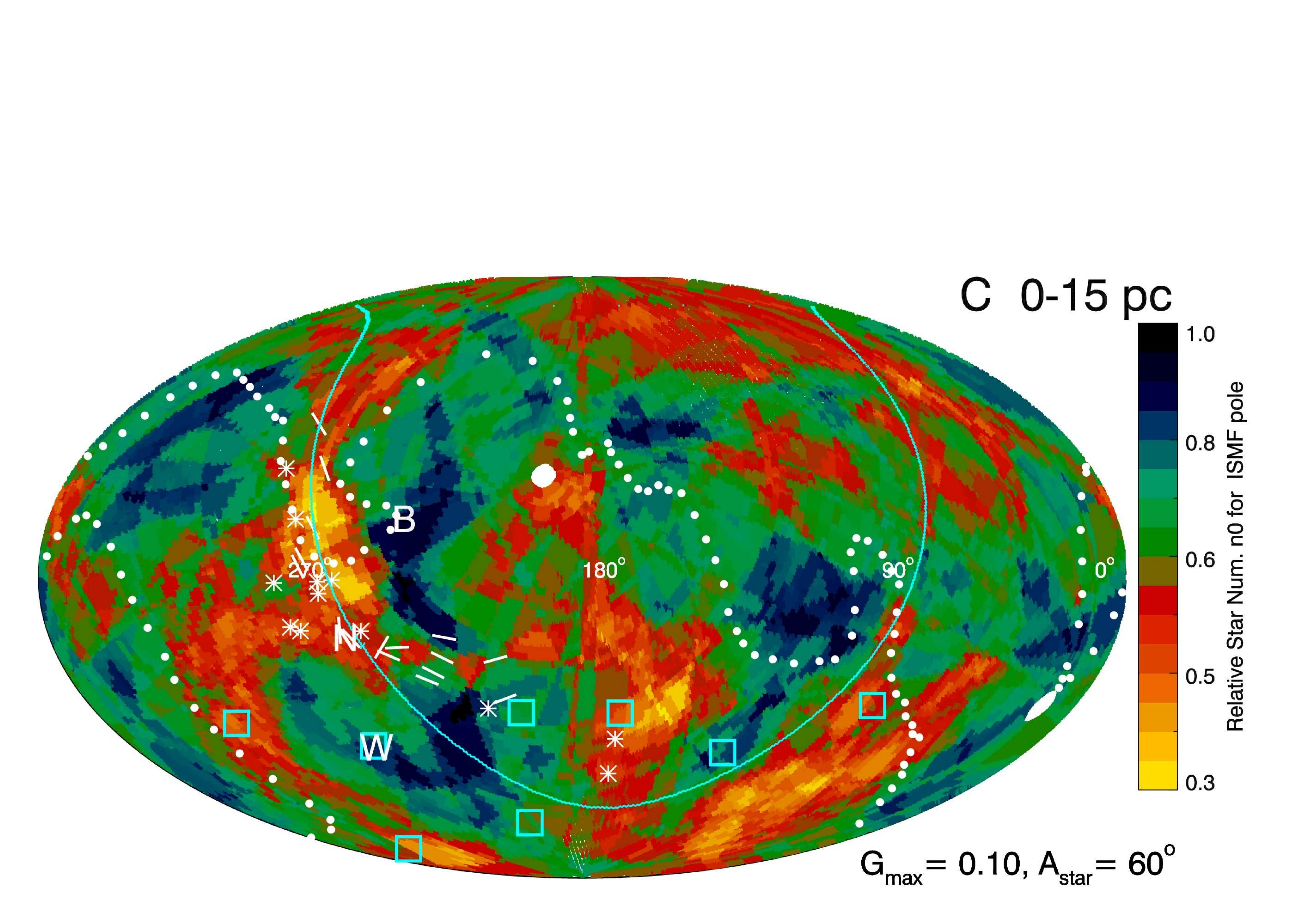}
\plottwo{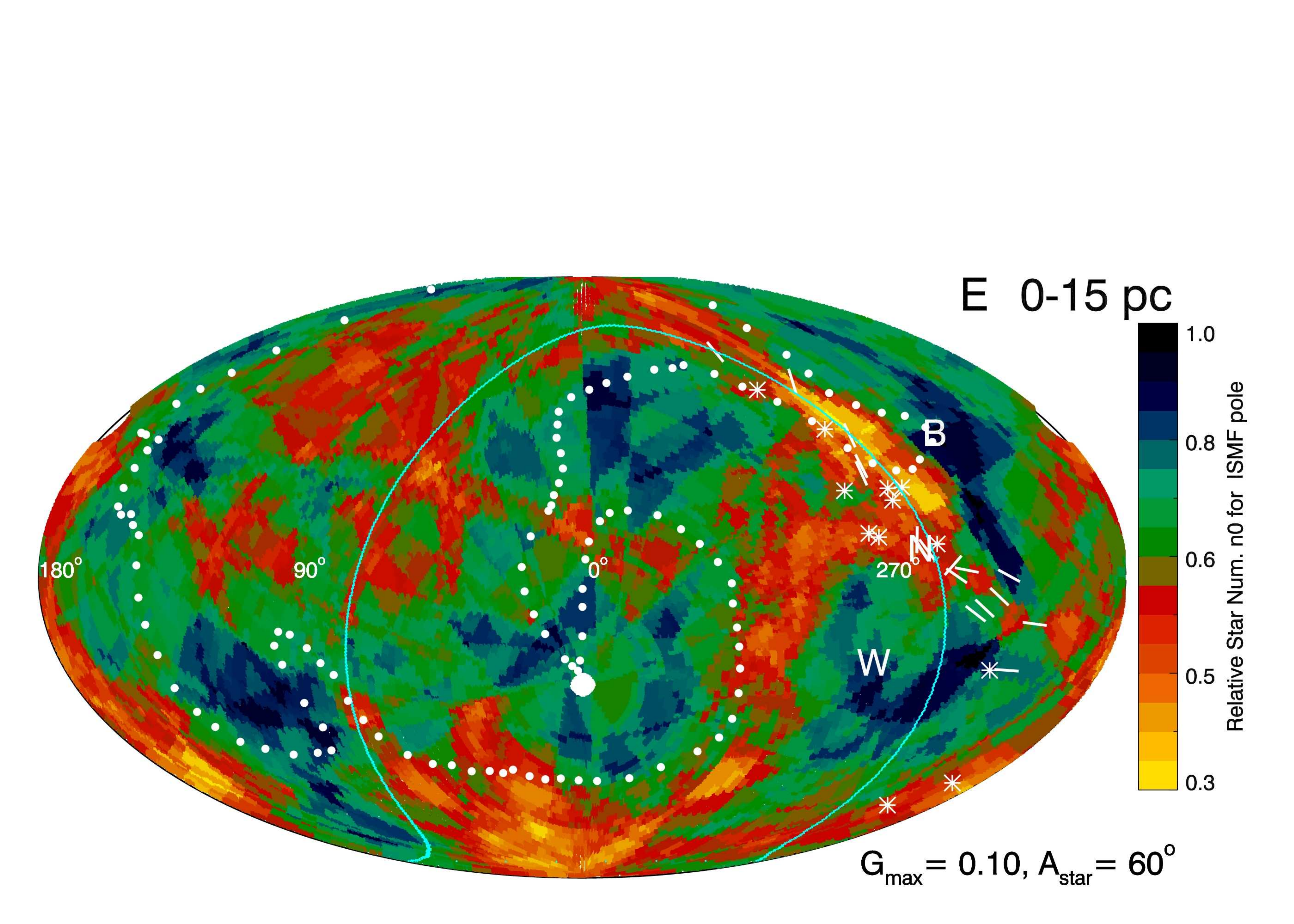}{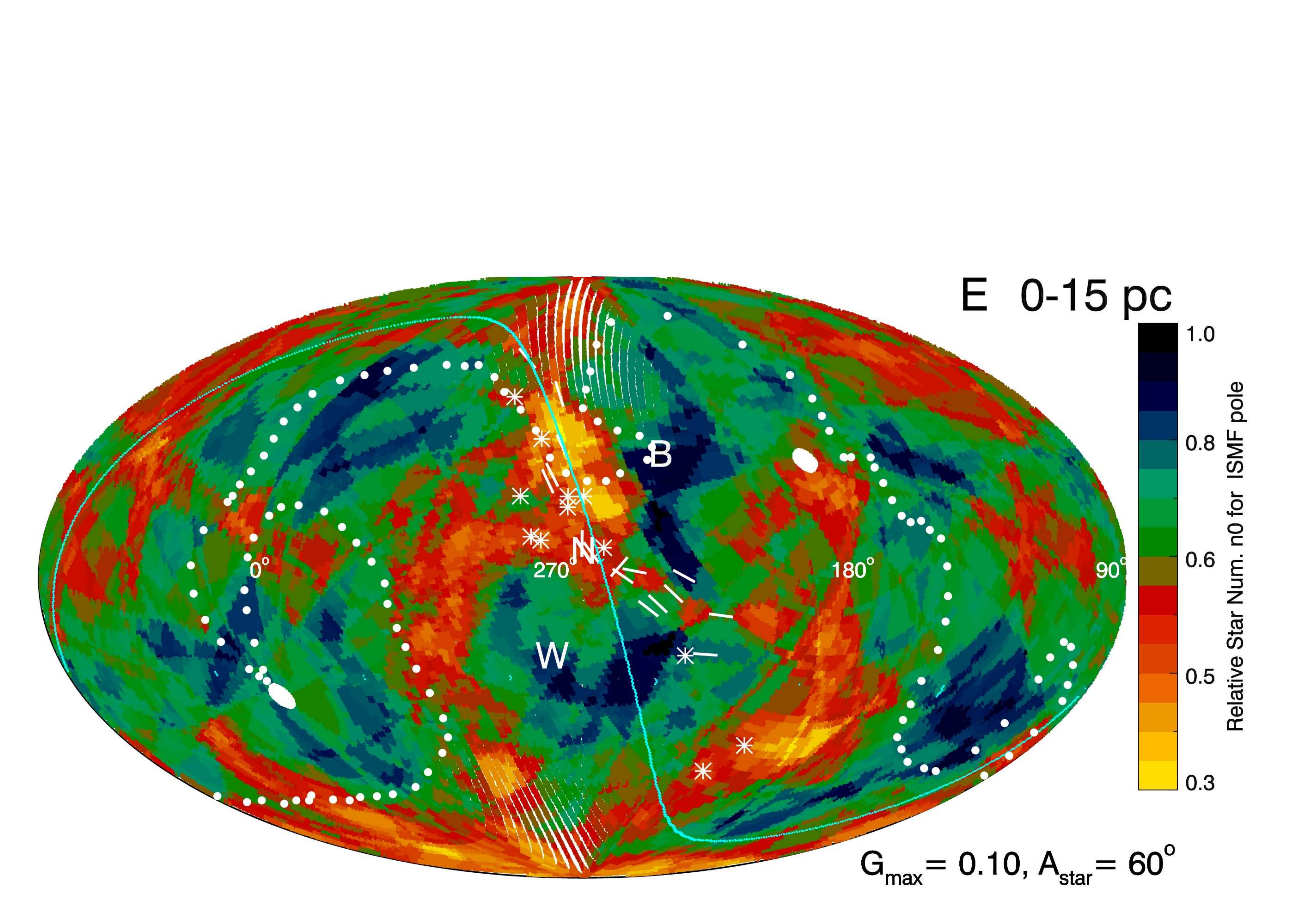}
\caption{ The three figure pairs show magnetic structure for
  statistical constraints \Grot$<$\Gmax=0.10 and angular sampling over
  intervals $\pm 60^\circ$ (\angmax=60\deeg) around each grid point.
  Numbers of statistically qualifying data points are normalized by
  the numbers of geometrically qualifying data points at each
  location.  Other features in figures are explained in Figure
  \ref{fig:ang60norm}.}
\label{fig:ang60}
\end{figure}

\begin{figure} % FIGURE 8 
\plottwo{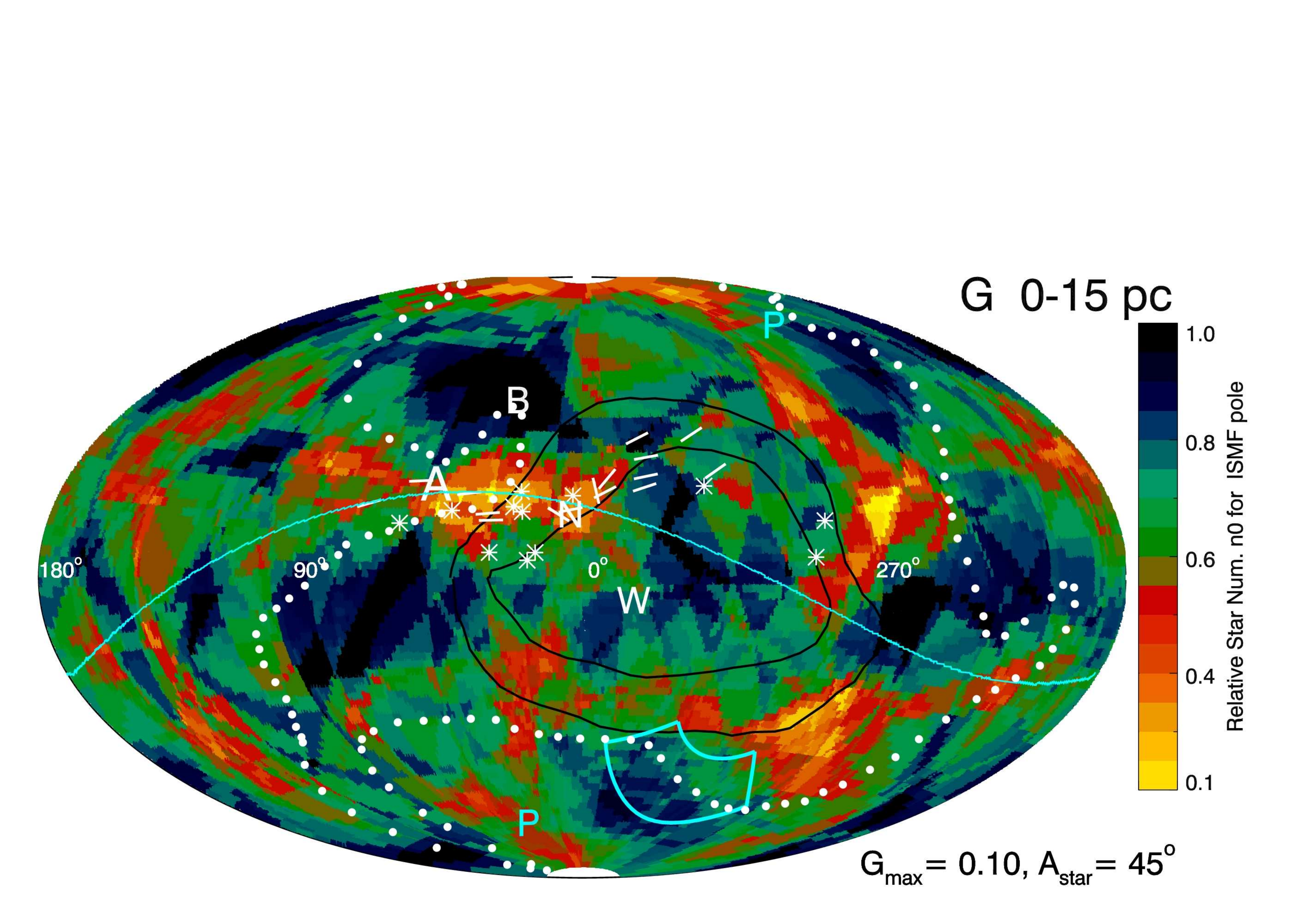}{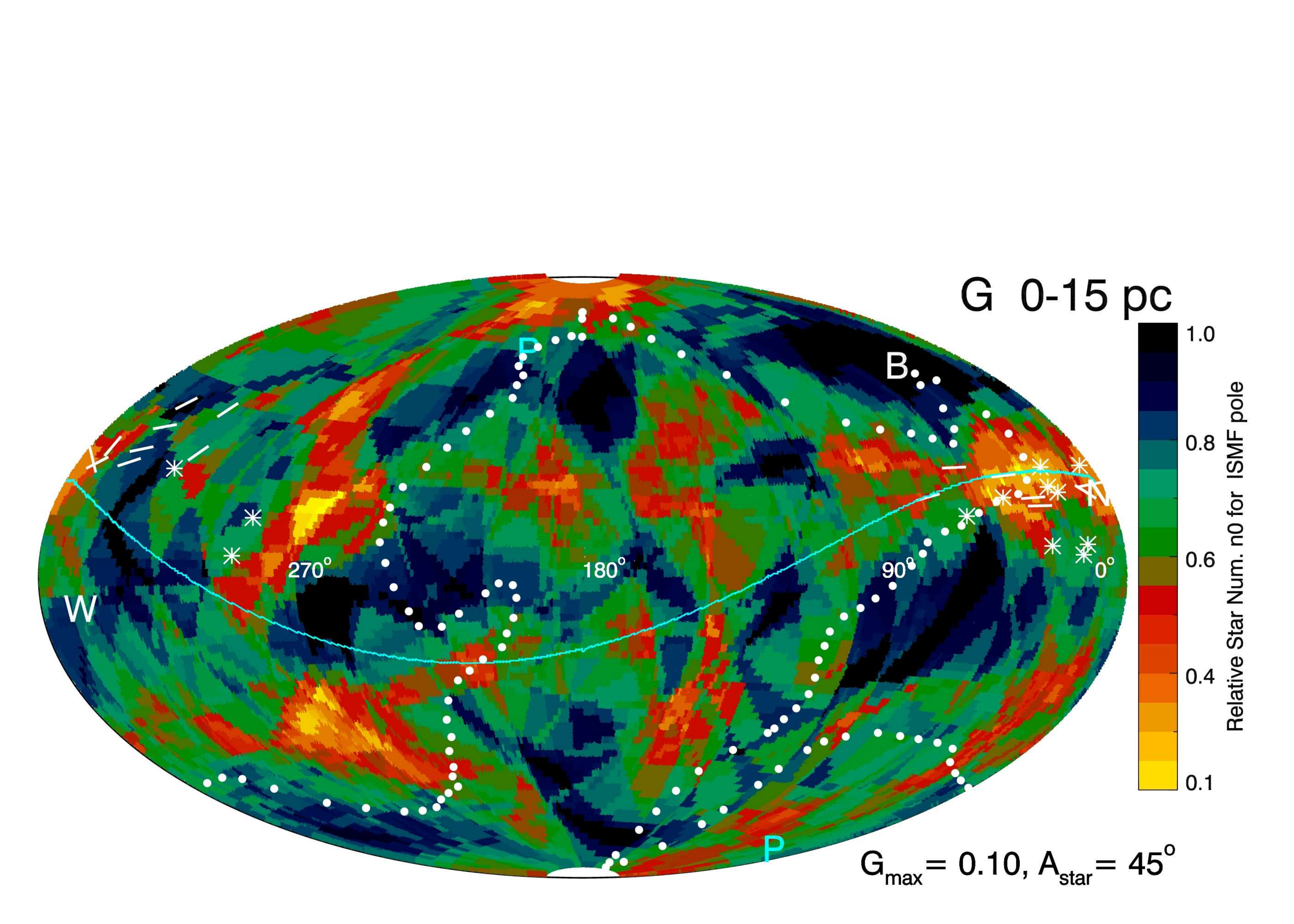}
\plottwo{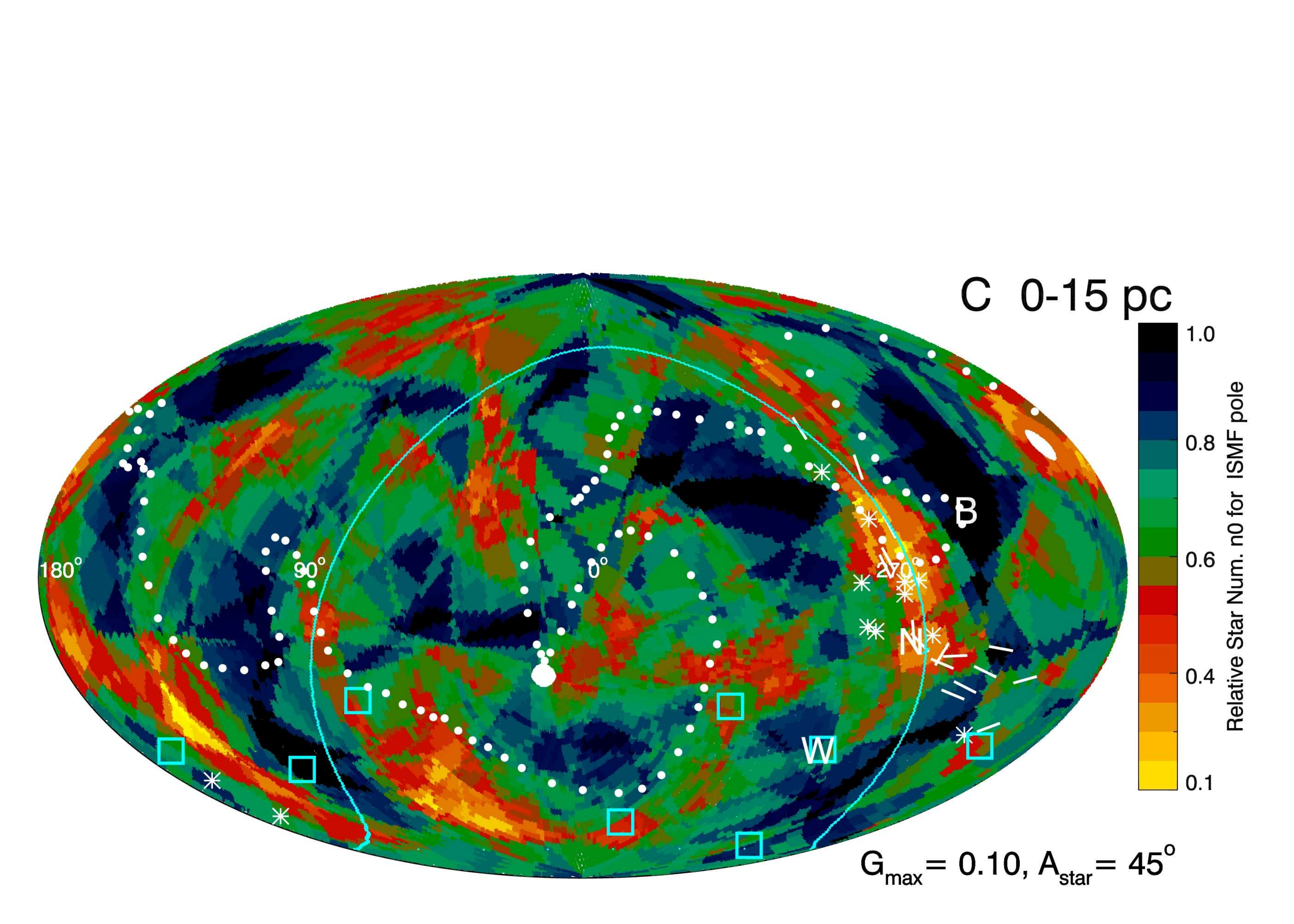}{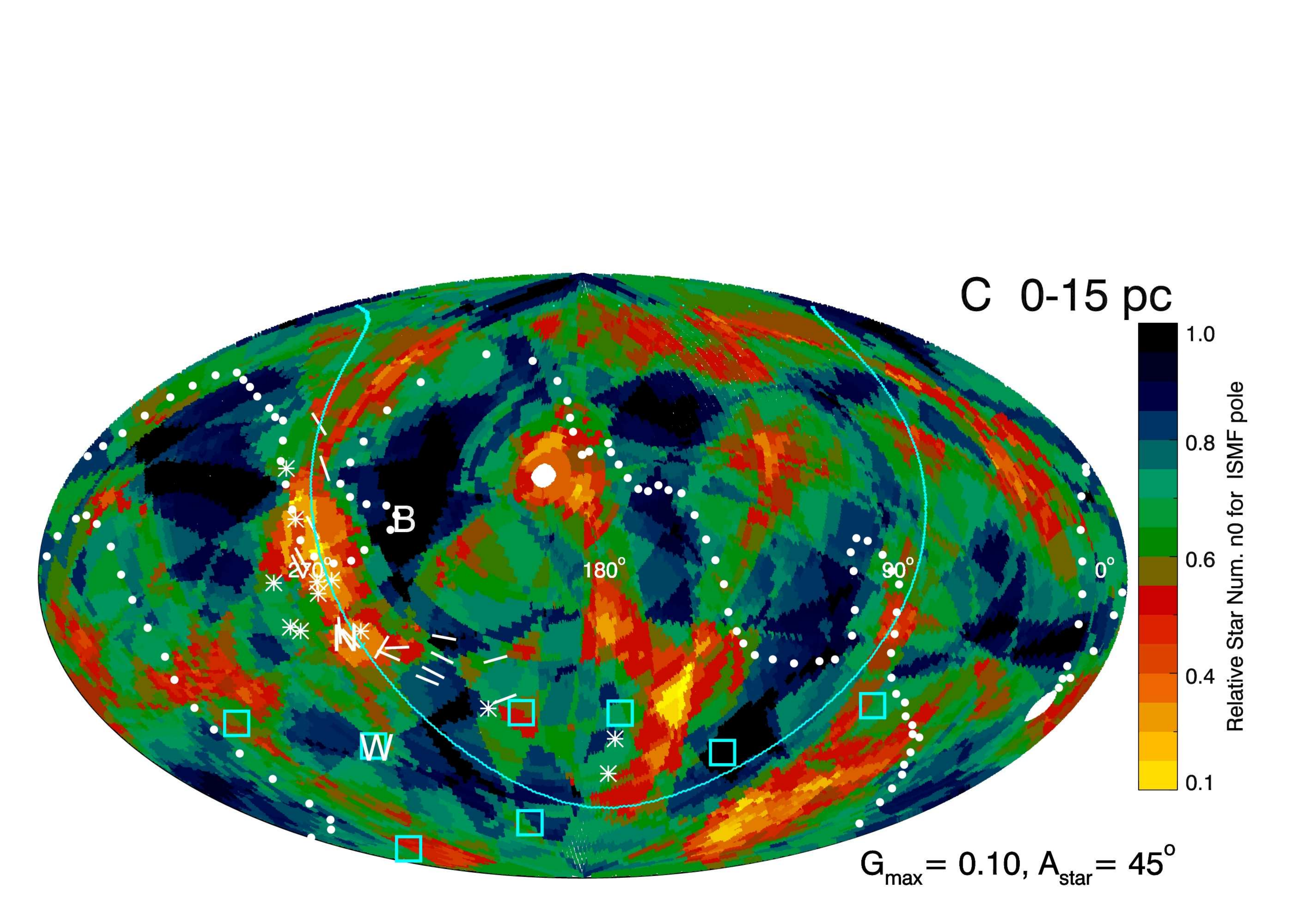}
\plottwo{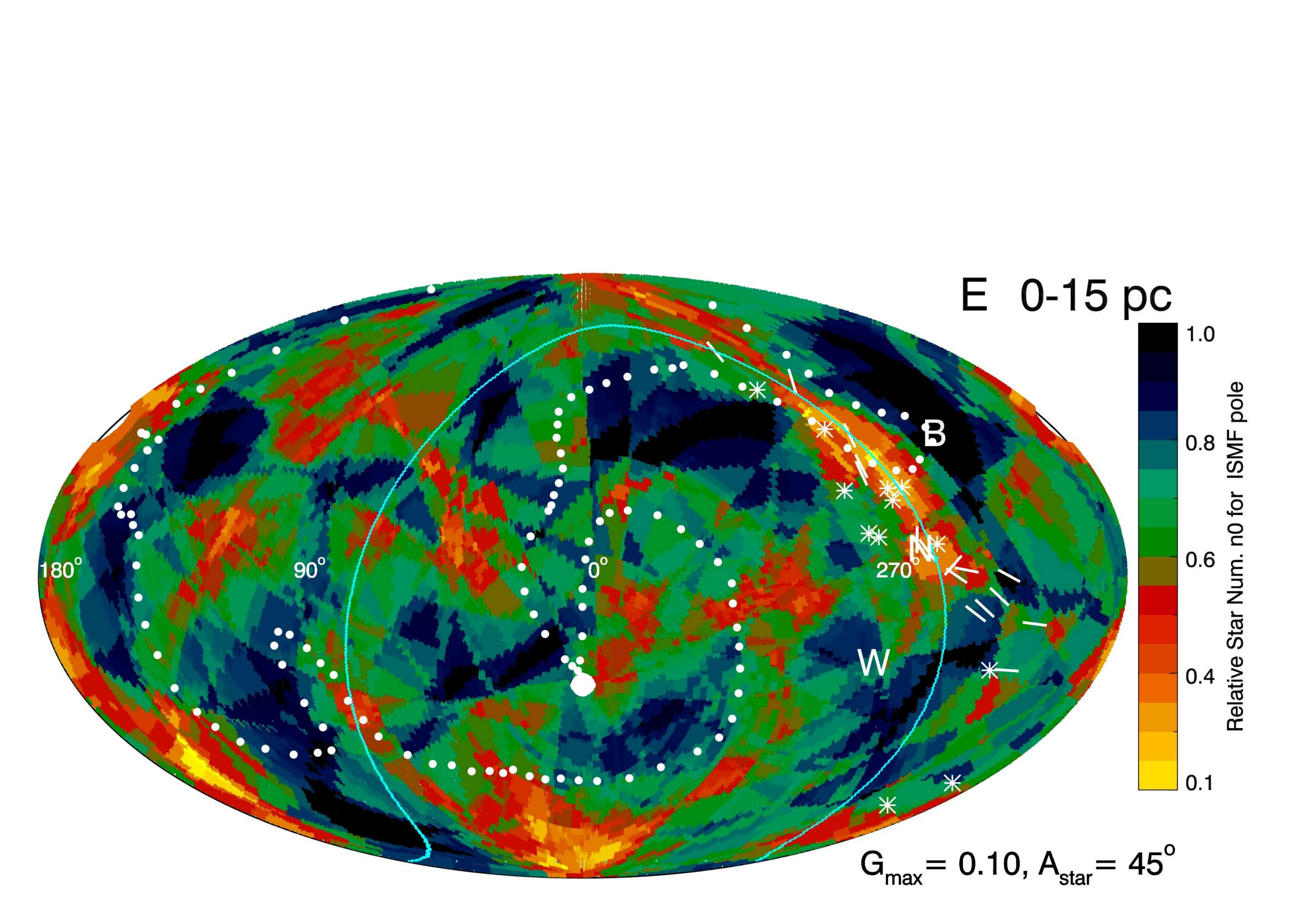}{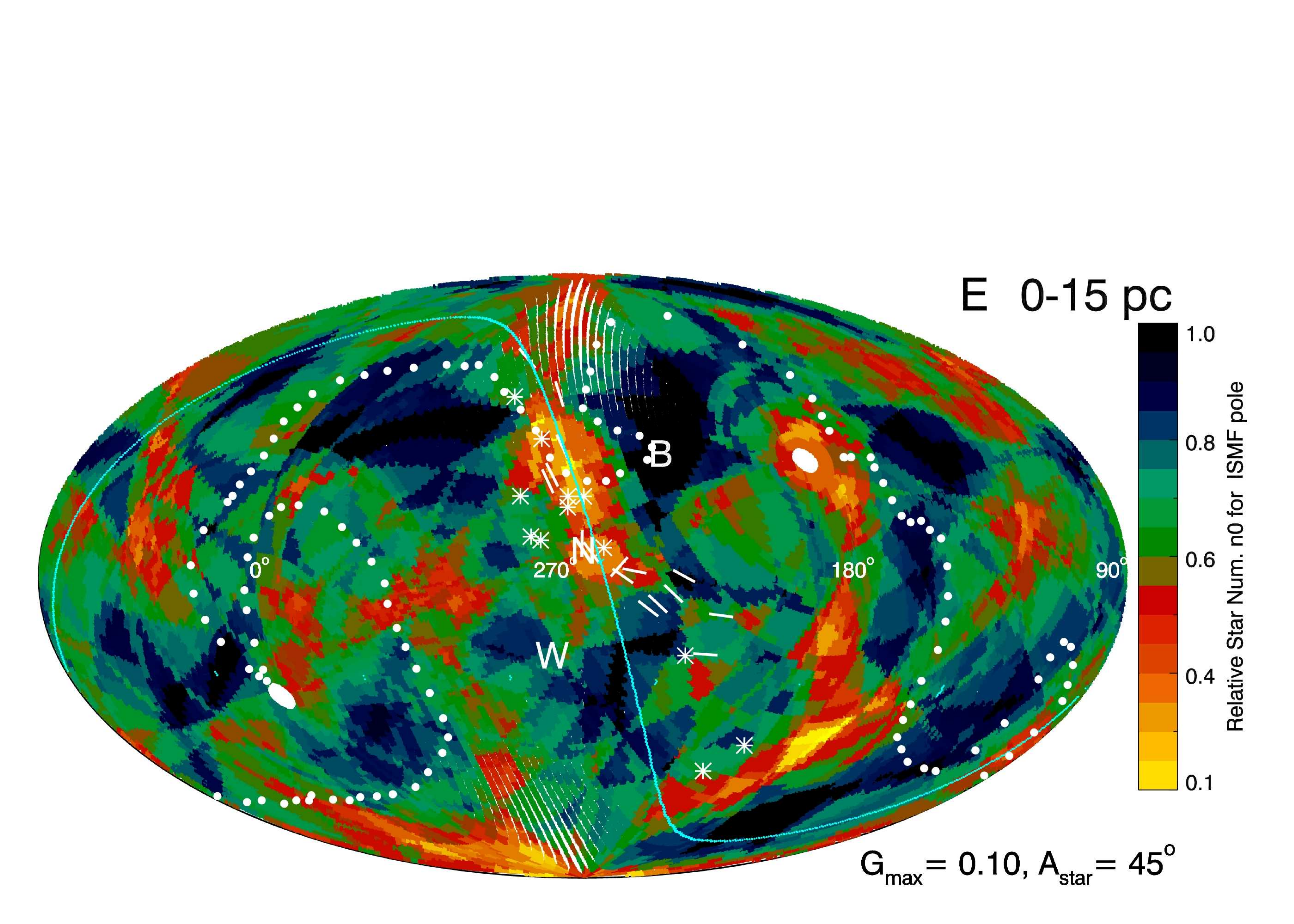}
\caption{Same as Figure \ref{fig:ang60} except that the angular
sampling is over intervals of \angmax=45\deeg.
}
\label{fig:ang45}
\end{figure}

\begin{figure}[t!] % FIGURE 9 
\plottwo{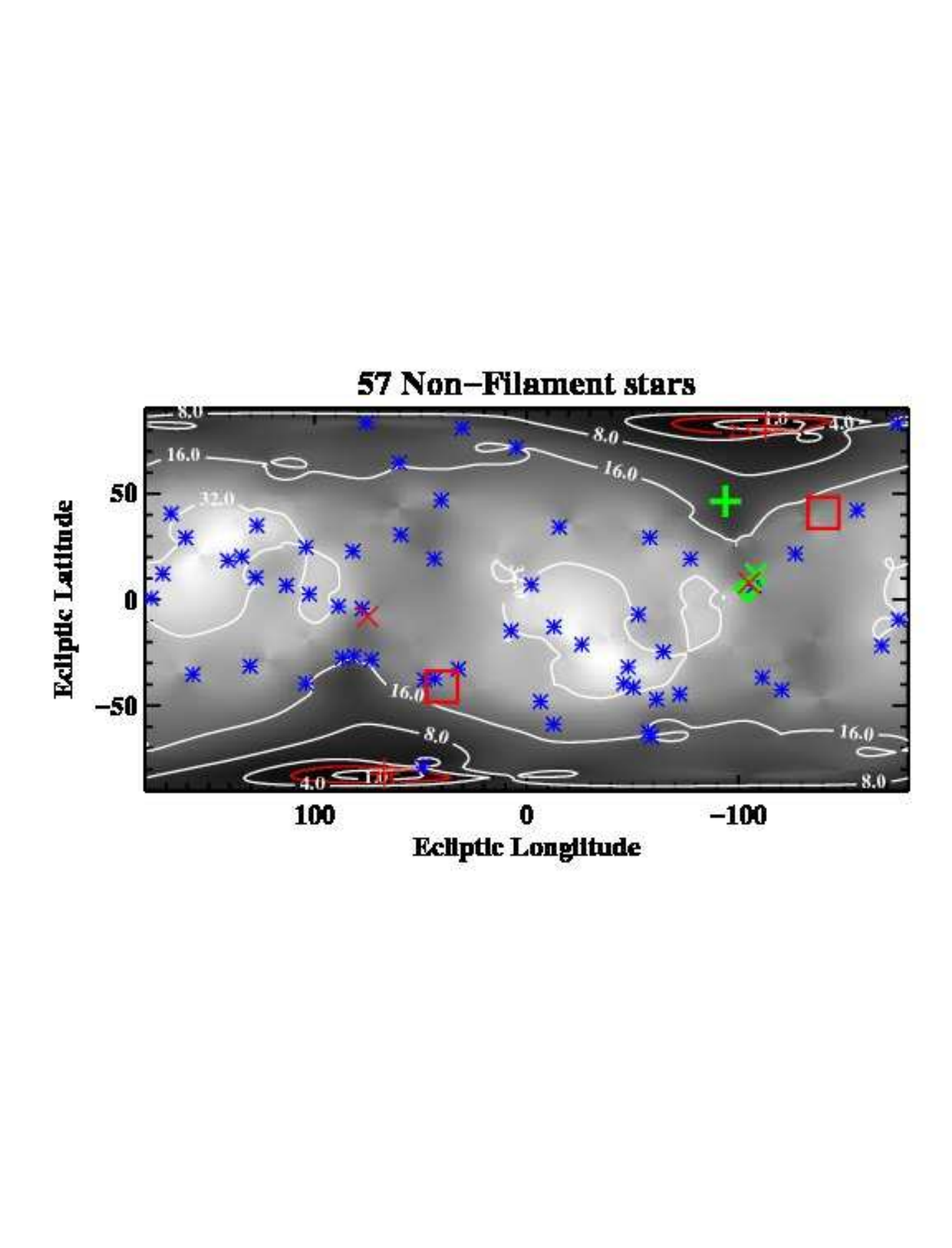}{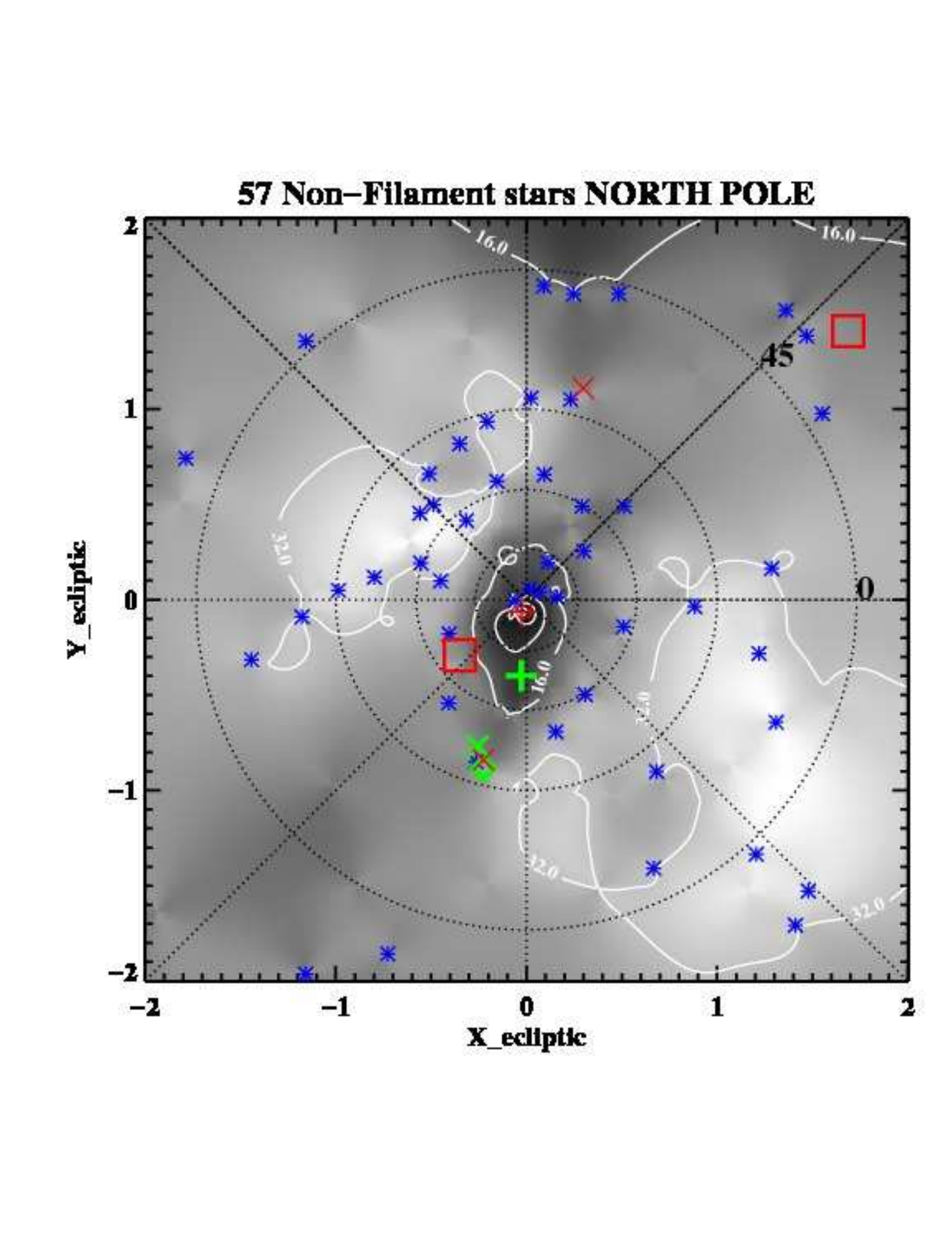}
\caption{Gray-scale image showing the \chisq\ probabilities that the
  best 57 stars creating the polarization band will trace a magnetic
  field at each location in the ecliptic, for a linear (left) and a
  stereographic projection centered on the north ecliptic pole
  (right).  The magnetic field direction traced by the polarization
  band (red ``X'') overlaps the north ecliptic pole (Table 1).  Blue
  asterisks show the locations of the 57 stars, tracing magnetic field
  directions within 10\deeg\ of the polarization band, used in this
  analysis.  Red squares show the opposite poles of the IBEX ribbon
  magnetic field.  Red plus symbols indicate locations of the poles of
  the dipole component of the magnetic field direction that is traced
  by the polarization band stars. Red ``Xs'' show the heliosphere nose
  and it's opposite direction. The locations of the stars used to
  obtain these results are plotted with blue asterisks.  The green
  diamonds, ``X'' and cross show the heliosphere nose location as
  defined by the interstellar \HeI\ wind, the warm breeze flowing into
  the heliosphere, and the ISMF direction found from polarization data
  in \citet{Frisch:2012ismf2}.  The red boxes indicates the ISMF
  direction from the IBEX ribbon and its opposite. }
\label{fig:heiles57}
\end{figure}

\begin{figure}[h!]  % FIGURE 10 
\plottwo{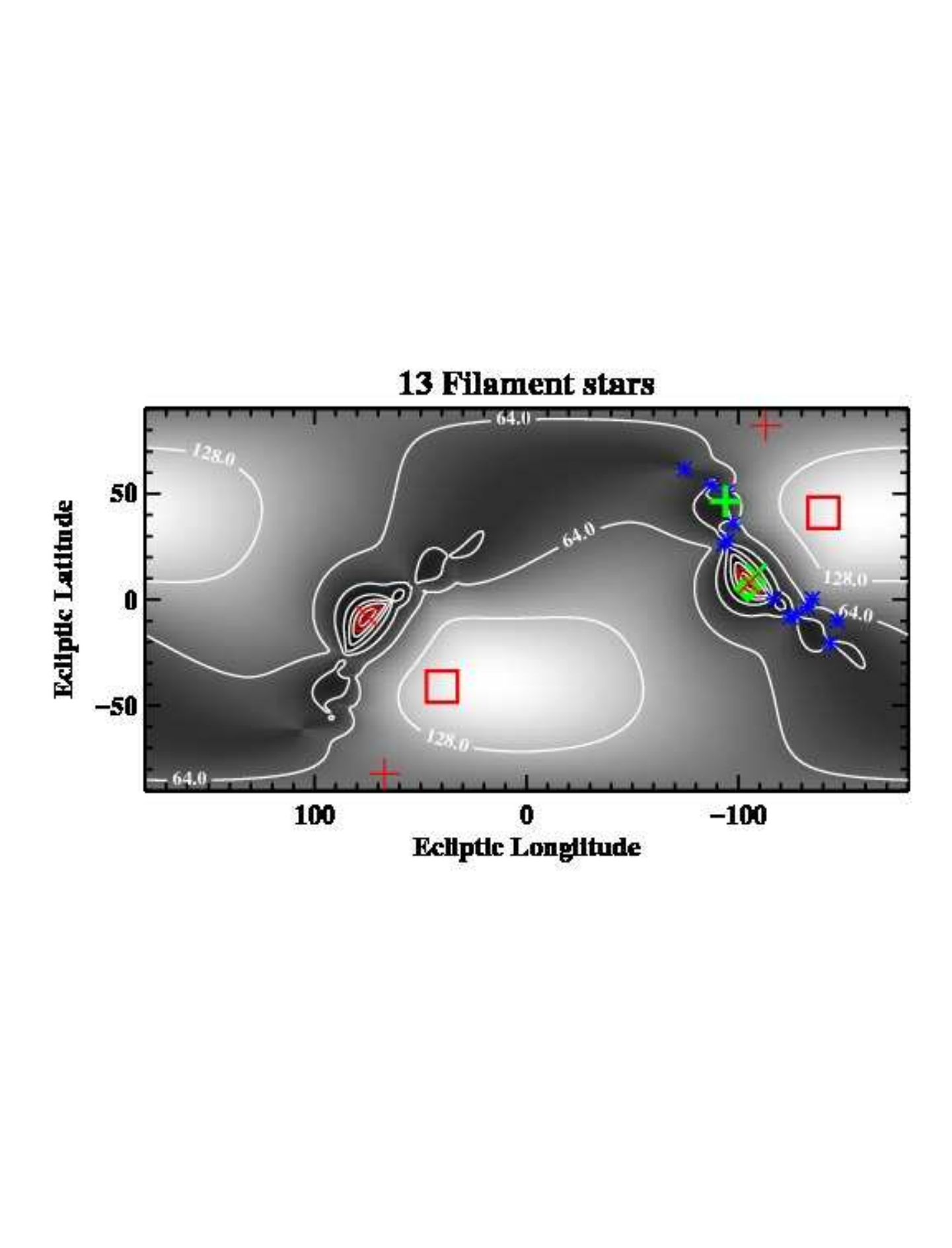}{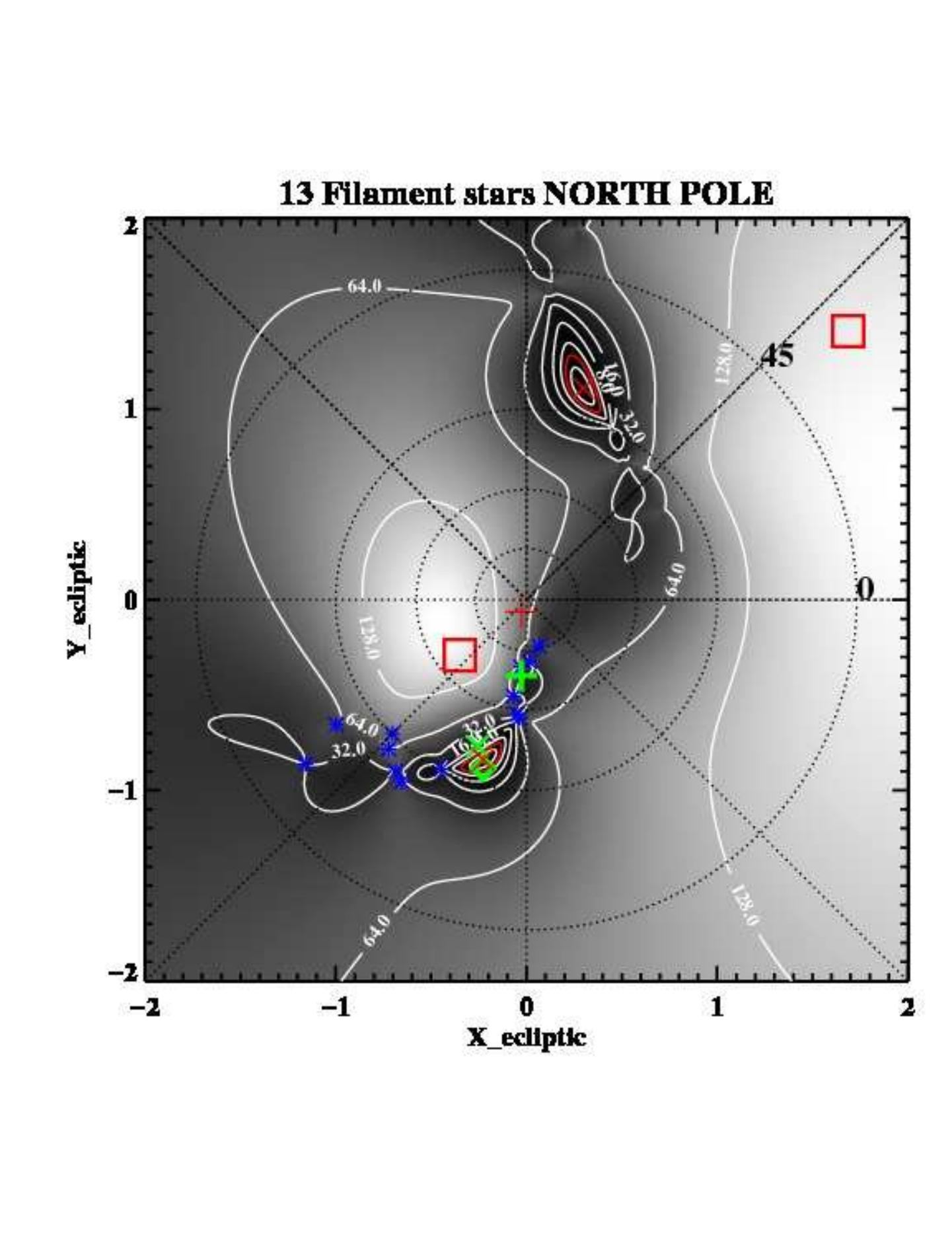}
\caption{Results of a \chisq ~analysis of the best-fitting magnetic
  pole to the 13 filament stars are plotted in a gray scale with
  contour values showing $\Delta \chi^2$.  The red contour shows
  \chisq ~=2.3, within which the magnetic pole has a 68\% of being
  located.  The Figures are plotted in ecliptic coordinates centered
  on \elon,\elat=0\deeg,0\deeg\ for a linear projection (left), and
  for a stereo projection centered on the north pole (right).  The low
  probability zones outlined by \chisq ~=128.0, where filament stars
  show no evidence of tracing a magnetic field, are located
  approximately downwind of the heliosphere nose.  The red ``X''
  symbols shows the opposite locations of the two magnetic poles for
  the filament stars.  See Figure \ref{fig:heiles57} for explanation
  of other symbols.  }
\label{fig:heilesfil}
\end{figure}

\begin{figure}[h!t]   %FIGURE 11
\plottwo{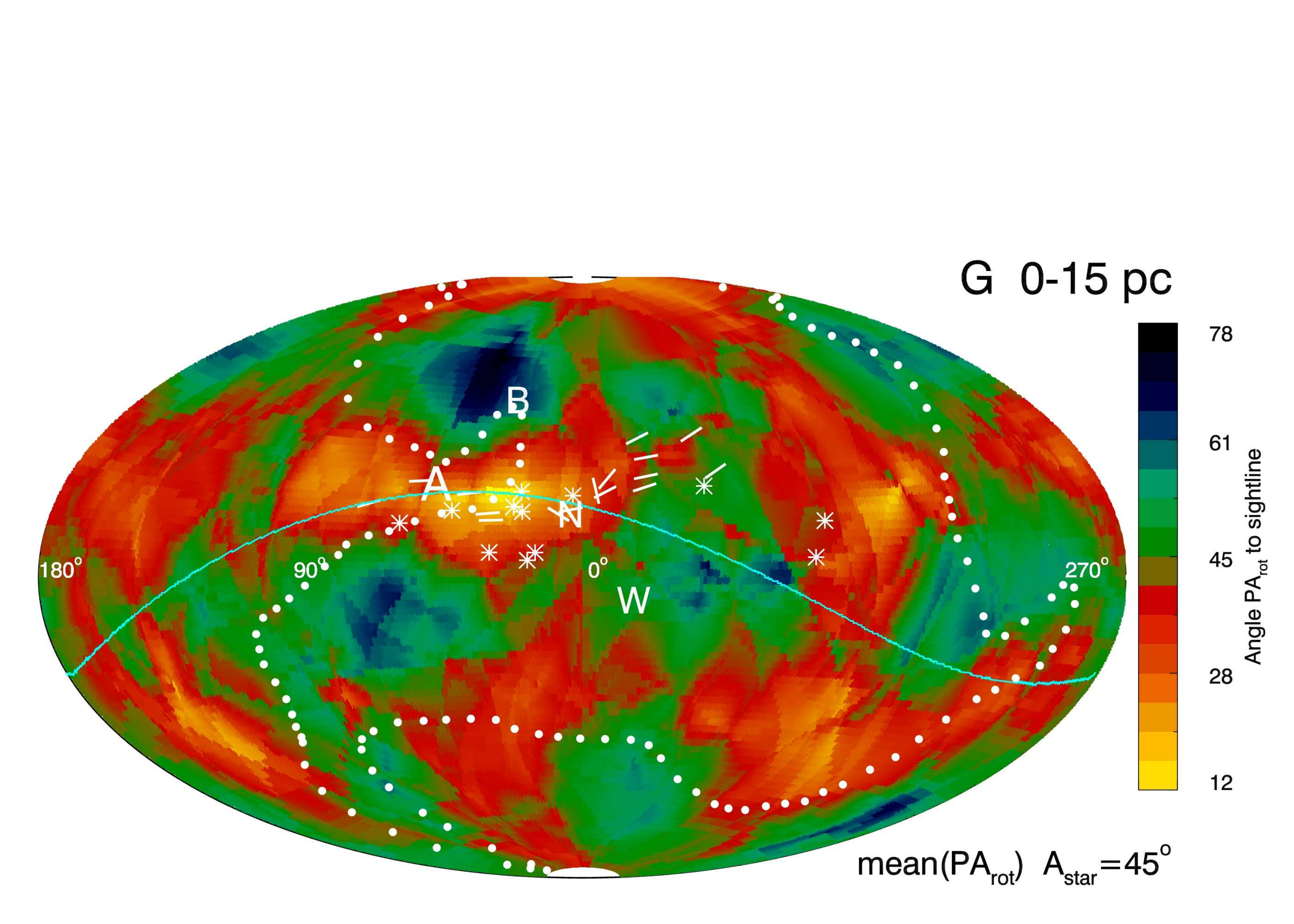}{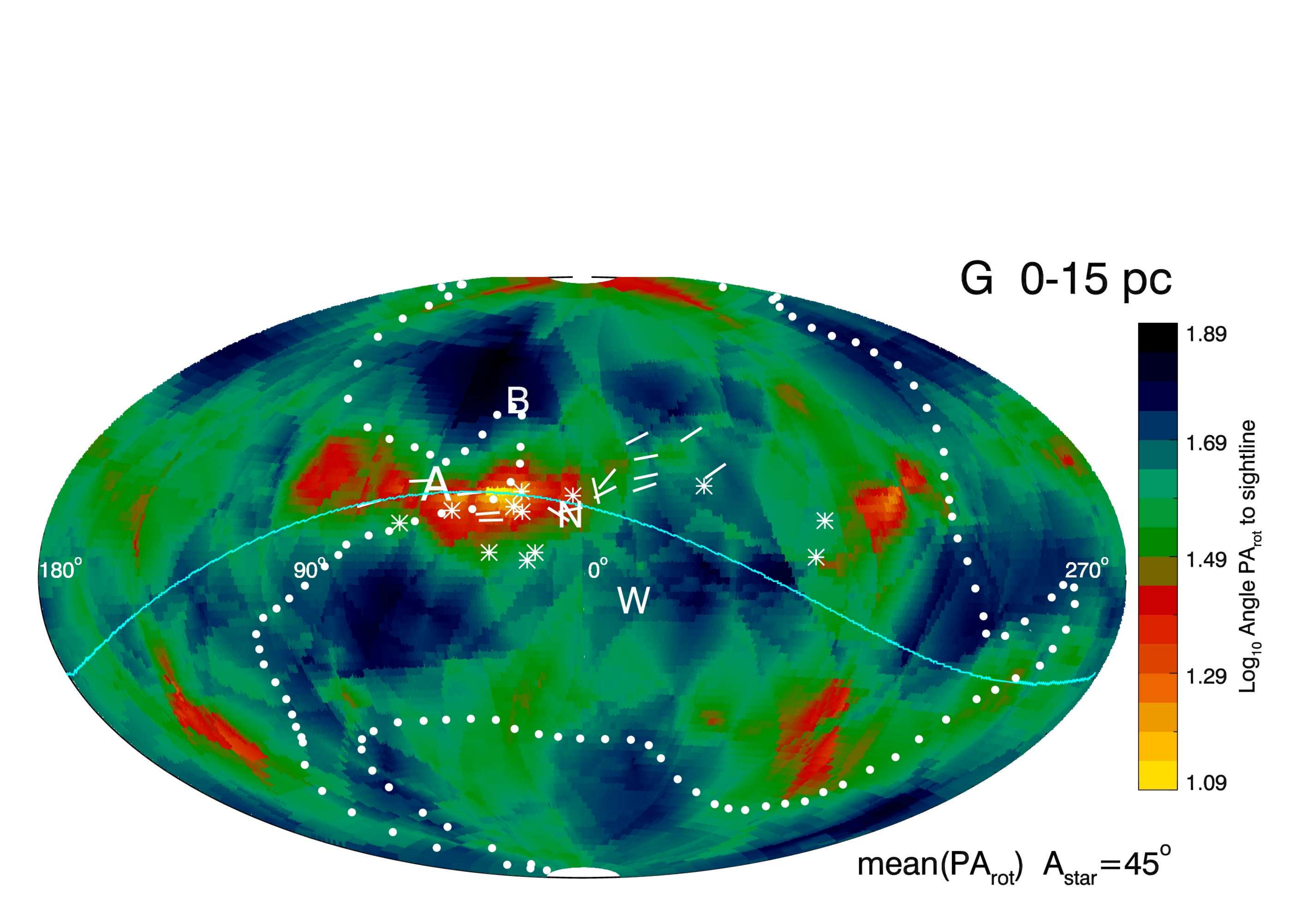}
\plottwo{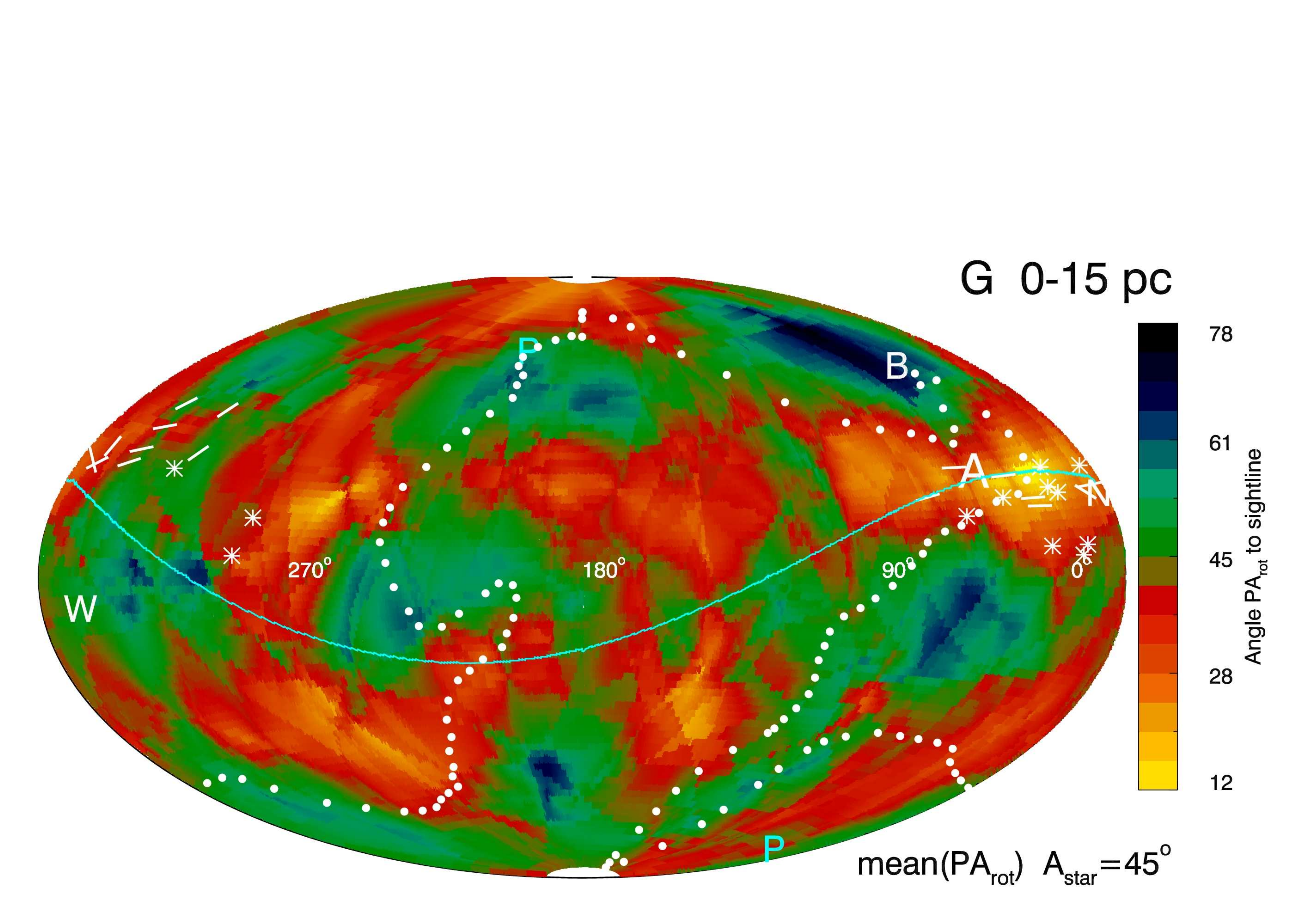}{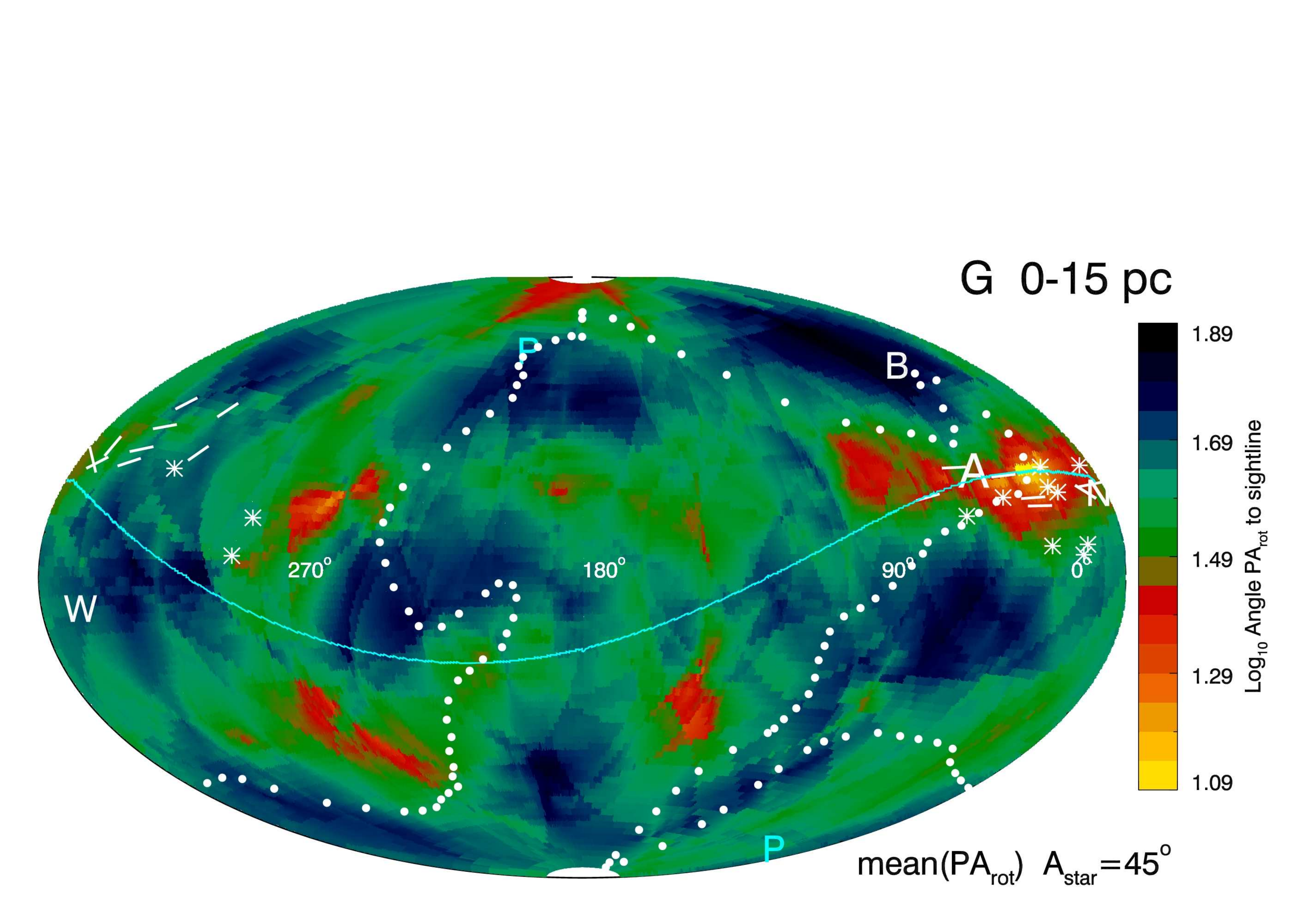}
\plottwo{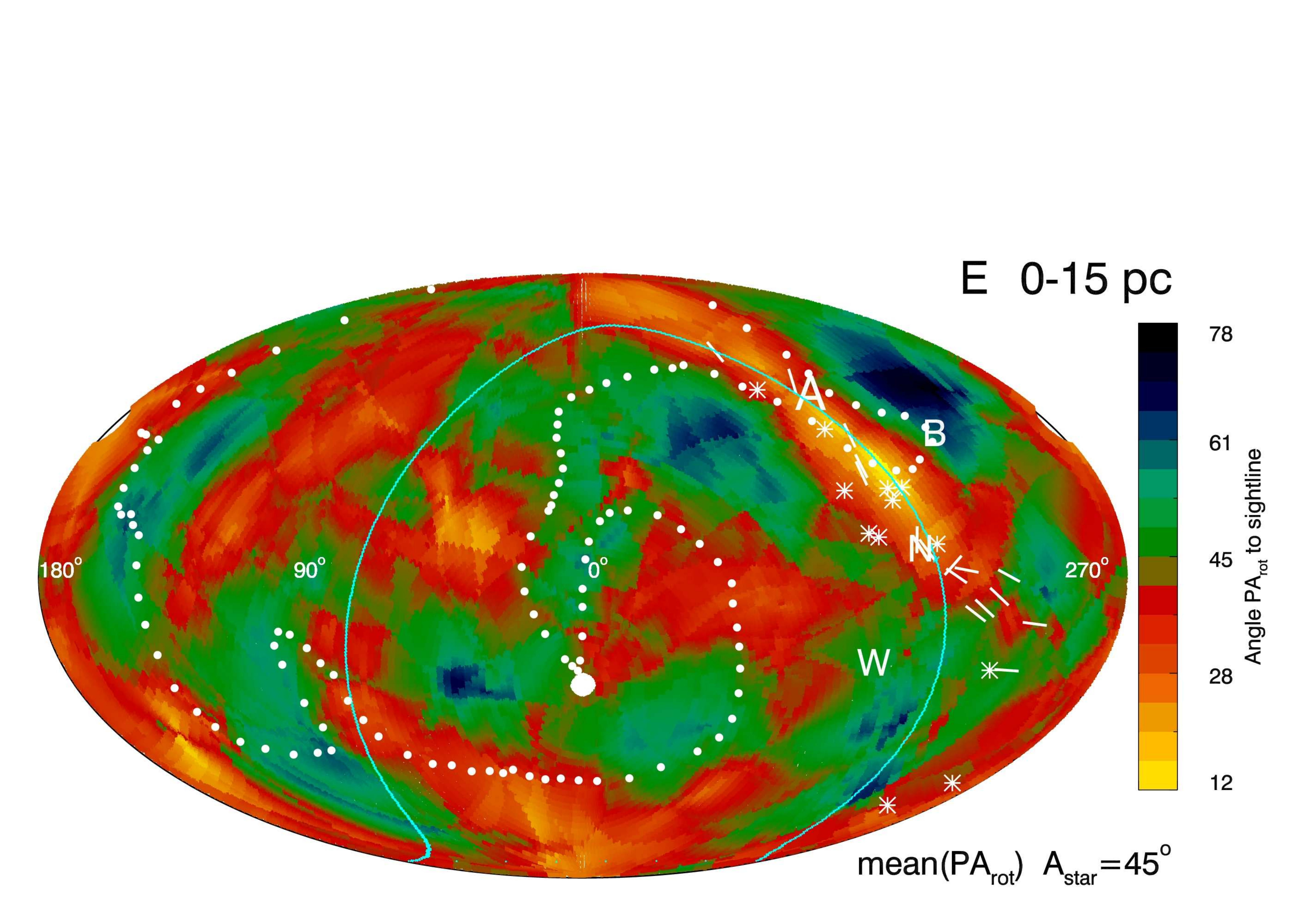}{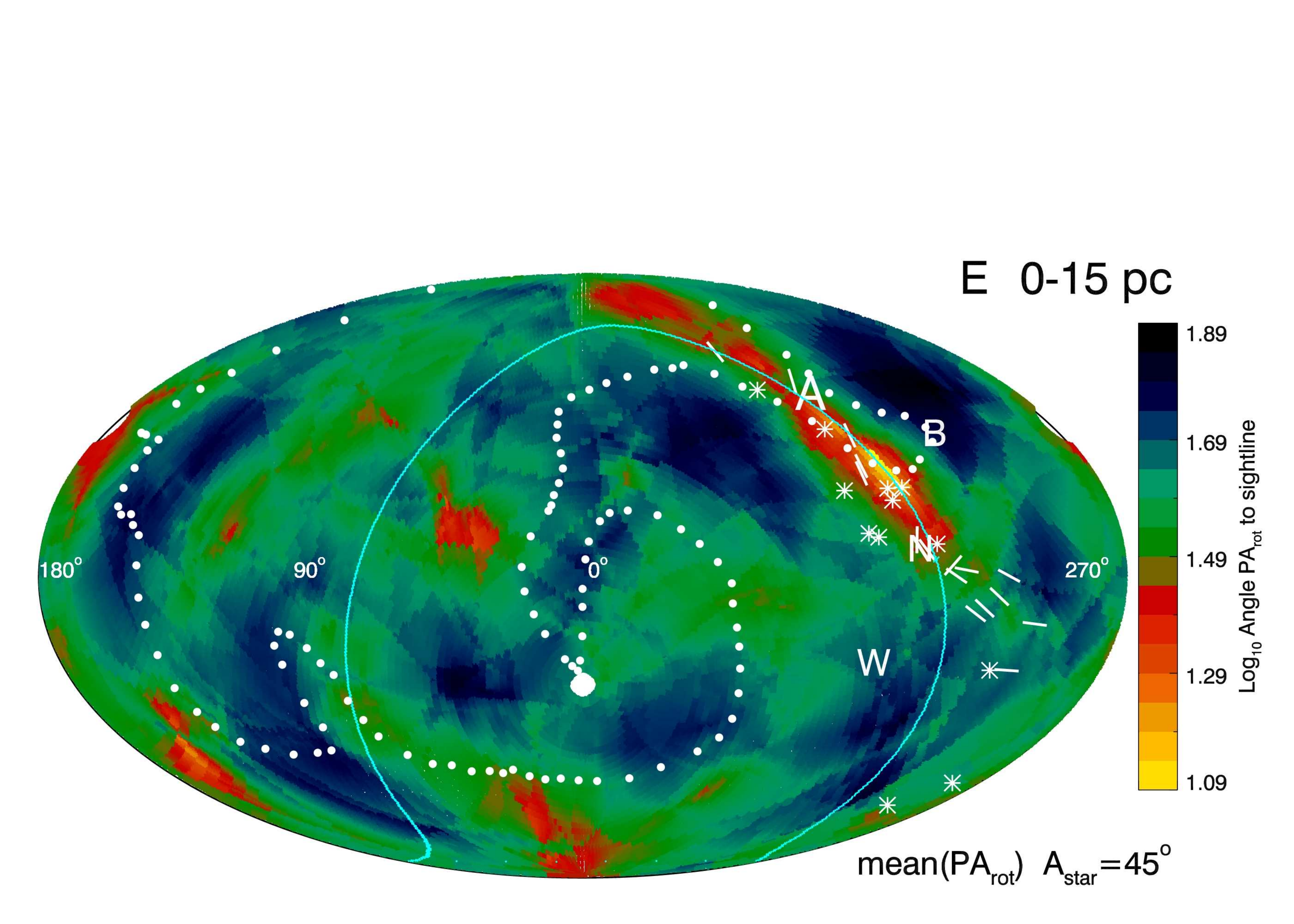}
\caption{Mean values of \PArot\ at at each grid location
\Lrot, \Brot\ are color-coded and plotted based on smoothing
radius of 45\deeg\ (\angmax=45\deeg).  Figures in the
left column are plotted on a linear color scale and
figures on the right column are plotted on a logarithmic
color scale (base 10).  Maps are displayed in galactic coordinates centered
on the galactic center (top), anti-center (middle),
and in ecliptic coordinates centered on 0\deeg\ (bottom).
See Figure \ref{fig:parot2} for additional projections.}
\label{fig:parot1}
\end{figure}

\begin{figure}[t!h]   %FIGURE 12
\plottwo{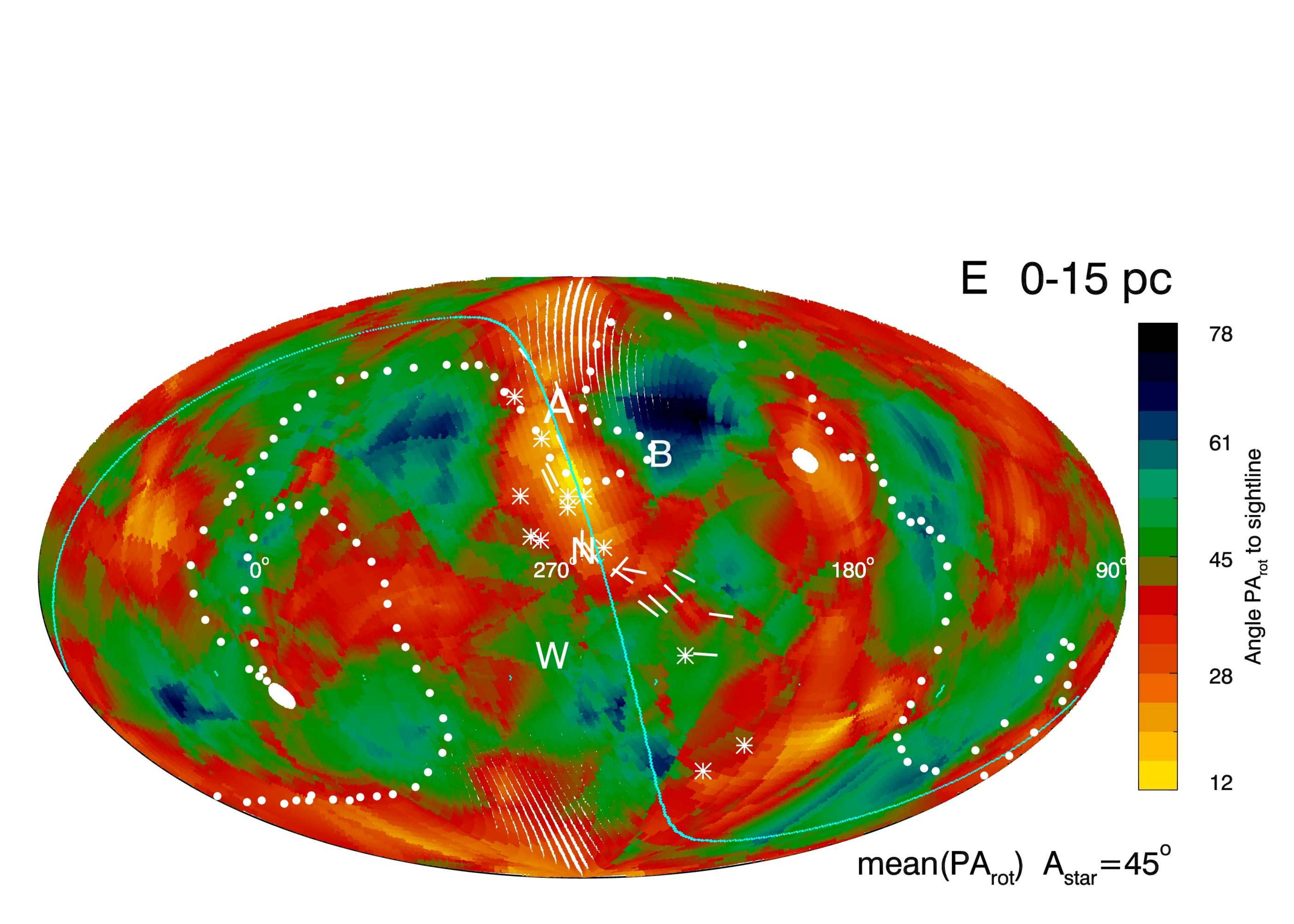}{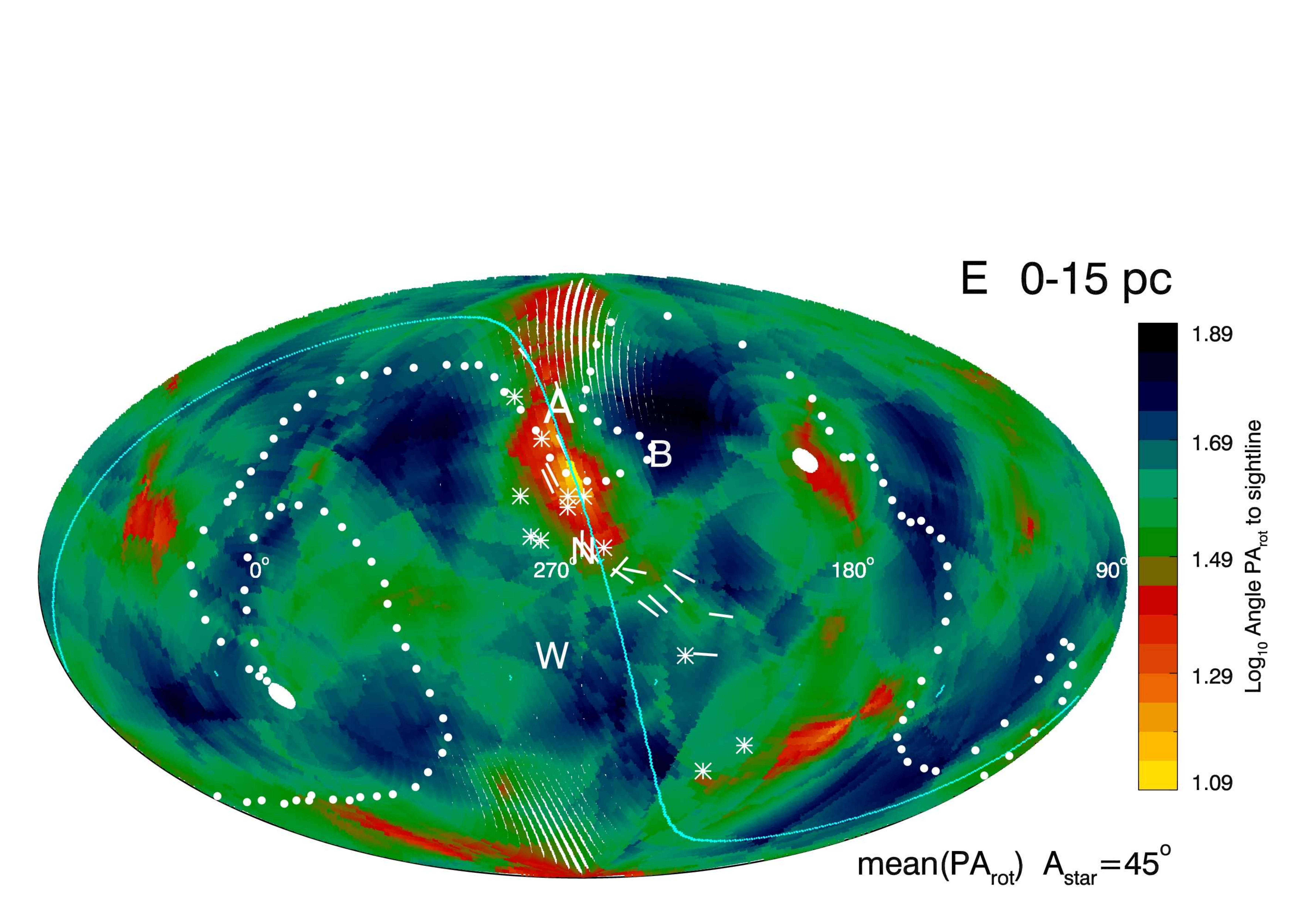}
\plottwo{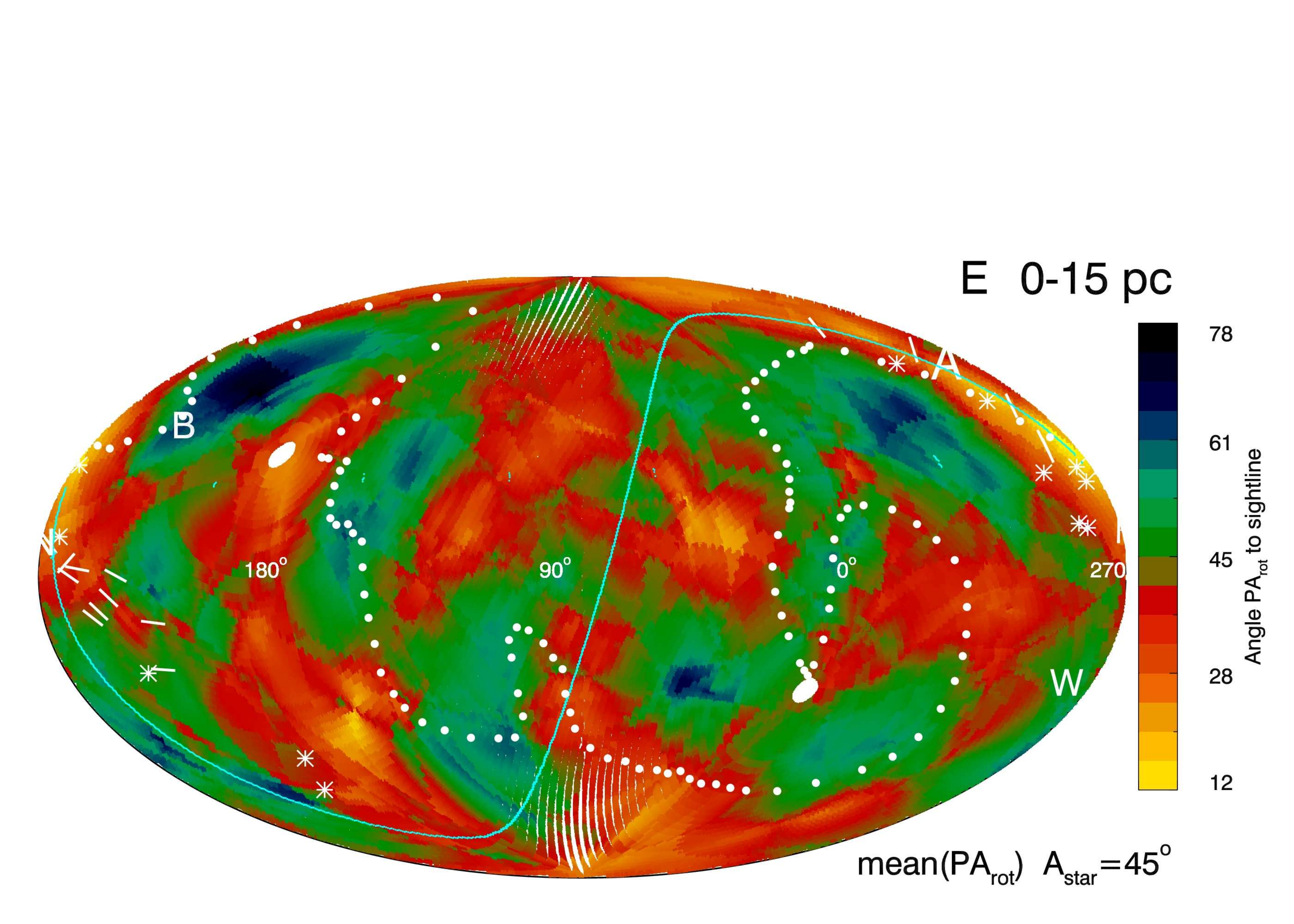}{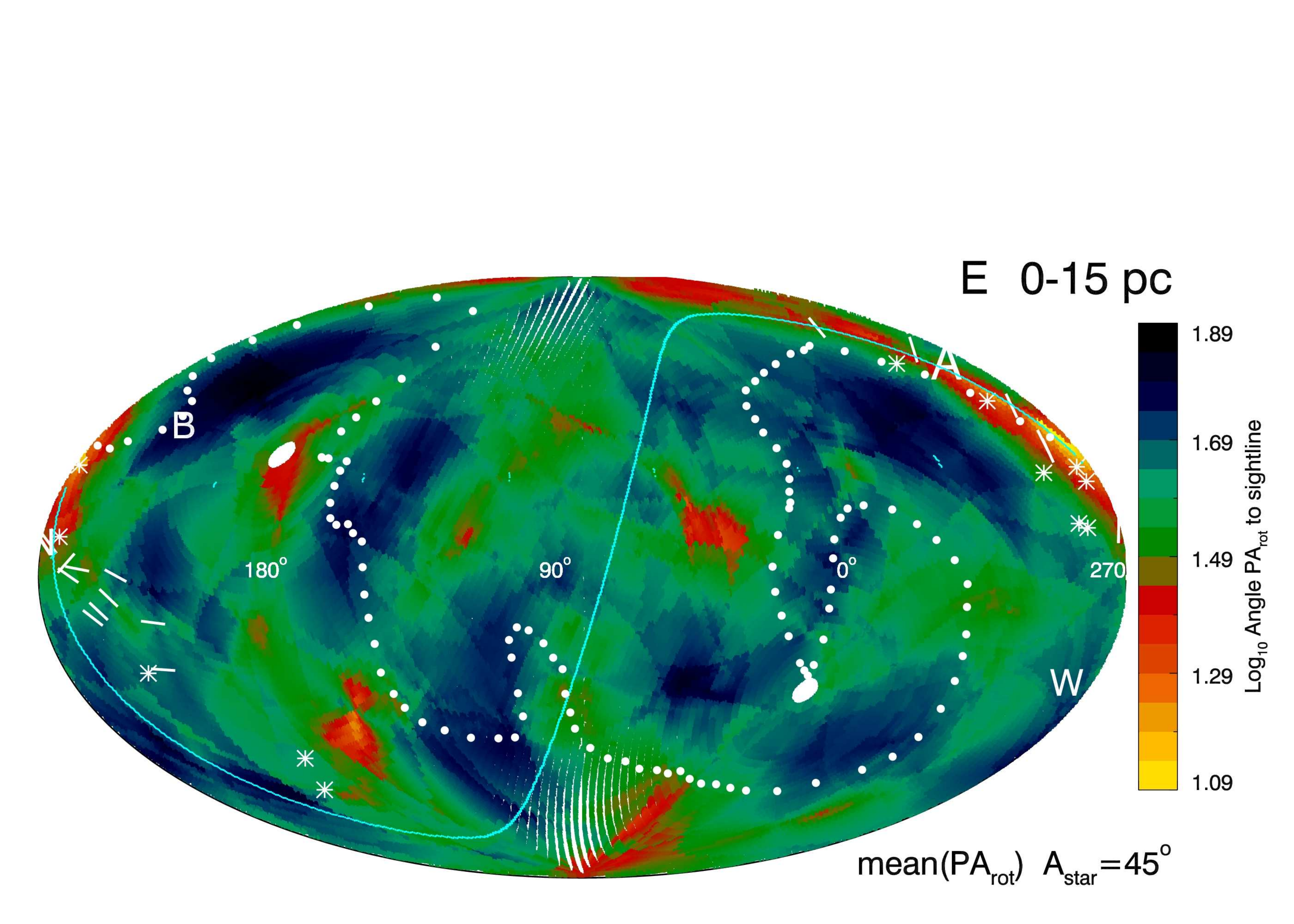}
\plottwo{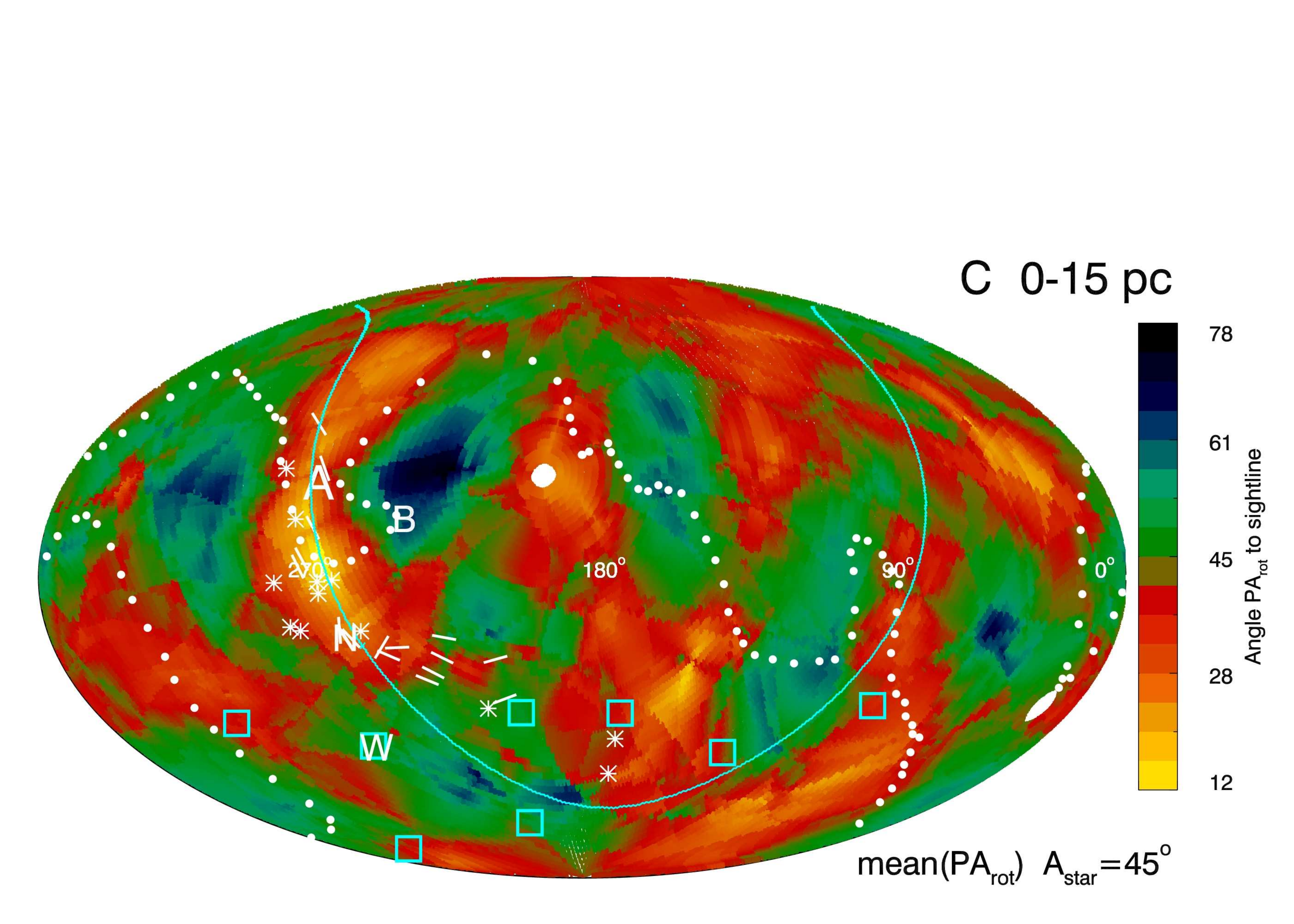}{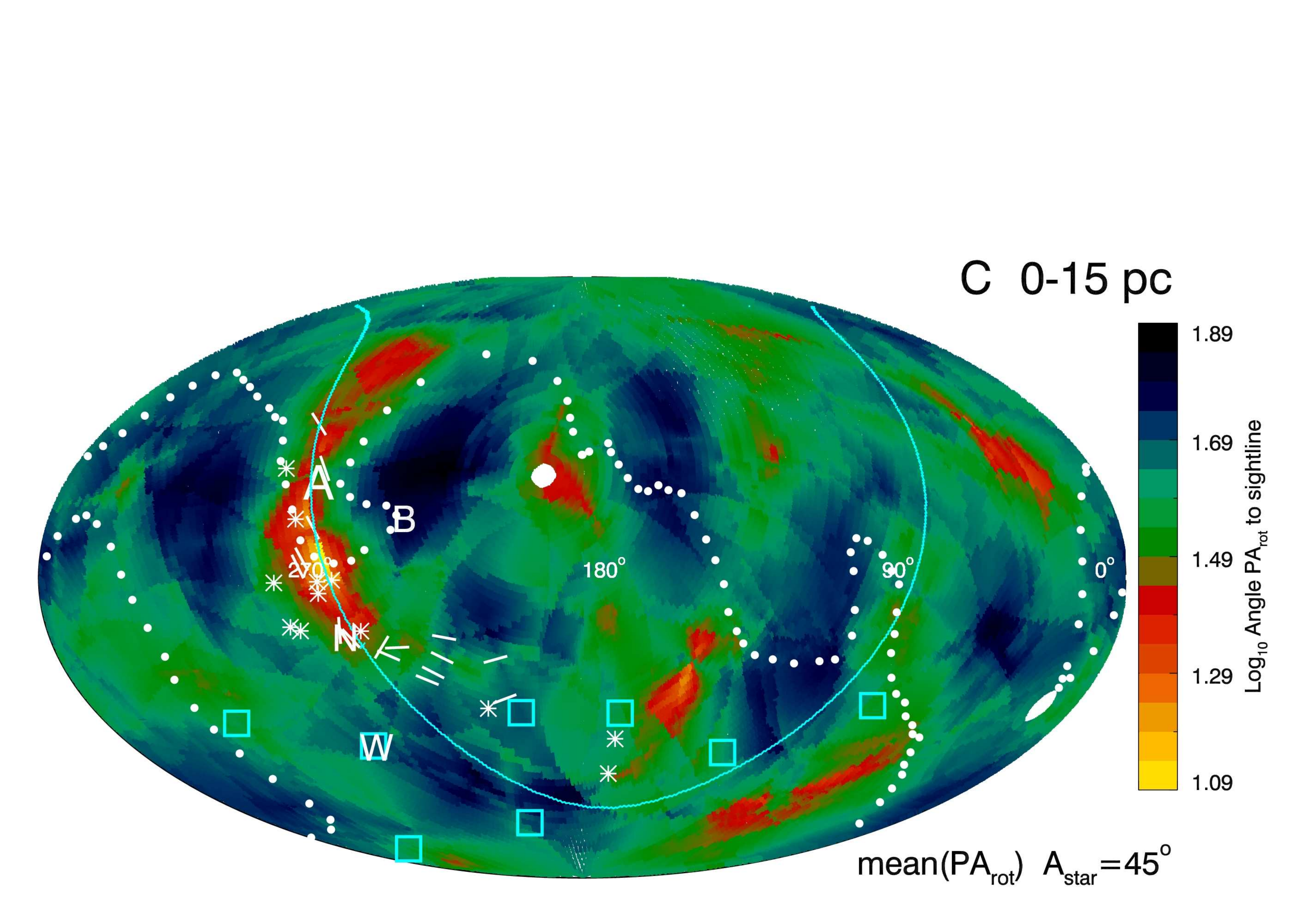}
\caption{Same as Figure \ref{fig:parot1} except that
projections are given in nose-centered and tail-centered
ecliptic coordinates (left, top and middle),
and equatorial coordinants centered on RA=12H.}
\label{fig:parot2}
\end{figure}

\begin{figure}[h!]  % FIGURE 13 
\plotone{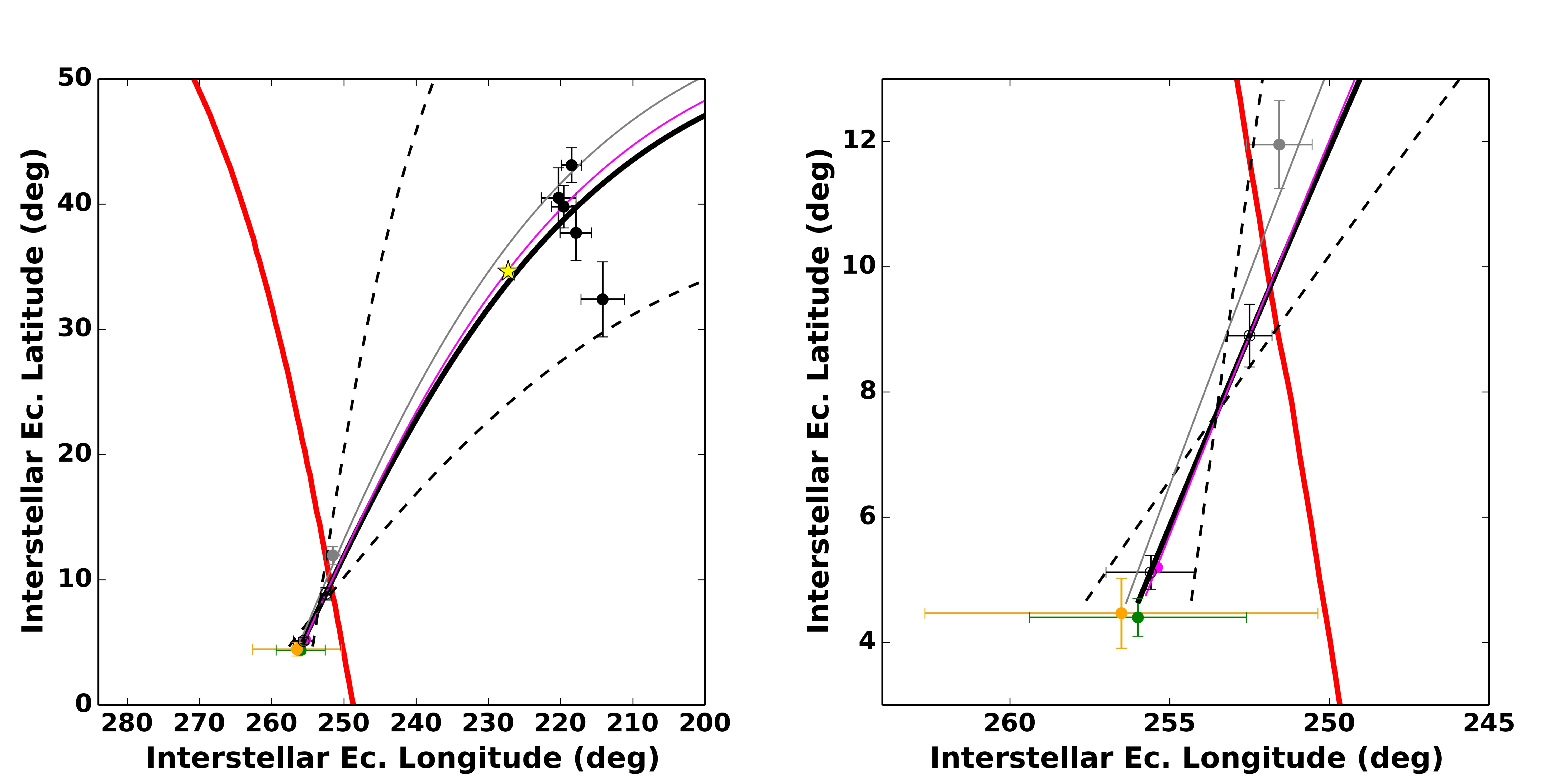}
%% \plotone{plotCentComb.pdf}
\caption{ The interstellar magnetic field direction \Bchm\
and the interstellar \HeI\ velocity \Vchm\ are offset from
  each other by 40\deeg\ and define the \BV ~plane that characterizes
  the observed offsets between primary neutral interstellar
  populations and those that include secondary neutrals created by
  charge-exchange between magnetically deflected ions and interstellar
  neutrals.  The thick red line shows the polarization band.
The \BV ~plane is shown for IBEX (thick black line and
  dashed lines) and Ulysses (pink line) \HeI\ velocities, and for the
  ISMF direction (star) obtained from MHD models of the ribbon
  \citep{Zirnstein:2016ismf}.  
The energy-dependent centers of the
  IBEX ribbon \citep{Funsten:2013} are shown with the set of five
  cross symbols, with lowest/highest ENA energies correspond to the
  ribbon centers at the highest/lowest latitudes.  
The polarization band (thick red line) is approximately parallel to the elongated
  configuration of the IBEX energy-dependent ribbon centers.  
  The warm secondary \HeI\ breeze (gray dot) and mixed primary and secondary
  \HI\ population (from SOHO/SWAN data, open gray box) are offset
  along the \BV\ plane, and away from the primary populations that
  anchor the \BV\ plane at the lowest latitudes.  The upwind
  directions of \HeI\ from IBEX (black circle) and Ulysses(pink dot),
  and IBEX results for the interstellar \OI\ wind (orange and green
  dots) define the \BV\ band at the lowest latitudes.  The right
  figure is a zoomed version of the left figure.  This figure is
  adapted from \citet{Schwadron:2015triBismf} and
  \citet{SchwadronMcComas:2017bv}.  }
\label{fig:ns}
\end{figure}

\begin{figure}[t!] % FIGURE 14 
\plotone{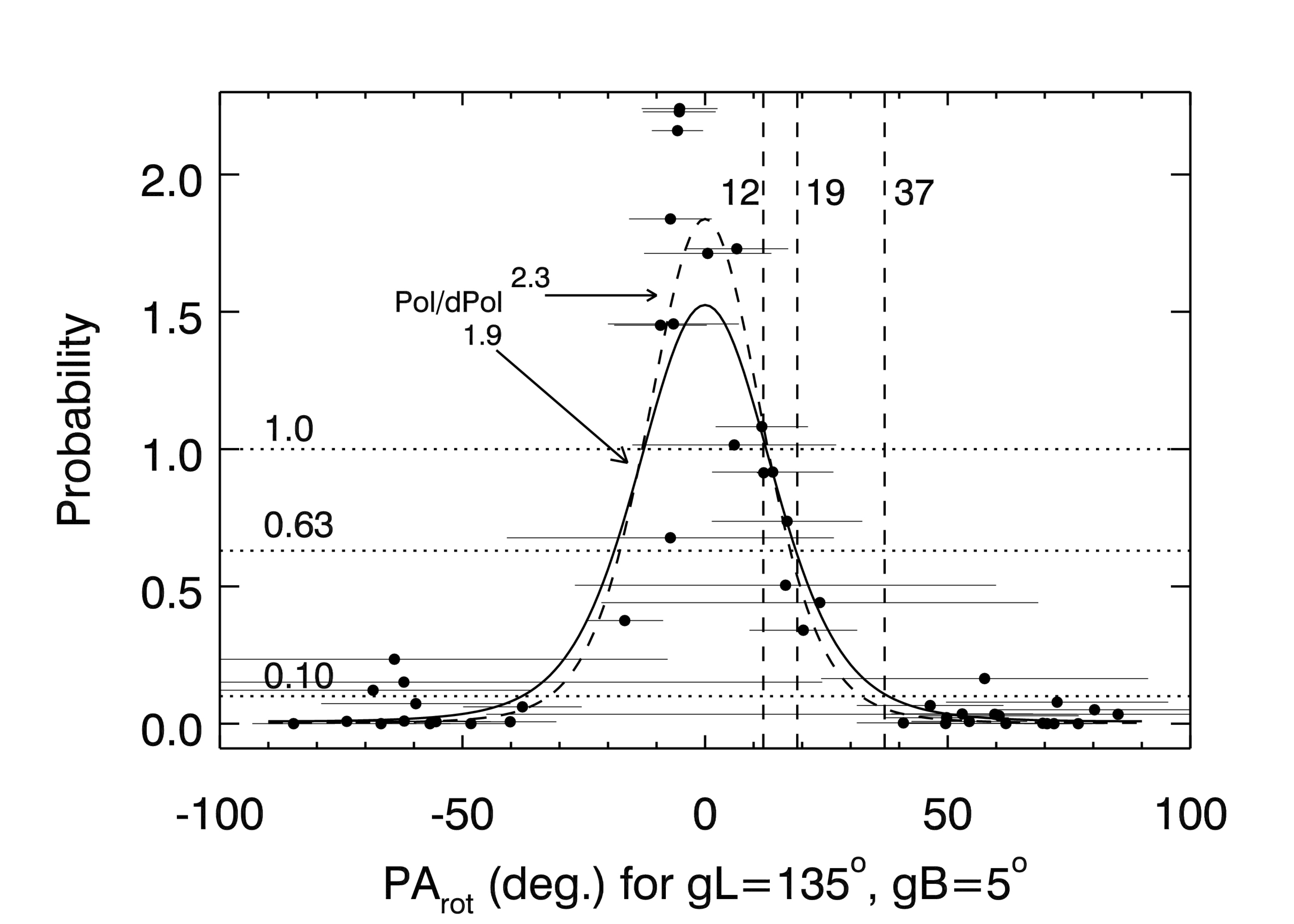}
\caption{The probabilities \Grot\ that stars within 90\deeg\ of the
  location \Lrot=135\deeg, \Brot=5\deeg\ (\elon,\elat=63\deeg,46\deeg,
RA,DEC=42\deeg,65\deeg) trace an ISMF toward that
  location (ordinate) are plotted against polarization position angles
  rotated to that location, \PArot\ (abscissa).  Values of \PArot\ for
  this set of stars are shown as dots.  The solid (dashed) curves show
  the theoretical probability distributions for \Pol/\dPol=1.9 (2.3)
  based on eqn. A1.  Probability values 0.10, 0.63, and 1.0 are
  plotted with dotted horizontal lines as an illustration of the
  effect of \Gmin\ (or \Gmax) on the selection of plotted data.  The
  vertical dashed lines show the intersections of these probability
  values with the probability distributions.  For example, stars with
  \Grot$>$\Gmin=0.63 would tend to trace magnetic field directions
  that are within 19\deeg\ of the sightline, although the detailed
  cutoff angle depends on \Pol/\dPol.  Stars with \Grot$<$\Gmax=0.1
  will tend to trace magnetic field directions with angles that are
  larger than 37\deeg\ with respect to the sightlines, or
  alternatively within 53\deeg\ of the plane of the sky.}
\label{fig:gfact1p9}
\end{figure}

\begin{figure}[t!] % FIGURE 15
\plotone{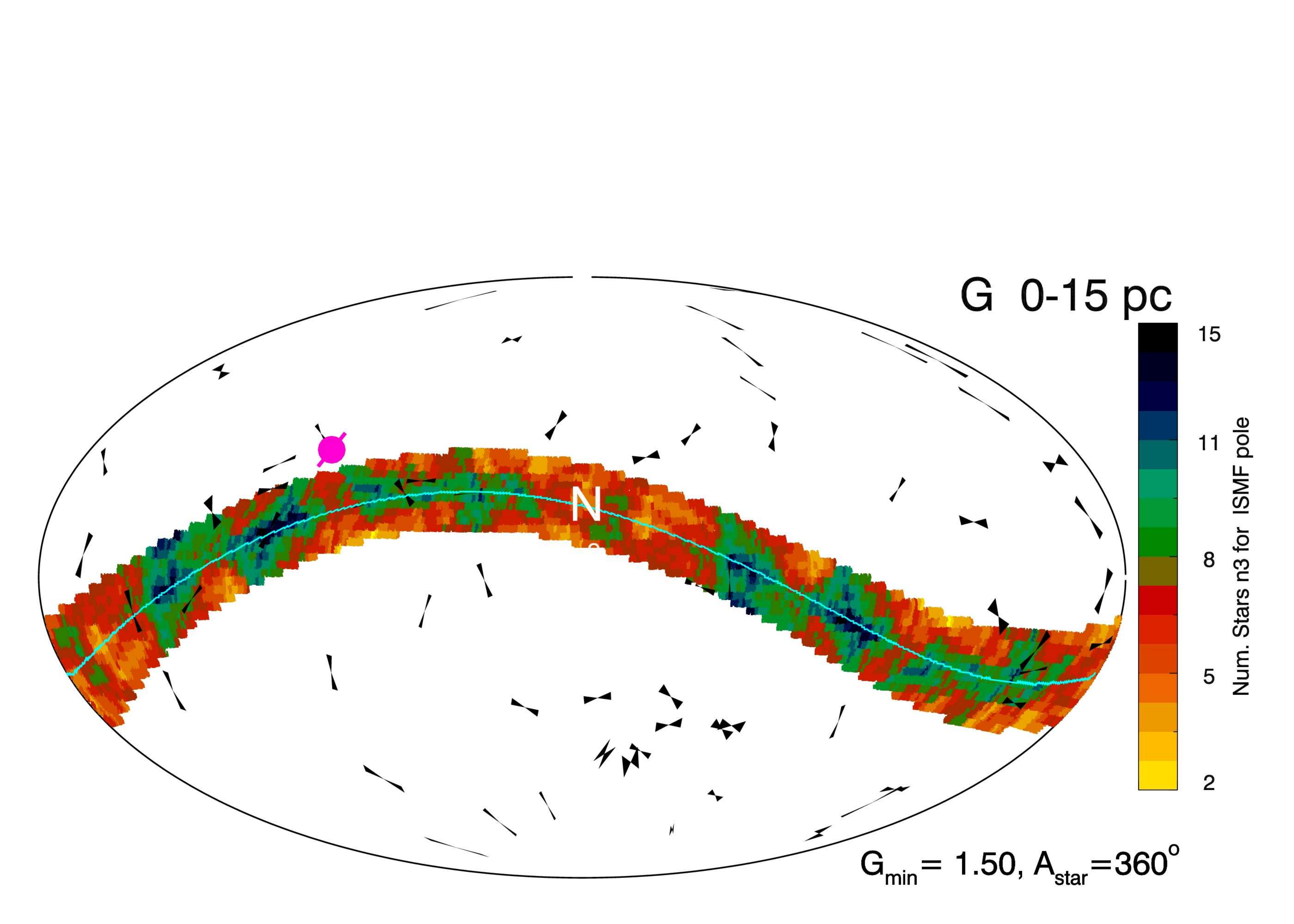}
\caption{The subset of the overlapping polarization position angle
  swaths that fall within $\pm 10^\circ$ of the polarization band
  great circle (\S \ref{sec:ang360}) are plotted.  The polarization
  band was defined by the great circle that is 90\deeg\ from an axis
  located at \glon=214\deeg, \glat=67\deeg\ (\S \ref{sec:ang360}) and
  is created by enhanced numbers of overlapping polarization position
  angle swaths compared to adjacent regions.  The triangular symbols
  show the locations and polarizations of all stars with polarization
  position angles that trace a magnetic field location within
  10\deeg\ of the polarization band great circle. 
  The highest quality subset of data that trace the polarization band 
  are used for the least-square fits to the magnetic dipole component (\S \ref{sec:heiles57}).  
The pink dot shows the location of this best-fitting magnetic field direction,
which is located close to the north ecliptic pole.}
\label{fig:polband}
\end{figure}

\begin{figure}[t!] % FIGURE 16
\plotone{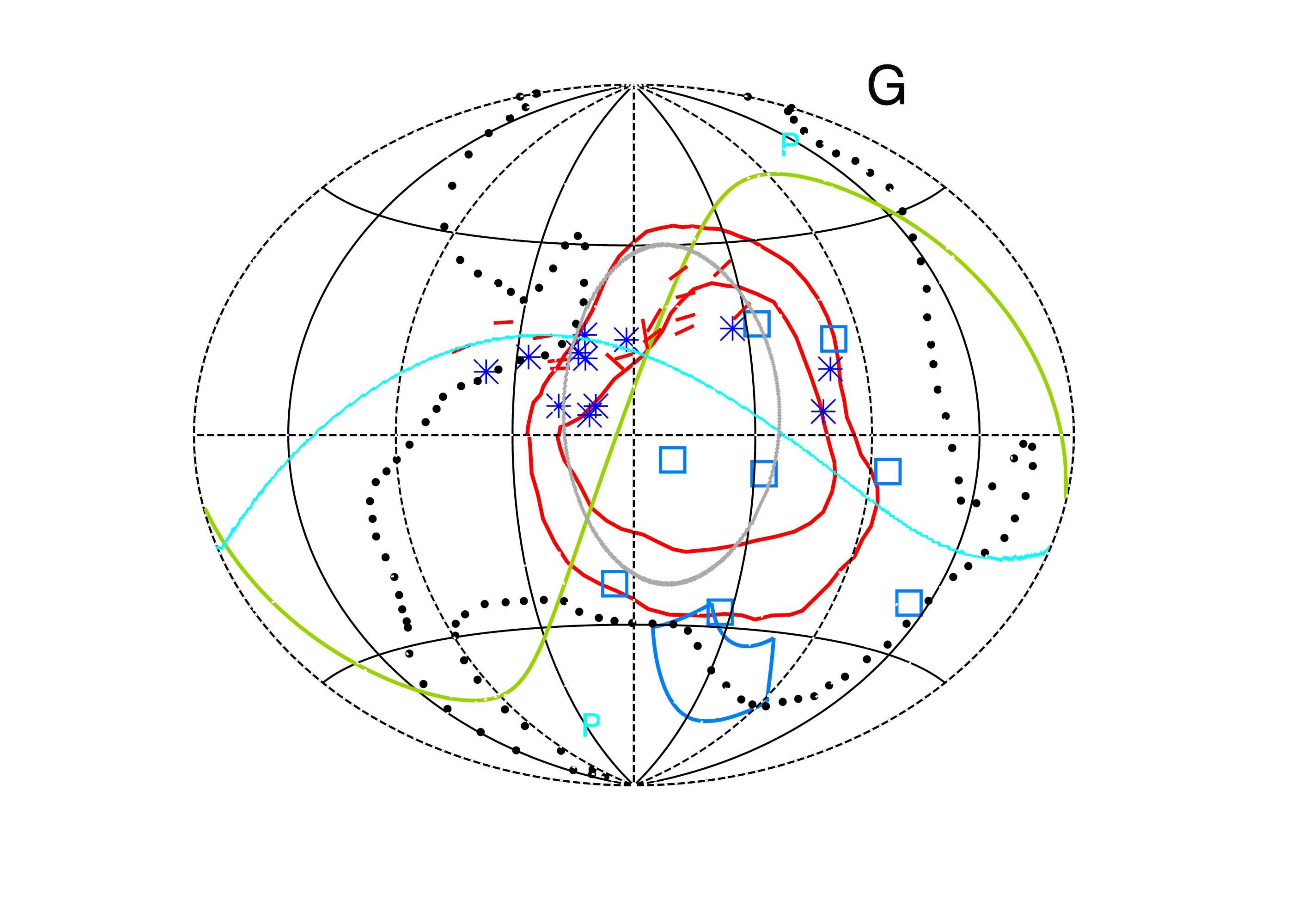}
\caption{The supplementary data discussed in the main text are plotted
in galactic coordinates and centered on the
center of the galaxy.  Phenomena displayed in this figure include 
the region of high heliosheath pressure found
by IBEX \citep[approximated with a gray circle, based on ][]{McComasSchwadron:2014V12pressure},
the LIC (dotted line), polarization band (cyan-colored curved line), ecliptic
plane (curved green line), BICEP2 region (compact structure with blue curved lines),
and outlines of the more distant regions of the Loop I shell 
\citep[semi-concentric red lines, from ][]{Santos:2011}. 
Blue squares show locations of IC59 small scale
cosmic ray sources.  The blue asterisks show locations of the Voyager 1, 2 low frequency emission events.
The red bars show the polarizations of a filament of interstellar dust grains draped over the
heliosphere.  
See the text for more information.}
\label{fig:supl}
\end{figure}

\begin{figure}[!t]   %FIGURE 17
\plotone{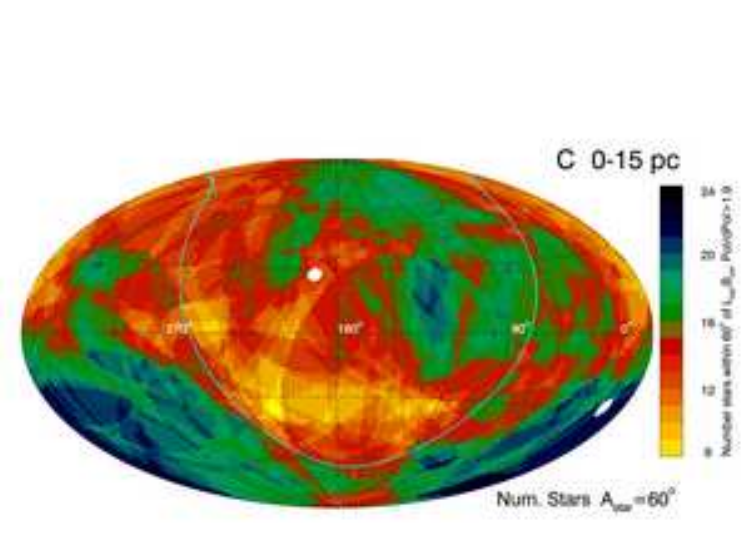}
\caption{The distributions of stars used in this study are shown in
equatorial coordinates, with zero degrees RA at the figure right. 
The figure shows counts of the number of stars within 60\deeg\ of each grid point
\Lrot, \Brot\ for stars with  \Pol/\dPol$>$1.9.  The target star grouping
shows that the magnetic structure identified in this
paper is unrelated to the spatial distribution of the target stars.}
\label{fig:starnum}
\end{figure}


\begin{thebibliography}{104}
\expandafter\ifx\csname natexlab\endcsname\relax\def\natexlab#1{#1}\fi

\bibitem[{Aartsen \& {IceCube Collaboration}(2016)}]{Aartsen_icecube:2016}
Aartsen, R. \& {IceCube Collaboration}. 2016, \apj, 826, 220

\bibitem[{{Abbassi} \& {IceCube Collaboration}(2012)}]{Abbassi:2012icecubeapj}
{Abbassi}, R. \& {IceCube Collaboration}. 2012, \apj, 746, 33

\bibitem[{{Ade} \& the BICEP2~team(2014)}]{bicep2:2014}
{Ade}, P.~A.~R. \& the BICEP2~team. 2014, Physical Review Letters, 112, 241101

\bibitem[{{Alexashov} {et~al.}(2016){Alexashov}, {Katushkina}, {Izmodenov}, \&
  {Akaev}}]{AlexashovKatushkinaIzmodenov:2016dust}
{Alexashov}, D.~B., {Katushkina}, O.~A., {Izmodenov}, V.~V., \& {Akaev}, P.~S.
  2016, \mnras, 458, 2553

\bibitem[{{Altobelli} {et~al.}(2016){Altobelli}, {Postberg}, \&
  et~al.}]{Altobelli:2016sci}
{Altobelli}, N., {Postberg}, F., \& et~al. 2016, Science, 352, 312

\bibitem[{{Andersson} {et~al.}(2015){Andersson}, {Lazarian}, \&
  {Vaillancourt}}]{Andersson:2015araa}
{Andersson}, B.-G., {Lazarian}, A., \& {Vaillancourt}, J.~E. 2015, \araa, 53,
  501

\bibitem[{{Bailey} {et~al.}(2017){Bailey}, {Cotton}, \&
  {Kedziora-Chudczer}}]{BaileyCotton:2017minihippi}
{Bailey}, J., {Cotton}, D.~V., \& {Kedziora-Chudczer}, L. 2017, \mnras, 465,
  1601

\bibitem[{{Bailey} {et~al.}(2015){Bailey}, {Kedziora-Chudczer}, {Cotton},
  {Bott}, {Hough}, \& {Lucas}}]{Bailey:2015hippi}
{Bailey}, J., {Kedziora-Chudczer}, L., {Cotton}, D.~V., {Bott}, K., {Hough},
  J.~H., \& {Lucas}, P.~W. 2015, \mnras, 449, 3064

\bibitem[{{Bailey} {et~al.}(2010){Bailey}, {Lucas}, \&
  {Hough}}]{planetpol:2010}
{Bailey}, J., {Lucas}, P.~W., \& {Hough}, J.~H. 2010, \mnras, 405, 2570

\bibitem[{{Berdyugin} {et~al.}(2014){Berdyugin}, {Piirola}, \&
  {Teerikorpi}}]{BerdyuginPiirola:2014s1}
{Berdyugin}, A., {Piirola}, V., \& {Teerikorpi}, P. 2014, \aap, 561, A24

\bibitem[{{Bohlin} {et~al.}(1978){Bohlin}, {Savage}, \&
  {Drake}}]{BohlinSavageDrake:1978}
{Bohlin}, R.~C., {Savage}, B.~D., \& {Drake}, J.~F. 1978, \apj, 224, 132

\bibitem[{{Burlaga} {et~al.}(2018){Burlaga}, {Florinski}, \&
  {Ness}}]{BurlagaFlorinskiNess:2018turb}
{Burlaga}, L.~F., {Florinski}, V., \& {Ness}, N.~F. 2018, \apj, 854, 20

\bibitem[{{Burlaga} \& {Ness}(2014)}]{Burlaga:2014ismf}
{Burlaga}, L.~F. \& {Ness}, N.~F. 2014, \apj, 784, 146

\bibitem[{{Burlaga} \& {Ness}(2016)}]{BurlagaNess:2016ismf}
---. 2016, \apj, 829, 136

\bibitem[{{Bzowski} {et~al.}(2017){Bzowski}, {Kubiak}, {Czechowski}, \&
  {Grygorczuk}}]{Bzowski:2017breeze}
{Bzowski}, M., {Kubiak}, M.~A., {Czechowski}, A., \& {Grygorczuk}, J. 2017,
  \apj, 845, 15

\bibitem[{{Bzowski} {et~al.}(2015){Bzowski}, {Swaczyna}, {Kubiak},
  {Sok{\'o}{\l}}, {Fuselier}, {Galli}, {Heirtzler}, {Kucharek}, {Leonard},
  {McComas}, {M{\"o}bius}, {Schwadron}, \& {Wurz}}]{Bzowski:2015isn}
{Bzowski}, M., {Swaczyna}, P., {Kubiak}, M.~A., {Sok{\'o}{\l}}, J.~M.,
  {Fuselier}, S.~A., {Galli}, A., {Heirtzler}, D., {Kucharek}, H., {Leonard},
  T.~W., {McComas}, D.~J., {M{\"o}bius}, E., {Schwadron}, N.~A., \& {Wurz}, P.
  2015, \apjs, 220, 28

\bibitem[{{Clark} {et~al.}(2016){Clark}, {Peek}, {Hill}, \&
  {Putman}}]{ClarkPeek:2016fibfilstriation}
{Clark}, S.~E., {Peek}, J.~E.~G., {Hill}, J.~C., \& {Putman}, M.~E. 2016, in
  IAU Symposium, Vol. 315, From Interstellar Clouds to Star-Forming Galaxies:
  Universal Processes?, ed. P.~{Jablonka}, P.~{Andr{\'e}}, \& F.~{van der Tak},
  E13

\bibitem[{{Cohen}(1995)}]{Cohen:1995sungalplane}
{Cohen}, M. 1995, \apj, 444, 874

\bibitem[{{Cotton} {et~al.}(2017{\natexlab{a}}){Cotton}, {Bailey}, {Howarth},
  {Bott}, {Kedziora-Chudczer}, {Lucas}, \& {Hough}}]{Cotton:2017natregulus}
{Cotton}, D.~V., {Bailey}, J., {Howarth}, I.-D., {Bott}, K.,
  {Kedziora-Chudczer}, L., {Lucas}, P.~W., \& {Hough}, J.~H.
  2017{\natexlab{a}}, Nature Astronomy, 1, 690

\bibitem[{{Cotton} {et~al.}(2016){Cotton}, {Bailey}, {Kedziora-Chudczer},
  {Bott}, {Lucas}, {Hough}, \& {Marshall}}]{Cotton:2016brightS}
{Cotton}, D.~V., {Bailey}, J., {Kedziora-Chudczer}, L., {Bott}, K., {Lucas},
  P.~W., {Hough}, J.~H., \& {Marshall}, J.~P. 2016, \mnras, 455, 1607

\bibitem[{{Cotton} {et~al.}(2017{\natexlab{b}}){Cotton}, {Marshall}, {Bailey},
  {Kedziora-Chudczer}, {Bott}, {Marsden}, \& {Carter}}]{Cotton:2017FGK}
{Cotton}, D.~V., {Marshall}, J.~P., {Bailey}, J., {Kedziora-Chudczer}, L.,
  {Bott}, K., {Marsden}, S.~C., \& {Carter}, B.~D. 2017{\natexlab{b}}, \mnras,
  467, 873

\bibitem[{{Dialynas} {et~al.}(2017){Dialynas}, {Krimigis}, {Mitchell},
  {Decker}, \& {Roelof}}]{DialynasKrimigis:2017nature}
{Dialynas}, K., {Krimigis}, S.~M., {Mitchell}, D.~G., {Decker}, R.~B., \&
  {Roelof}, E.~C. 2017, Nature Astronomy, 1, 0115

\bibitem[{{Dialynas} {et~al.}(2013){Dialynas}, {Krimigis}, {Mitchell},
  {Roelof}, \& {Decker}}]{Dialynas:2013}
{Dialynas}, K., {Krimigis}, S.~M., {Mitchell}, D.~G., {Roelof}, E.~C., \&
  {Decker}, R.~B. 2013, \apj, 778, 40

\bibitem[{{Egger} \& {Aschenbach}(1995)}]{Egger:1995}
{Egger}, R.~J. \& {Aschenbach}, B. 1995, \aap, 294, L25

\bibitem[{{Fosalba} {et~al.}(2002){Fosalba}, {Lazarian}, {Prunet}, \&
  {Tauber}}]{FosalbaLazarian:2002apj}
{Fosalba}, P., {Lazarian}, A., {Prunet}, S., \& {Tauber}, J.~A. 2002, \apj,
  564, 762

\bibitem[{{Frisch} \& {Dwarkadas}(2018)}]{FrischDwarkadas:2018}
{Frisch}, P. \& {Dwarkadas}, V.~V. 2018, ArXiv e-prints

\bibitem[{{Frisch} {et~al.}(2005){Frisch}, {Gr\"un}, \&
  {Hoppe}}]{FrischGruenHoppe:2005}
{Frisch}, P., {Gr\"un}, E., \& {Hoppe}, P. 2005, {Interstellar and Presolar
  Grains in the {G}alaxy and in the {S}olar {S}ystem } (The Solar System and
  Beyond: Ten Years of ISSI: Eds. J. Geiss and B. Hultqvist, ESTEC), 193--196

\bibitem[{{Frisch} {et~al.}(2012){Frisch}, {Andersson}, {Berdyugin}, {Piirola},
  {DeMajistre}, {Funsten}, {Magalhaes}, {Seriacopi}, {McComas}, {Schwadron},
  {Slavin}, \& {Wiktorowicz}}]{Frisch:2012ismf2}
{Frisch}, P.~C., {Andersson}, B.-G., {Berdyugin}, A., {Piirola}, V.,
  {DeMajistre}, R., {Funsten}, H.~O., {Magalhaes}, A.~M., {Seriacopi}, D.~B.,
  {McComas}, D.~J., {Schwadron}, N.~A., {Slavin}, J.~D., \& {Wiktorowicz},
  S.~J. 2012, \apj, 760, 106

\bibitem[{{Frisch} {et~al.}(2015{\natexlab{a}}){Frisch}, {Andersson},
  {Berdyugin}, {Piirola}, {Funsten}, {Magalhaes}, {Seriacopi}, {McComas},
  {Schwadron}, {Slavin}, \& {Wiktorowicz}}]{Frisch:2015fil}
{Frisch}, P.~C., {Andersson}, B.-G., {Berdyugin}, A., {Piirola}, V., {Funsten},
  H.~O., {Magalhaes}, A.~M., {Seriacopi}, D.~B., {McComas}, D.~J., {Schwadron},
  N.~A., {Slavin}, J.~D., \& {Wiktorowicz}, S.~J. 2015{\natexlab{a}}, \apj,
  805, 60(8pp)

\bibitem[{{Frisch} {et~al.}(2015{\natexlab{b}}){Frisch}, {Berdyugin},
  {Piirola}, {Magalhaes}, {Seriacopi}, {Wiktorowicz}, {Andersson}, {Funsten},
  {McComas}, {Schwadron}, {Slavin}, {Hanson}, \& {Fu}}]{Frisch:2015ismf3}
{Frisch}, P.~C., {Berdyugin}, A., {Piirola}, V., {Magalhaes}, A.~M.,
  {Seriacopi}, D.~B., {Wiktorowicz}, S.~J., {Andersson}, B., {Funsten}, H.~O.,
  {McComas}, D.~J., {Schwadron}, N.~A., {Slavin}, J.~D., {Hanson}, A.~J., \&
  {Fu}, C.-W. 2015{\natexlab{b}}, \apj, 814, 112

\bibitem[{{Frisch} {et~al.}(2016){Frisch}, {Berdyugin}, {Piirola}, {Magalhaes},
  {Seriacopi}, {Ferrari}, {Santos}, {Schwadron}, {Funsten}, {McComas}, \&
  {Heiles}}]{Frisch:2016method2}
{Frisch}, P.~C., {Berdyugin}, A.~B., {Piirola}, V., {Magalhaes}, A.~M.,
  {Seriacopi}, D.~B., {Ferrari}, T., {Santos}, F.~P., {Schwadron}, N.~A.,
  {Funsten}, H.~O., {McComas}, D.~J., \& {Heiles}, C.~E. 2016, in Journal of
  Physics Conference Series, Vol. 767, Journal of Physics Conference Series,
  012010

\bibitem[{{Frisch} {et~al.}(1999){Frisch}, {Dorschner}, {Geiss}, {Greenberg},
  {Gr\"un}, {Landgraf}, {Hoppe}, {Jones}, {Kr{\"{a}}tschmer}, {Linde},
  {Morfill}, {Reach}, {Slavin}, {Svestka}, {Witt}, \& {Zank}}]{Frisch:1999}
{Frisch}, P.~C., {Dorschner}, J.~M., {Geiss}, J., {Greenberg}, J.~M., {Gr\"un},
  E., {Landgraf}, M., {Hoppe}, P., {Jones}, A.~P., {Kr{\"{a}}tschmer}, W.,
  {Linde}, T.~J., {Morfill}, G.~E., {Reach}, W., {Slavin}, J.~D., {Svestka},
  J., {Witt}, A.~N., \& {Zank}, G.~P. 1999, \apj, 525, 492

\bibitem[{{Frisch} {et~al.}(2011){Frisch}, Redfield, \&
  Slavin}]{Frisch:2011araa}
{Frisch}, P.~C., Redfield, S., \& Slavin, J. 2011, \araa, 49

\bibitem[{{Frisch} \& {Schwadron}(2014)}]{FrischSchwadron:2014icns}
{Frisch}, P.~C. \& {Schwadron}, N.~A. 2014, in Astronomical Society of the
  Pacific Conference Series, Vol. 484, {Outstanding Problems in Heliophysics:
  From Coronal Heating to the Edge of the Heliosphere}, ed. Q.~{Hu} \& G.~P.
  {Zank}, 42

\bibitem[{{Frisch} \& {Slavin}(2006)}]{FrischSlavin:2006book}
{Frisch}, P.~C. \& {Slavin}, J.~D. 2006, {\rm Short Term Variations in the
  Galactic Environment of the {S}un, in \ { \it Solar Journey: The Significance
  of Our Galactic Environment for the Heliosphere and Earth}, ed. P. C. Frisch}
  (Springer), 133--193

\bibitem[{{Funsten} {et~al.}(2009){Funsten}, {Allegrini}, {Crew}, {DeMajistre},
  {Frisch}, {Fuselier}, {Gruntman}, {Janzen}, {McComas}, {M{\"o}bius},
  {Randol}, {Reisenfeld}, {Roelof}, \& {Schwadron}}]{Funsten:2009sci}
{Funsten}, H.~O., {Allegrini}, F., {Crew}, G.~B., {DeMajistre}, R., {Frisch},
  P.~C., {Fuselier}, S.~A., {Gruntman}, M., {Janzen}, P., {McComas}, D.~J.,
  {M{\"o}bius}, E., {Randol}, B., {Reisenfeld}, D.~B., {Roelof}, E.~C., \&
  {Schwadron}, N.~A. 2009, Science, 326, 964

\bibitem[{{Funsten} {et~al.}(2013){Funsten}, {DeMajistre}, {Frisch},
  {Heerikhuijsen}, {Higdon}, {Janzen}, {Larsen}, {Livadiotis}, {McComas},
  {M{\"o}bius}, {Reese}, {Reisenfeld}, {Schwadron}, \&
  {Zirnstein}}]{Funsten:2013}
{Funsten}, H.~O., {DeMajistre}, R., {Frisch}, P.~C., {Heerikhuijsen}, J.,
  {Higdon}, D.~M., {Janzen}, P., {Larsen}, B., {Livadiotis}, G., {McComas},
  D.~J., {M{\"o}bius}, E., {Reese}, C., {Reisenfeld}, D.~B., {Schwadron},
  N.~A., \& {Zirnstein}, E.~J. 2013, \apj, 776, 30

\bibitem[{{Fuselier} {et~al.}(2009){Fuselier}, {Allegrini}, {Funsten},
  {Ghielmetti}, {Heirtzler}, {Kucharek}, {Lennartsson}, {McComas},
  {M{\"o}bius}, {Moore}, {Petrinec}, {Saul}, {Scheer}, {Schwadron}, \&
  {Wurz}}]{Fuselier:2009sci}
{Fuselier}, S.~A., {Allegrini}, F., {Funsten}, H.~O., {Ghielmetti}, A.~G.,
  {Heirtzler}, D., {Kucharek}, H., {Lennartsson}, O.~W., {McComas}, D.~J.,
  {M{\"o}bius}, E., {Moore}, T.~E., {Petrinec}, S.~M., {Saul}, L.~A., {Scheer},
  J.~A., {Schwadron}, N., \& {Wurz}, P. 2009, Science, 326, 962

\bibitem[{{Gayley} {et~al.}(1997){Gayley}, {Zank}, {Pauls}, {Frisch}, \&
  {Welty}}]{Gayleyetal:1997}
{Gayley}, K.~G., {Zank}, G.~P., {Pauls}, H.~L., {Frisch}, P.~C., \& {Welty},
  D.~E. 1997, \apj, 487, 259

\bibitem[{{Gry} \& {Jenkins}(2014{\natexlab{a}})}]{GryJenkins:2014clic}
{Gry}, C. \& {Jenkins}, E.~B. 2014{\natexlab{a}}, ArXiv e-prints

\bibitem[{{Gry} \& {Jenkins}(2014{\natexlab{b}})}]{GryJenkins:2014}
---. 2014{\natexlab{b}}, \aap, 567, A58

\bibitem[{{Gry} \& {Jenkins}(2017)}]{GryJenkins:2017leo}
---. 2017, \aap, 598, A31

\bibitem[{{Gurnett} {et~al.}(1993){Gurnett}, {Kurth}, {Allendorf}, \&
  {Poynter}}]{GurnettKurthetal:1993}
{Gurnett}, D.~A., {Kurth}, W.~S., {Allendorf}, S.~C., \& {Poynter}, R.~L. 1993,
  Science, 262, 199

\bibitem[{{Gurnett} {et~al.}(2013){Gurnett}, {Kurth}, {Burlaga}, \&
  {Ness}}]{Gurnett:2013sci}
{Gurnett}, D.~A., {Kurth}, W.~S., {Burlaga}, L.~F., \& {Ness}, N.~F. 2013,
  Science, 341, 1489

\bibitem[{{Heger}(1919)}]{Heger:1919}
{Heger}, M.~L. 1919, \pasp, 31, 304

\bibitem[{{Heiles}(1996)}]{Heiles:1996curve}
{Heiles}, C. 1996, \apj, 462, 316

\bibitem[{{Heiles}(2000)}]{Heiles:2000}
---. 2000, \aj, 119, 923

\bibitem[{{Hoang} \& {Lazarian}(2016)}]{HoangLazarian:2016rat}
{Hoang}, T. \& {Lazarian}, A. 2016, \apj, 831, 159

\bibitem[{{Kimura} {et~al.}(2003){Kimura}, {Mann}, \&
  {Jessberger}}]{KimuraMann:2003clicvel}
{Kimura}, H., {Mann}, I., \& {Jessberger}, E.~K. 2003, \apj, 582, 846

\bibitem[{{Krimigis} {et~al.}(2009){Krimigis}, {Mitchell}, {Roelof}, {Hsieh},
  \& {McComas}}]{Krimigis:2009sci}
{Krimigis}, S.~M., {Mitchell}, D.~G., {Roelof}, E.~C., {Hsieh}, K.~C., \&
  {McComas}, D.~J. 2009, Science, 326, 971

\bibitem[{{Kr{\"u}ger} {et~al.}(2015){Kr{\"u}ger}, {Strub}, {Gruen}, \&
  {Sterken}}]{Krueger:2015}
{Kr{\"u}ger}, H., {Strub}, P., {Gruen}, E., \& {Sterken}, V.~J. 2015, \apj, 812

\bibitem[{{Kubiak} {et~al.}(2014){Kubiak}, {Bzowski}, {Sok{\'o}{\l}},
  {Swaczyna}, {Grzedzielski}, {Alexashov}, {Izmodenov}, {Moebius}, {Leonard},
  {Fuselier}, {Wurz}, \& {McComas}}]{KubiakBzowski:2014breeze}
{Kubiak}, M.~A., {Bzowski}, M., {Sok{\'o}{\l}}, J.~M., {Swaczyna}, P.,
  {Grzedzielski}, S., {Alexashov}, D.~B., {Izmodenov}, V.~V., {Moebius}, E.,
  {Leonard}, T., {Fuselier}, S.~A., {Wurz}, P., \& {McComas}, D.~J. 2014, ArXiv
  e-prints

\bibitem[{{Kubiak} {et~al.}(2016){Kubiak}, {Swaczyna}, {Bzowski},
  {Sok{\'o}{\l}}, {Fuselier}, {Galli}, {Heirtzler}, {Kucharek}, {Leonard},
  {McComas}, {M{\"o}bius}, {Park}, {Schwadron}, \& {Wurz}}]{Kubiak:2016breeze}
{Kubiak}, M.~A., {Swaczyna}, P., {Bzowski}, M., {Sok{\'o}{\l}}, J.~M.,
  {Fuselier}, S.~A., {Galli}, A., {Heirtzler}, D., {Kucharek}, H., {Leonard},
  T.~W., {McComas}, D.~J., {M{\"o}bius}, E., {Park}, J., {Schwadron}, N.~A., \&
  {Wurz}, P. 2016, \apjs, 223, 25

\bibitem[{{Kurth} \& {Gurnett}(2003)}]{KurthGurnett:2003pos3khz}
{Kurth}, W.~S. \& {Gurnett}, D.~A. 2003, \jgr, 108, 2

\bibitem[{{Lallement} {et~al.}(1995){Lallement}, {Ferlet}, {Lagrange},
  {Lemoine}, \& {Vidal-Madjar}}]{Lallement_etal_1995}
{Lallement}, R., {Ferlet}, R., {Lagrange}, A.~M., {Lemoine}, M., \&
  {Vidal-Madjar}, A. 1995, \aap, 304, 461

\bibitem[{{Lallement} {et~al.}(2010){Lallement}, {Qu{\'e}merais}, {Koutroumpa},
  {Bertaux}, {Ferron}, {Schmidt}, \& {Lamy}}]{Lallement:2010soho}
{Lallement}, R., {Qu{\'e}merais}, E., {Koutroumpa}, D., {Bertaux}, J.,
  {Ferron}, S., {Schmidt}, W., \& {Lamy}, P. 2010, Twelfth International Solar
  Wind Conference, 1216, 555

\bibitem[{{Landgraf}(2000{\natexlab{a}})}]{Landgraf:2000}
{Landgraf}, M. 2000{\natexlab{a}}, \jgr, 105, 10303

\bibitem[{{Landgraf}(2000{\natexlab{b}})}]{Landgraf_2000}
---. 2000{\natexlab{b}}, \jgr, 105, 10303

\bibitem[{{Landgraf} {et~al.}(1999){Landgraf}, {Augustsson}, {Gr{\"u}n}, \&
  {Gustafson}}]{Landgrafetal:1999}
{Landgraf}, M., {Augustsson}, K., {Gr{\"u}n}, E., \& {Gustafson}, B.~{\AA}.~S.
  1999, Science, 286, 2319

\bibitem[{{Landgraf} {et~al.}(2000){Landgraf}, {Baggaley}, {Gr{\" u}n}, {Kr{\"
  u}ger}, \& {Linkert}}]{Landgrafetal:2000}
{Landgraf}, M., {Baggaley}, W.~J., {Gr{\" u}n}, E., {Kr{\" u}ger}, H., \&
  {Linkert}, G. 2000, \jgr, 105, 10343

\bibitem[{{Lazarian}(2007)}]{Lazarian:2007rev}
{Lazarian}, A. 2007, \jqsrt, 106, 225

\bibitem[{{Leroy}(1993)}]{Leroy:1993lism}
{Leroy}, J.~L. 1993, \aaps, 101, 551

\bibitem[{{Ma} {et~al.}(2013){Ma}, {Matthews}, {Land}, \&
  {Hyde}}]{Ma:2013fluffyinism}
{Ma}, Q., {Matthews}, L.~S., {Land}, V., \& {Hyde}, T.~W. 2013, \apj, 763, 77

\bibitem[{{Magalhaes} {et~al.}(1996){Magalhaes}, {Rodrigues}, {Margoniner},
  {Pereyra}, \& {Heathcote}}]{Magalhaes:1996iagpol}
{Magalhaes}, A.~M., {Rodrigues}, C.~V., {Margoniner}, V.~E., {Pereyra}, A., \&
  {Heathcote}, S. 1996, in Astronomical Society of the Pacific Conference
  Series, Vol.~97, Polarimetry of the Interstellar Medium, ed. {W.~G.~Roberge
  \& D.~C.~B.~Whittet}, 118--+

\bibitem[{{Mann}(2010)}]{Mann:2010araa}
{Mann}, I. 2010, \araa, 48, 173

\bibitem[{{Mann} \& {Czechowski}(2004)}]{MannCzechowski:2004dustdefl}
{Mann}, I. \& {Czechowski}, A. 2004, in AIP Conf. Proc. 719: Physics of the
  Outer Heliosphere, 53--58

\bibitem[{{Marshall} {et~al.}(2016){Marshall}, {Cotton}, {Bott}, {Ertel},
  {Kennedy}, {Wyatt}, {del Burgo}, {Absil}, {Bailey}, \&
  {Kedziora-Chudczer}}]{MarshallCotton:2016hotdustlism}
{Marshall}, J.~P., {Cotton}, D.~V., {Bott}, K., {Ertel}, S., {Kennedy}, G.~M.,
  {Wyatt}, M.~C., {del Burgo}, C., {Absil}, O., {Bailey}, J., \&
  {Kedziora-Chudczer}, L. 2016, \apj, 825, 124

\bibitem[{{Mathewson} \& {Ford}(1970)}]{MathewsonFord:1970}
{Mathewson}, D.~S. \& {Ford}, V.~L. 1970, \memras, 74, 139

\bibitem[{{McComas} {et~al.}(2009){McComas}, {Allegrini}, {Bochsler},
  {Bzowski}, {Christian}, {Crew}, {DeMajistre}, {Fahr}, {Fichtner}, {Frisch},
  {Funsten}, {Fuselier}, {Gloeckler}, {Gruntman}, {Heerikhuisen}, {Izmodenov},
  {Janzen}, {Knappenberger}, {Krimigis}, {Kucharek}, {Lee}, {Livadiotis},
  {Livi}, {MacDowall}, {Mitchell}, {M{\"o}bius}, {Moore}, {Pogorelov},
  {Reisenfeld}, {Roelof}, {Saul}, {Schwadron}, {Valek}, {Vanderspek}, {Wurz},
  \& {Zank}}]{McComas:2009sci}
{McComas}, D.~J., {Allegrini}, F., {Bochsler}, P., {Bzowski}, M., {Christian},
  E.~R., {Crew}, G.~B., {DeMajistre}, R., {Fahr}, H., {Fichtner}, H., {Frisch},
  P.~C., {Funsten}, H.~O., {Fuselier}, S.~A., {Gloeckler}, G., {Gruntman}, M.,
  {Heerikhuisen}, J., {Izmodenov}, V., {Janzen}, P., {Knappenberger}, P.,
  {Krimigis}, S., {Kucharek}, H., {Lee}, M., {Livadiotis}, G., {Livi}, S.,
  {MacDowall}, R.~J., {Mitchell}, D., {M{\"o}bius}, E., {Moore}, T.,
  {Pogorelov}, N.~V., {Reisenfeld}, D., {Roelof}, E., {Saul}, L., {Schwadron},
  N.~A., {Valek}, P.~W., {Vanderspek}, R., {Wurz}, P., \& {Zank}, G.~P. 2009,
  Science, 326, 959

\bibitem[{{McComas} \& {Schwadron}(2014)}]{McComasSchwadron:2014V12pressure}
{McComas}, D.~J. \& {Schwadron}, N.~A. 2014, \apj, 795, L17

\bibitem[{{McComas} {et~al.}(2017){McComas}, {Zirnstein}, {Bzowski}, {Dayeh},
  {Funsten}, {Fuselier}, {Janzen}, {Kubiak}, {Kucharek}, {M{\"o}bius},
  {Reisenfeld}, {Schwadron}, {Sok{\'o}{\l}}, {Szalay}, \&
  {Tokumaru}}]{McComas:2017yr7}
{McComas}, D.~J., {Zirnstein}, E.~J., {Bzowski}, M., {Dayeh}, M.~A., {Funsten},
  H.~O., {Fuselier}, S.~A., {Janzen}, P.~H., {Kubiak}, M.~A., {Kucharek}, H.,
  {M{\"o}bius}, E., {Reisenfeld}, D.~B., {Schwadron}, N.~A., {Sok{\'o}{\l}},
  J.~M., {Szalay}, J.~R., \& {Tokumaru}, M. 2017, \apjs, 229, 41

\bibitem[{{Mitchell} {et~al.}(2004){Mitchell}, {Cairns}, \&
  {Robinson}}]{MitchellCairnsetal:2004}
{Mitchell}, J.~J., {Cairns}, I.~H., \& {Robinson}, P.~A. 2004, Journal of
  Geophysical Research (Space Physics), 109, 6108

\bibitem[{{Naghizadeh-Khouei} \& {Clarke}(1993)}]{NaghizadehClarke:1993stat}
{Naghizadeh-Khouei}, J. \& {Clarke}, D. 1993, \aap, 274, 968

\bibitem[{{Pereyra} \& {Magalh{\~a}es}(2007)}]{PereyraMagalhaes:2007vela}
{Pereyra}, A. \& {Magalh{\~a}es}, A.~M. 2007, \apj, 662, 1014

\bibitem[{{Perryman}(1997)}]{Perryman_etal_1997}
{Perryman}, M.~A.~C. 1997, \aap, 323, L49

\bibitem[{{Piirola}(1973)}]{Piirola:1973}
{Piirola}, V. 1973, \aap, 27, 383

\bibitem[{{Piirola}(1977)}]{Piirola:1977}
---. 1977, \aaps, 30, 213

\bibitem[{{Piirola} {et~al.}(2014){Piirola}, {Berdyugin}, \&
  {Berdyugina}}]{Piirola:2014spie}
{Piirola}, V., {Berdyugin}, A., \& {Berdyugina}, S. 2014, in Society of
  Photo-Optical Instrumentation Engineers (SPIE) Conference Series, Vol. 9147,
  Society of Photo-Optical Instrumentation Engineers (SPIE) Conference Series,
  8

\bibitem[{{Plaszczynski} {et~al.}(2014){Plaszczynski}, {Montier}, {Levrier}, \&
  {Tristram}}]{Plazwiktor}
{Plaszczynski}, S., {Montier}, L., {Levrier}, F., \& {Tristram}, M. 2014,
  \mnras, 462, 4048=4056

\bibitem[{{Pogorelov} {et~al.}(2011){Pogorelov}, {Heerikhuisen}, {Zank},
  {Borovikov}, {Frisch}, \& {McComas}}]{PogorelovFrisch:2011}
{Pogorelov}, N.~V., {Heerikhuisen}, J., {Zank}, G.~P., {Borovikov}, S.~N.,
  {Frisch}, P.~C., \& {McComas}, D.~J. 2011, \apj, 742, 104

\bibitem[{{Redfield} \& {Linsky}(2008{\natexlab{a}})}]{RLIV:2008}
{Redfield}, S. \& {Linsky}, J.~L. 2008{\natexlab{a}}, \apj, 673, 283

\bibitem[{{Redfield} \& {Linsky}(2008{\natexlab{b}})}]{RLIV}
---. 2008{\natexlab{b}}, \apj, 673, 283

\bibitem[{{Santos} {et~al.}(2011){Santos}, {Corradi}, \& {Reis}}]{Santos:2011}
{Santos}, F.~P., {Corradi}, W., \& {Reis}, W. 2011, \apj, 728, 104

\bibitem[{{Schwadron} {et~al.}(2014{\natexlab{a}}){Schwadron}, {Adams},
  {Christian}, {Desiati}, {Frisch}, {Funsten}, {Jokipii}, {McComas}, {Moebius},
  \& {Zank}}]{Schwadron:2014sci}
{Schwadron}, N.~A., {Adams}, F.~C., {Christian}, E.~R., {Desiati}, P.,
  {Frisch}, P., {Funsten}, H.~O., {Jokipii}, J.~R., {McComas}, D.~J.,
  {Moebius}, E., \& {Zank}, G.~P. 2014{\natexlab{a}}, Science, 343, 988

\bibitem[{{Schwadron} {et~al.}(2009){Schwadron}, {Bzowski}, {Crew}, {Gruntman},
  {Fahr}, {Fichtner}, {Frisch}, {Funsten}, {Fuselier}, {Heerikhuisen},
  {Izmodenov}, {Kucharek}, {Lee}, {Livadiotis}, {McComas}, {Moebius}, {Moore},
  {Mukherjee}, {Pogorelov}, {Prested}, {Reisenfeld}, {Roelof}, \&
  {Zank}}]{Schwadron:2009sci}
{Schwadron}, N.~A., {Bzowski}, M., {Crew}, G.~B., {Gruntman}, M., {Fahr}, H.,
  {Fichtner}, H., {Frisch}, P.~C., {Funsten}, H.~O., {Fuselier}, S.,
  {Heerikhuisen}, J., {Izmodenov}, V., {Kucharek}, H., {Lee}, M., {Livadiotis},
  G., {McComas}, D.~J., {Moebius}, E., {Moore}, T., {Mukherjee}, J.,
  {Pogorelov}, N.~V., {Prested}, C., {Reisenfeld}, D., {Roelof}, E., \& {Zank},
  G.~P. 2009, Science, 326, 966

\bibitem[{{Schwadron} \& {McComas}(2017)}]{SchwadronMcComas:2017bv}
{Schwadron}, N.~A. \& {McComas}, D.~J. 2017, \apjl, 135, 135

\bibitem[{{Schwadron} {et~al.}(2016){Schwadron}, {M{\"o}bius}, {McComas},
  {Bochsler}, {Bzowski}, {Fuselier}, {Livadiotis}, {Frisch}, {M{\"u}ller},
  {Heirtzler}, {Kucharek}, \& {Lee}}]{Schwadron:2016oxy}
{Schwadron}, N.~A., {M{\"o}bius}, E., {McComas}, D.~J., {Bochsler}, P.,
  {Bzowski}, M., {Fuselier}, S.~A., {Livadiotis}, G., {Frisch}, P.,
  {M{\"u}ller}, H.-R., {Heirtzler}, D., {Kucharek}, H., \& {Lee}, M.~A. 2016,
  \apj, 828, 81

\bibitem[{{Schwadron} {et~al.}(2014{\natexlab{b}}){Schwadron}, {Moebius}, \&
  {Fuselier}}]{Schwadron:2014sep2}
{Schwadron}, N.~A., {Moebius}, E., \& {Fuselier}, S.~A. e.~a.
  2014{\natexlab{b}}, \apj, in press

\bibitem[{{Schwadron} {et~al.}(2015{\natexlab{a}}){Schwadron}, {Moebius},
  {Leonard}, {Fuselier}, {McComas}, {Heirtzler}, {Kucharek}, {Rahmanifard},
  {Bzowski}, {Kubiak}, {Sokol}, {Swaczyna}, \& {Frisch}}]{Schwadron:2015He}
{Schwadron}, N.~A., {Moebius}, E., {Leonard}, T., {Fuselier}, S.~A., {McComas},
  D.~J., {Heirtzler}, D., {Kucharek}, H., {Rahmanifard}, F., {Bzowski}, M.,
  {Kubiak}, M.~A., {Sokol}, J., {Swaczyna}, P., \& {Frisch}, P.
  2015{\natexlab{a}}, \apjs, 220, 25

\bibitem[{{Schwadron} {et~al.}(2015{\natexlab{b}}){Schwadron}, {Richardson},
  {Burlaga}, {McComas}, \& {Moebius}}]{Schwadron:2015triBismf}
{Schwadron}, N.~A., {Richardson}, J.~D., {Burlaga}, L.~F., {McComas}, D.~J., \&
  {Moebius}, E. 2015{\natexlab{b}}, \apjl, 813, L20

\bibitem[{{Serkowski} {et~al.}(1975){Serkowski}, {Mathewson}, \&
  {Ford}}]{Serkowski:1975ebvpol}
{Serkowski}, K., {Mathewson}, D.~S., \& {Ford}, V.~L. 1975, \apj, 196, 261

\bibitem[{{Slavin} \& {Frisch}(2008)}]{SlavinFrisch:2008}
{Slavin}, J.~D. \& {Frisch}, P.~C. 2008, \aap, 491, 53

\bibitem[{{Slavin} {et~al.}(2012){Slavin}, {Frisch}, {M{\"u}ller},
  {Heerikhuisen}, {Pogorelov}, {Reach}, \& {Zank}}]{SlavinFrisch:2012}
{Slavin}, J.~D., {Frisch}, P.~C., {M{\"u}ller}, H.-R., {Heerikhuisen}, J.,
  {Pogorelov}, N.~V., {Reach}, W.~T., \& {Zank}, G. 2012, \apj, 760, 46

\bibitem[{{Sterken} {et~al.}(2013){Sterken}, {Altobelli}, {Kempf},
  {Kr{\"u}ger}, {Srama}, {Strub}, \& {Gr{\"u}n}}]{Sterken:2013filter}
{Sterken}, V.~J., {Altobelli}, N., {Kempf}, S., {Kr{\"u}ger}, H., {Srama}, R.,
  {Strub}, P., \& {Gr{\"u}n}, E. 2013, \aap, 552, A130

\bibitem[{{Sterken} {et~al.}(2015){Sterken}, {Strub}, {Kr{\"u}ger}, {von
  Steiger}, \& {Frisch}}]{Sterken:2015}
{Sterken}, V.~J., {Strub}, P., {Kr{\"u}ger}, H., {von Steiger}, R., \&
  {Frisch}, P. 2015, \apj, 812

\bibitem[{{Stone} \& {Cummings}(2013)}]{StoneCummings:2013sci}
{Stone}, E.~C. \& {Cummings}, A.~C. e.~a. 2013, Science, 341, 150

\bibitem[{Tinbergen(1982)}]{Tinbergen:1982}
Tinbergen, J. 1982, \aap, 105, 53

\bibitem[{{Wiktorowicz} \& {Matthews}(2008)}]{WiktorowiczMatthews:2008polish2}
{Wiktorowicz}, S.~J. \& {Matthews}, K. 2008, \pasp, 120, 1282

\bibitem[{{Wiktorowicz} \& {Nofi}(2015)}]{Wiktorowicz:2015vesta}
{Wiktorowicz}, S.~J. \& {Nofi}, L.~A. 2015, \apjl, 800, L1

\bibitem[{{Wood} {et~al.}(2000){Wood}, {Linsky}, \& {Zank}}]{Wood:20036Oph}
{Wood}, B.~E., {Linsky}, J.~L., \& {Zank}, G.~P. 2000, \apj, 537, 304

\bibitem[{{Wood} {et~al.}(2017){Wood}, {M{\"u}ller}, \&
  {Witte}}]{WoodMuellerWitte:2017hebreeze}
{Wood}, B.~E., {M{\"u}ller}, H.-R., \& {Witte}, M. 2017, \apj, 851, 35

\bibitem[{{Wood} {et~al.}(2005){Wood}, {Redfield}, {Linsky}, {M{\"u}ller}, \&
  {Zank}}]{Wood_etal_2005}
{Wood}, B.~E., {Redfield}, S., {Linsky}, J.~L., {M{\"u}ller}, H.-R., \& {Zank},
  G.~P. 2005, \apjs, 159, 118

\bibitem[{{Yu}(1974)}]{Yu:1974}
{Yu}, G. 1974, \apj, 194, 187

\bibitem[{{Zirnstein} {et~al.}(2016){Zirnstein}, {Heerikhuisen}, {Funsten},
  {Livadiotis}, {McComas}, \& {Pogorelov}}]{Zirnstein:2016ismf}
{Zirnstein}, E.~J., {Heerikhuisen}, J., {Funsten}, H.~O., {Livadiotis}, G.,
  {McComas}, D.~J., \& {Pogorelov}, N.~V. 2016, \apjl, 818, L18

\end{thebibliography}
\end{document}